\documentclass[traditabstract]{aa}
\pdfoutput=1

\usepackage{savesym}
\usepackage{amsmath}
\savesymbol{iint}
\usepackage[varg]{txfonts}
\restoresymbol{TXF}{iint}
\usepackage{natbib}
\usepackage{graphicx}
\usepackage{multirow}
\usepackage{comment} 
\usepackage{color}

\bibpunct{(}{)}{;}{a}{}{,}

\def\ie{{\it i.e.,}\,}
\def\eg{{\it e.g.,}\,}

\begin{document}

\title{Three-dimensional simulations of core-collapse supernovae: from
  shock revival to shock breakout} 
\titlerunning{3D CCSN simulations}

\author{A. Wongwathanarat 
        \thanks{RIKEN, Astrophysical Big Bang Laboratory, 2-1 Hirosawa,
                Wako, Saitama 351-0198, Japan} \and
        E. M\"uller \and
        H.-Th. Janka}
\authorrunning{A. Wongwathanarat et al.}

\institute{Max-Planck Institut f\"{u}r Astrophysik,
  Karl-Schwarzschild-Stra\ss e 1, D-85748 Garching, Germany}

\abstract{ We present three-dimensional hydrodynamic simulations of
  the evolution of core-collapse supernovae from blast-wave initiation
  by the neutrino-driven mechanism to shock breakout from the stellar
  surface, using an axis-free Yin-Yang grid and considering two
  15\,$M_\odot$ red supergiants (RSG) and two blue supergiants (BSG)
  of 15\,$M_\odot$ and 20\,$M_\odot$. We demonstrate that the
  metal-rich ejecta in homologous expansion still carry fingerprints
  of asymmetries at the beginning of the explosion, but the final
  metal distribution is massively affected by the detailed progenitor
  structure. The most extended and fastest metal fingers and clumps
  are correlated with the biggest and fastest-rising plumes of
  neutrino-heated matter, because these plumes most effectively seed
  the growth of Rayleigh-Taylor (RT) instabilities at the C+O/He and
  He/H composition-shell interfaces after the passage of the SN shock.
  The extent of radial mixing, global asymmetry of the metal-rich
  ejecta, RT-induced fragmentation of initial plumes to smaller-scale
  fingers, and maximal Ni and minimal H velocities do not only depend
  on the initial asphericity and explosion energy (which determine the
  shock and initial Ni velocities) but also on the density profiles
  and widths of C+O core and He shell and on the density gradient at
  the He/H transition, which lead to unsteady shock propagation and
  the formation of reverse shocks. Both RSG explosions retain a great
  global metal asymmetry with pronounced clumpiness and substructure,
  deep penetration of Ni fingers into the H-envelope (with maximum
  velocities of 4000--5000\,km\,s$^{-1}$ for an explosion energy
  around 1.5 bethe) and efficient inward H-mixing.  While the
  15\,$M_\odot$ BSG shares these properties (maximum Ni speeds up to
  $\sim$3500\,km\,s$^{-1}$), the 20\,$M_\odot$ BSG develops a much
  more roundish geometry without pronounced metal fingers (maximum Ni
  velocities only $\sim$2200\,km\,s$^{-1}$) because of reverse-shock
  deceleration and insufficient time for strong RT growth and
  fragmentation at the He/H interface.}


\maketitle


\section{Introduction}
Current state-of-the-art simulations of neutrino-driven CCSN predict
that hydrodynamic instabilities play a crucial role in the explosion
mechanism that leads to the ejection of the stellar mantle and
envelope of exploding massive stars \citep[see, \eg][for a recent
  review]{Janka12}.  Convective instability occurs in the region of
net neutrino heating (gain region) behind the stalled shock wave
\citep{Bethe90, Herantetal92, Herantetal94, Burrowsetal95,
  JankaMueller95, JankaMueller96}, and the standing accretion shock
instability \citep[SASI;][]{Blondinetal03, Foglizzo02,
  BlondinMezzacappa06, Ohnishietal06, Foglizzoetal07, Schecketal08}
causes an oscillatory growth of non-radial shock deformation with a
dominance of low-order spherical harmonic modes. The combined effects
of both instabilities give rise to large-scale asymmetries and
non-radial flow in the neutrino-heated post-shock matter.

Once the shock is revived by neutrino heating, it resumes its
propagation through the onion shell-like composition structure of the
progenitor star. It sweeps up matter causing local density inversions
because the shock propagates non-steadily through the stellar envelope
due to progenitor dependent variations of the density gradient. The
blast wave accelerates in gradients steeper than $r^{-3}$ and
decelerates otherwise \citep{Sedov59}. Such variations are
particularly large near the C+O/He and He/H composition interfaces.
These layers are prone to Rayleigh-Taylor (RT) instability
\citep{Chevalier76}, which are seeded by the asymmetries generated in
the innermost part of the ejecta, \ie in the O-core, during the first
seconds of the explosion \citep{Kifonidisetal03}. Multi-dimensional
hydrodynamic simulations in the 1990s, triggered by observations of
SN\,1987\,A \citep[for a review, see e.g.,][]{Arnettetal89rev}, showed
that the RT instabilities cause large-scale spatial mixing, whereby
heavy elements are dredged up from the interior of the exploding star
and hydrogen is mixed inwards \citep[e.g.,][]{Fryxelletal91,
  Muelleretal91}.

Until now most studies simulating RT instabilities in SN envelopes
have disregarded the early-time asymmetries generated by
neutrino-driven convection and the SASI during the first second of the
explosion. They relied on explosions that were initiated assuming
spherical symmetry either by a point-like thermal explosion
\citep[\eg][]{Arnettetal89, Fryxelletal91, Muelleretal91,
  Hachisuetal90, Hachisuetal92, Hachisuetal94, YamadaSato90,
  YamadaSato91, HerantBenz91, HerantBenz92, HerantWoosley94,
  Iwamotoetal97, Nagatakietal98, Kaneetal00}, by piston-driven
explosions \citep[\eg][]{Hungerfordetal03, Hungerfordetal05,
  Joggerstetal09, Joggerstetal10a, Joggerstetal10b}, or by aspherical
injection of kinetic and thermal energy at a chosen location
\citep{Couchetal09, Couchetal11, Onoetal13}. An alternative method was
used by \citet{Ellingeretal12, Ellingeretal13}, who initiated 1D
explosions by means of one-dimensional (1D) Lagrangian simulations
with an approximate grey neutrino transport scheme. The 1D "initial"
data were mapped to a multi-dimensional grid after some time, \eg when
the SN shock left the iron core \citep{Ellingeretal12,
  Ellingeretal13}, or when it had already propagated further into the
envelope of the progenitor star \citep[e.g.,][]{Arnettetal89}. In
order to trigger the growth of the RT instabilities small asymmetries
were either imposed "by hand" and/or were intrinsically present in the
grid-based or smoothed particle hydrodynamics (SPH) simulations.

The first attempt to consistently evolve the asymmetries created by
neutrino-driven convection and the SASI during the first second of the
explosion until shock breakout from the progenitor's surface was made
by \citet{Kifonidisetal00} and in a subsequent work by
\citet{Kifonidisetal03}. These authors combined the ``early-time'' ($t
\lesssim 1$\,s) and ``late-time'' ($t \gtrsim 1$\,s) evolution by
performing two-dimensional (2D) simulations that were split into these
two phases. The early-time evolution, \ie the onset of the explosion,
was simulated with all the physics (EoS, neutrino matter interactions,
self-gravity) necessary to properly capture the development of the
early-time asymmetries.  \citet{Kifonidisetal03} stopped their
calculations of the explosion phase at 885\,ms after core bounce when
the explosion energy had eventually saturated. The late-time
evolution, \ie the phase after shock revival and saturation of the
explosion energy, was simulated on a much larger computational domain
at higher spatial resolution using adaptive mesh refinement (AMR), but
neglecting the effects of neutrinos and self-gravity.

\citet{Kifonidisetal03} found that the development of the RT
instabilities differs greatly from that seen in previous simulations
relying on 1D explosion models that ignored the presence of
asymmetries associated with the explosion mechanism.
\citet{Kifonidisetal06} improved the explosion models by replacing the
neutrino light-bulb scheme used in \citet{Kifonidisetal03} with a
ray-by-ray gray neutrino transport approximation \citep[see][for
  details]{Schecketal06}. With this improvement the models exploded
both later and with larger-scale asymmetries at the onset of the
explosions than in \citet{Kifonidisetal03}, leading to dramatic
differences in the development of subsequent mixing instabilities. The
simulations showed that the peak velocity of iron-group elements can
reach $\sim$3300\,km\,s$^{-1}$ in cases where a low-$l$ mode
instability is able to grow during the first second of the
explosion. Such a mode was not observed in the models of
\citet{Kifonidisetal03} since the explosions developed too quickly
resulting only in high-$l$ mode asymmetries.  Using the same initial
model as \citet{Kifonidisetal06}, \citet{Gawryszczaketal10} followed
the evolution in 2D until the ejecta were expanding homologously after
about 7\,days. They observed a strong SASI-induced lateral expansion
of matter away from the equator towards the polar regions, which
affects the evolution from minutes to hours after the onset of the
explosion.

\citet{Hammeretal10} (henceforth referred to as HJM10) studied the
effects of dimensionality on the evolution of the ejecta performing a
set of 2D simulations and a three-dimensional (3D) simulation using a
3D neutrino-driven explosion model calculated by \citet{Scheck07} as
initial data.  They followed the evolution until the time of shock
breakout from the progenitor. Their simulations demonstrated that the
results obtained with 2D axisymmetric models differ crucially from
those of 3D models without any symmetry constraint. Because clumps of
heavy elements have toroidal topology in 2D and bubble-sphere topology
in 3D, they experience less drag in 3D models than in 2D ones.
Consequently, they can retain higher velocities in the 3D model, and
are able to reach the hydrogen envelope before the formation of the
dense decelerating helium ``wall'' that builds up as a consequence of
the non-steady shock propagation. HJM10 also found that inward radial
mixing of hydrogen into the (former) metal core is more efficient in
3D than in 2D. This result agrees with that of earlier work by
\citet{Kaneetal00}, who investigated differences in the growth rate of
a single-mode perturbation at composition interfaces in 2D and 3D
models.

In contrast, \citet{Joggerstetal10b} found no qualitative differences
of the mixing efficiency between 2D and 3D models. They performed a
set of 2D and 3D simulations starting from 1D piston-driven
explosions. They argued that instabilities do grow faster in 3D than
in 2D initially, but this effect is compensated because mixing ceases
earlier in 3D simulations. Interactions between small-scale RT fingers
cause the flow to become turbulent in 3D models, resulting in a
decrease of the local Atwood number and consequently in a reduced
growth of the instabilities. This is different from the 3D simulation
of HJM10 where asymmetries of low-order spherical harmonics modes
dominate, and the resulting large-scale RT fingers do not interact
with each other. In addition, \citet{Joggerstetal10b} concluded that
the inverse cascade by which small-scale RT fingers merge into
larger-scale ones should be truncated before the wavelengths of the
instabilities become large enough to produce the large-scale
asymmetries observed in SN remnants. Therefore, the observed
large-scale asymmetries must be a result of the explosion mechanism
itself.

When modeling instabilities in CCSN, besides the dimensionality and
the asymmetries created at the onset of the explosion, the SN
progenitor structure is highly relevant, too. The locations where the
stellar layers become RT unstable depend on the progenitor, and the
growth rate of the instability is tightly connected to the progenitor
density structure. \citet{HerantBenz91} observed different
interactions between the RT instabilities occurring at the He/H and
C+O/He interface in two different progenitor stars. In one case, RT
mushrooms grown from the C+O/He interface merged with the mushrooms
grown from the He/H interface. In another progenitor, RT instabilities
gave rise to two distinct sets of RT mushrooms. \citet{HerantBenz92}
pointed out that the growth rate of the instability depends on how
strongly the SN shock accelerates when it crosses a composition
interface, \ie it depends on the steepness of the density gradient
near the respective interfaces. Considering explosions of both red
supergiant (RSG) and blue supergiant (BSG) stars
\citet{Joggerstetal10b} noticed differences between the growth rates
of RT instabilities in these two types of progenitors in their 3D
simulations. They also found that mixing of heavy elements lasts 5 to
10 times longer in a RSG than in a BSG star.

In summary, all these previous studies have demonstrated the
importance of the early-time asymmetries created by the SN mechanism,
the effects of dimensionality, and the influence of the progenitor
structure. 

In the present work, we proceed along
the line of HJM10 including a number of improvements. We
present a set of 3D CCSN simulations considering four different
progenitor stars, for which we compute the evolution
from about 1\,s after bounce until the SN shock breaks out from
the stellar surface hours later. As initial data for
our simulations we use the 3D explosion models presented
in our previous studies (Wongwathanarat et al. 2010b, 2013).
These 3D neutrino-driven explosion models were started from 
initially spherically symmetric progenitor models as provided
by 1D stellar evolution modeling until the onset of stellar
core collapse. Such a choice is still the standard approach in
computational supernovamodeling, although it is an unrealistic
idealization \citep[e.g.,][]{ArnettMeakin11}. In view of the lack
of multi-D progenitor models evolved to the onset of core collapse,
however, it is presently unclear in which aspects and to
how large an extent spherically symmetric progenitor structures
fail to describe the true conditions in dying stars. In
order to seed the growth of nonradial instabilities, we artificially
perturbed the spherical progenitor models by imposing random 
perturbations of the radial velocity on the whole
computational grid at the beginning of our simulations.

Together both explosion and late-time simulations of
\citet{Wongwathanaratetal10b, Wongwathanaratetal13} and this paper,
resepctively, span the evolution of CCSN from shortly ($\sim 15\,$ms)
after core bounce until hours later in full 3D, \ie without any
symmetry restrictions and covering the full $4\pi$ solid angle. It is
the goal of our study to explore how the explosion asymmetries
associated with the explosion mechanism connect to the large-scale
radial mixing and the ejecta asymmetries that are present hours later,
taking into account the dependence on the progenitor properties. We
will demonstrate that the ejecta structure observed by HJM10 can be
reproduced by our improved models simulated on an axis-free Yin-Yang
grid. In addition, we will show that the morphological structure of
iron/nickel-rich ejecta at late times reflects the initial asymmetry
of the neutrino-heated bubble layer rather than being a result of
stochastically growing interactions of secondary RT instabilities at
the progenitor's composition interfaces. We also will demonstrate that
the interaction of the early-time morphological structures with later,
secondary instabilities in the outer SN layers depends sensitively on
the ratio of the shock speed to the expansion velocity of the heavy
elements, which in turn is a sensitive function of the progenitor
density structure.

The paper is organized as follows: In Sect.\,2 we describe the initial
models, numerics and input physics used to perform our simulations.
We compare our approach to that of HJM10 in
Sect.\,\ref{sec:compareHJM10} by discussing the results of two 3D
simulations which use the same initial data as the one performed by
HJM10.  In Sect.\,\ref{sec:1dmodels} we present the results of a
linear stability analysis, which we conducted to obtain RT growth
factors for our initial models.  We discuss the results of our 3D
simulations in Sect.\,\ref{sec:3Dresults}. Finally, we summarize our
findings and consider possible implications in
Sect.\,\ref{sec:conclusions}.


\section{Models, numerics, and input physics}
\label{sec:methods}

\subsection{Models}
\label{subsec:models}

The simulations presented here are initialized from 3D neutrino-driven
explosion models (see Table\,\ref{tab:expmod}) discussed in our
previous studies \citep{Wongwathanaratetal10b, Wongwathanaratetal13},
which covered the evolution from 11--15\,ms to 1.1--1.4\,s after core
bounce. The explosions were initiated imposing suitable values for
neutrino luminosities (and mean energies) at time-dependent neutrino
optical depths of $\sim$10--1000. Neutrino transport and
neutrino-matter interactions were treated by the ray-by-ray gray
neutrino transport approximation of \citet{Schecketal06} including a
slight modification of the boundary condition for the mean neutrino
energies, which we chose to be a fixed multiple of the time-evolving
gas temperature in the innermost radial grid zone
\citep{Uglianoetal12}. At the end of the explosion simulations the SN
shock had reached a radius of $10^9$ to 2$\times10^9$\,cm, \ie it
still resided inside the C+O core of the progenitor star.

\begin{table}
\caption{Some properties of the explosion models used as input for our
  3D simulations (see text for details).}
\centering
\renewcommand{\arraystretch}{1.4}
\setlength{\tabcolsep}{.2em}
\begin{tabular}{cccccccc}
\hline
\hline
   explosion & \multicolumn{4}{c}{progenitor\phantom{M}} 
             & $t_\mathrm{exp}$ & $t_\mathrm{map}$ 
             & $E_\mathrm{map}$ \\ 
   model     & name & type & mass [$M_\odot$] & $R_\ast [10^6$\,km] 
             & [ms] & [s] & [B] \\
\hline

 W15-1 & \multirow{2}{*}{W15} & \multirow{2}{*}{RSG} & \multirow{2}{*}{15} 
       & \multirow{2}{*}{339} & 246 & \multirow{2}{*}{1.3} & 1.12 \\ 
 W15-2 &                      &                      &                     
       & & 248 &              &        1.13 \\
 L15-1 & \multirow{2}{*}{L15} & \multirow{2}{*}{RSG} & \multirow{2}{*}{15} 
       & \multirow{2}{*}{434} & 422 & \multirow{2}{*}{1.4} & 1.13 \\
 L15-2 &                      &                      &       
       & & 382 &              &        1.74 \\
 N20-4 & N20                  & BSG                  & 20                  
       & 33.8 & 334 & 1.3      &        1.35 \\
 B15-1 & \multirow{2}{*}{B15} & \multirow{2}{*}{BSG} & \multirow{2}{*}{15} 
       & \multirow{2}{*}{39.0} & 164 & \multirow{2}{*}{1.1} & 1.25 \\
 B15-3 &                      &                      &
       & & 175 &              &        1.04 \\
\hline
\end{tabular}
\renewcommand{\arraystretch}{1}
\label{tab:expmod}
\end{table}

%
\begin{figure*}
\resizebox{\hsize}{!}{\includegraphics*{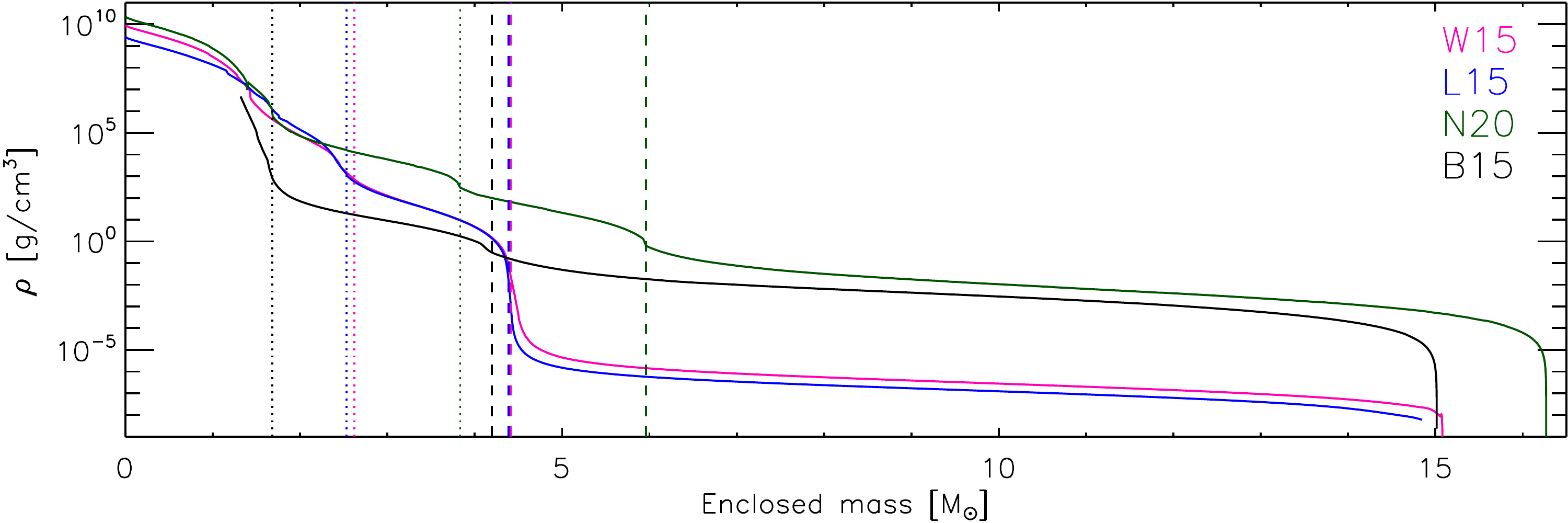}}\\
\caption{Density as a function of enclosed mass for all four
  considered progenitor stars at the onset of core collapse. The
  vertical dotted and dashed lines mark the locations of the C+O/He and
  He/H composition interfaces,respectively, for each progenitor star.}
\label{fig:rhom-progenitor}
\end{figure*}
%

%
\begin{figure*}
\resizebox{\hsize}{!}{\includegraphics*{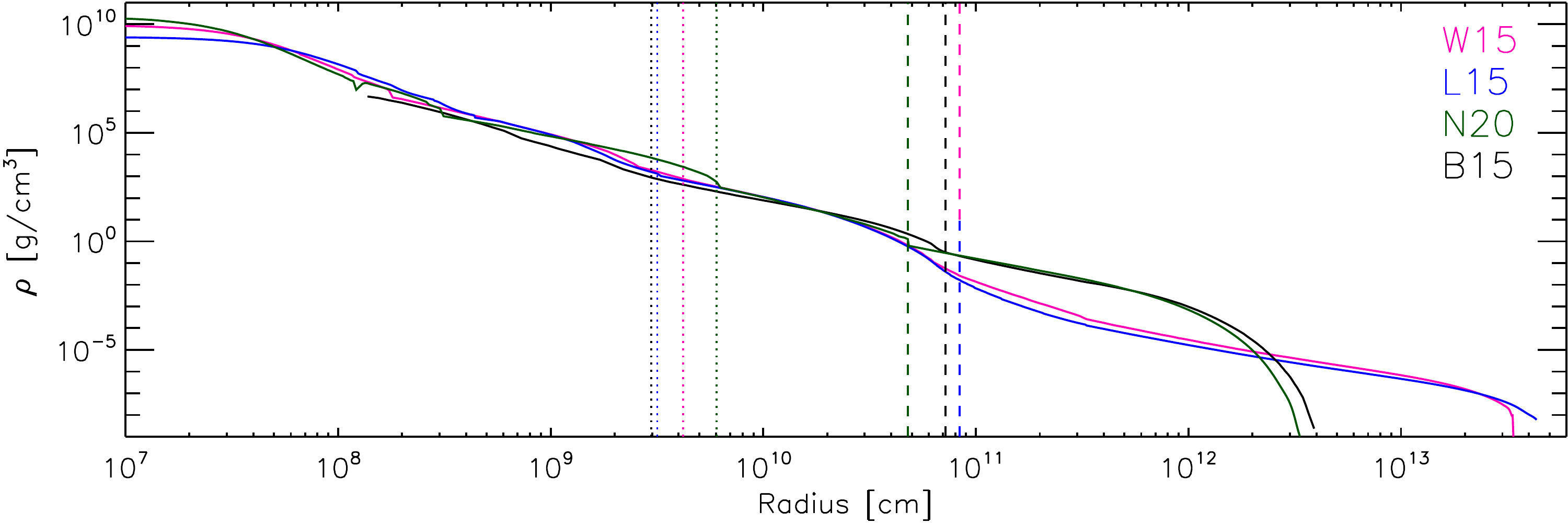}}\\
\caption{Same as Fig.\,\ref{fig:rhom-progenitor}, but as a function of
  radius.}
\label{fig:rhor-progenitor}
\end{figure*}
%

In the present study, we consider only a subset of the explosion
models of \citet{Wongwathanaratetal13}, namely models W15-1, W15-2,
L15-1, L15-2, N20-4, B15-1, and B15-3 (see Tab.\,\ref{tab:expmod}),
where the first three characters of the model name indicate the
respective progenitor star. W15 denotes the 15\,$M_\odot$ RSG model
s15s7b2 of \citet{WoosleyWeaver95}, L15 a 15\,$M_\odot$ RSG model
evolved by \citet{Limongi00}, N20 a 20\,$M_\odot$ BSG progenitor model
for SN\,1987A by \citet{ShigeyamaNomoto90}, and B15 a corresponding
15\,$M_\odot$ BSG model by \citet{Woosleyetal88}.  We note that the
mass of model N20 is reduced from a main-sequence value of
$20\,M_\odot$ to $16.3\,M_\odot$ at the onset of core collapse due to
mass loss.

Figs.\,\ref{fig:rhom-progenitor} and \ref{fig:rhor-progenitor} show
the density as a function of enclosed mass and radius, respectively,
for all four progenitors at the onset of core collapse. We define the
locations of the C+O/He and He/H composition interfaces as those
positions at the bottom of the He and H layers of the star where the
He and H mass fractions drop below half of their maximum values in the
respective layer. The radial coordinates of the C+O/He and He/H
interfaces are denoted as $R_\mathrm{C+O/He}$ and $R_\mathrm{He/H}$,
respectively.

The density profiles of the two RSG progenitors W15 and L15 are
overall quite similar, but the density drops more steeply at the He/H
composition interface in model L15 and the hydrogen envelope of this
model is more dilute, \ie its envelope is more extended than that of
model W15.  The density of the two BSG models B15 and N20 decreases
much less steeply at the He/H interface than in the RSG models.  Model
N20 has the largest densities inside the C+O core which is also the
most massive C+O core ($\approx 3.8\,M_\odot$) of all models, while
model B15 possesses the least massive C+O core ($\approx
1.7\,M_\odot$).

In order to simulate the evolution of the SN explosion well beyond the
epoch of shock revival we mapped the data of the seven 3D explosion
models described above onto a larger computational grid at time
$t_\mathrm{map}$ (see Table\,\ref{tab:expmod}). Defining the
(time-dependent) explosion energy of a model as the sum of the total
(\ie internal plus kinetic plus gravitational) energy of all grid
cells where this energy is positive, the explosion energy at the time
of mapping $E_\mathrm{map}$ ranges from 1.12\,B to 1.74\,B (where 1B$
\equiv 10^{51}$\,erg; see Table\,\ref{tab:expmod}).  The explosion
time $t_{\mathrm{exp}}$, also provided in Table\,\ref{tab:expmod}, is
defined as the moment when the explosion energy exceeds a value of
$10^{48}$\,erg, which roughly coincides with the time when the average
shock radius becomes larger than 400--500\,km.

Because the hydrodynamic timescale varies strongly through the
exploding star we also move the inner grid boundary (located close to
a radius of 15\,km in models W15, N20 and B15, and 25\,km in model L15
at time $t_\textrm{map}$; see Table\,1 in \citet{Wongwathanaratetal13})
to a larger radius of $R_\textrm{ib}=500$\,km at the time of
mapping. This relaxes the CFL time step constraint imposed by the high
sound speed in the vicinity of the proto-neutron star. Moreover,
during the course of the simulations we moved the inner grid boundary
repeatedly to larger radii, thereby further relaxing the CFL
condition. This procedure (see Sect.\,\ref{subsec:numerics}) allowed
us to complete the otherwise computationally too expensive simulations
with a reasonable amount of computing time.

For comparison with the work of HJM10 we performed two additional 3D
simulations where the numerical setup differed slightly from the
standard one described above (see Sect.\,\ref{subsec:H15setup}).

\subsection{Numerics}
\label{subsec:numerics}

We performed our simulations with {\sc Prometheus}, an explicit,
finite-volume Eulerian multi-fluid hydrodynamics code
\citep{Fryxelletal91, Muelleretal91, Muelleretal91b}. It solves the
multi-dimensional Newtonian hydrodynamic equations using dimensional
splitting \citep{Strang68}, piecewise parabolic reconstruction
\citep[PPM;][]{ColellaWoodward84}, a Riemann solver for real gases
\citep{CollelaGlaz85}, and the consistent multi-fluid advection scheme
\citep[CMA;][]{PlewaMueller99}. To prevent odd-even decoupling inside
grid zones with strong grid-aligned shocks \citep{Quirk94}, we used
the AUSM+ fluxes of \citet{Liou96} instead of the fluxes computed with
the Riemann solver.  We further utilized the ``Yin-Yang'' overlapping
grid technique \citep{KageyamaSato04, Wongwathanaratetal10} instead of
a spherical polar (latitude-longitude) grid to cover the computational
domain, because it alleviates the CFL condition resulting from the
coordinate singularity of the polar grid at the poles.  In this way we
could use in our simulations time steps that were several ten times
larger than for the case of a spherical polar computational grid with
similar lateral resolution. The Yin-Yang grid also prevents numerical
artifacts which might arise near the polar axis, where a boundary
condition has to be imposed when a spherical polar grid is used.

Newtonian self-gravity was taken into account by solving the integral
form of Poisson's equation with a multipole expansion method as
described in \citet{MuellerSteinmetz95}. In addition, a point mass was
placed at the coordinate origin to account for the (monopole part of
the) gravity of the proto-neutron star and of the matter excised
successively from our computational domain.

The computational grid covers the full $4\pi$ solid angle. Initially,
it extends in radius from $R_\mathrm{ib}$=500\,km to the stellar
radius $R_\ast$ (see Table\,\ref{tab:expmod}). The radial grid is
logarithmically spaced and initially has a resolution of 5\,km at the
inner grid boundary. $R_\mathrm{ib}$ is fixed during the first 2\,s of
the simulations and thereafter successively shifted to larger radii,
removing the respective innermost radial grid zone whenever
$R_\mathrm{ib}$ becomes smaller than 2\% of the minimum radius of the
(aspherical) SN shock. This reduces the number of radial grid zones in
the course of a simulation, thereby speeding it up. We cease the
movement of the inner radial boundary at 10\,000\,s and 60\,000\,s for
the BSG and RSG models, respectively. When the SN shock approaches the
stellar surface we extend the computational domain to $R_\mathrm{ob} =
10^9$\,km in order to continue the simulation beyond shock breakout.
The circumstellar conditions assumed at $r>R_\ast$ will be described
in the next subsection.

In the simulations performed with the W15 and L15 progenitors we used
2590 radial zones for $500\,\mathrm{km} \leq r \leq 10^9$\,km, while
we used 3034 and 2957 zones in the cases of the N20 and B15
progenitors, respectively. This yields a relative radial resolution
$\Delta r/r$ of better than 1\% at all radii. The angular resolution
is 2$^\circ$ in all models. At the end of a simulation the number of
radial grid zones is more than halved, while only $\sim 10^{-6}$ of
the initial computational volume is cut off.

\subsection{Input physics}
\label{subsec:inputphysics}

We performed our longtime simulations with the tabulated equation of
state (EoS) of \citet{TimmesSwesty00} taking into account arbitrarily
degenerate and relativistic electrons and positrons, photons, and a
set of nuclei.  The nuclei, consisting of protons, $\alpha$-nuclei,
and a tracer nucleus $X$ which traces the production of a neutron-rich
(non-alpha) nucleus in neutron-rich environments ($Y_e < 0.49$), are
treated as a mixture of ideal Boltzmann gases which are advected by
the flow. Moreover, the mass fractions of the $\alpha$-nuclei and of
the tracer nucleus may also change during the evolution due to nuclear
reactions, which are described by an $\alpha$-chain reaction network
\citep{Kifonidisetal03, Wongwathanaratetal13}. 

Initially, there are no protons in the matter inside the shock (the
proton mass fraction $X_\mathrm{p}$ is set to a floor value of
$10^{-10}$). Outside of the shock the proton abundance is given by
that of the respective progenitor model, and is only of relevance in
the hydrogen envelope.  The mass fractions of the $\alpha$-nuclei and
the tracer $X$ are the ones of the corresponding explosion simulation
at the time of mapping (see Sect.\,\ref{subsec:models}).  We note here
that in the explosion simulations we also integrated an
$\alpha$-network with the flow. Its only purpose was, however, to
provide abundance information, \ie in these simulations the network
was neither energetically nor through composition changes coupled to
the hydrodynamics and the EoS
\footnote{The explosion model simulations were performed with the
  tabulated EoS of \citet{JankaMueller96}, which includes arbitrarily
  degenerate and arbitrarily relativistic electrons and positrons,
  photons, and four predefined nuclear species (n, p, $\alpha$, and a
  representative Fe-group nucleus) in nuclear statistical
  equilibrium.}.

Two versions of the network are available in our code: one involves
the 13 $\alpha$-nuclei from $^4$He to $^{56}$Ni plus the tracer
nucleus $X$, and a second, smaller network which does not contain the
$\alpha$-nuclei $^{32}$S, $^{36}$Ar, $^{48}$Cr, and $^{52}$Fe. The
smaller network was used to simulate the W15 and L15 models.  The
network is solved in grid cells whose temperature is within the range
of $10^8$\,K -- $8\times10^9$\,K.  Once the temperature drops below
$10^8\,$K, all nuclear reactions are switched off because they become
too slow to change the nuclear composition on the explosion timescale.
To prevent the nuclear burning timestep becoming too small, we also do
not perform any network calculations in zones with temperatures above
$8\times 10^9\,$K. We assume a pure $\alpha$-particle composition in
these zones.

We neglect the feedback of the nuclear energy release in the network
calculations on the hydrodynamic flow. This energy release is of minor
relevance for the dynamics because the production of $\approx
0.1\,M_\odot$ of $^{56}$Ni means a contribution of only
$~10^{50}\,$erg to the explosion energy.  It is important to note that
our model parameters are chosen to give energetic explosions already
by neutrino energy input.

When the density (temperature) in a zone drops below
$10^{-10}$\,g\,cm$^{-3}$ ($10^4$\,K), which is the smallest value
given in the EoS table of \citet{TimmesSwesty00}, we switch to a
simpler EoS taking into account only a set of ideal Boltzmann gases
and blackbody radiation. The pressure $p$ and specific internal energy
$e$ are then given by
\begin{eqnarray}
  p &=& \frac{1}{3}aT^4 + \frac{k_\mathrm{B}}{\mu m_\mathrm{H}}\rho T\,, \\ 
  e &=& \frac{aT^4}{\rho} + \frac{3}{2}\frac{k_\mathrm{B}}{\mu m_\mathrm{H}}T\,, 
\end{eqnarray}
where $a, \rho, T, k_\mathrm{B}, \mu$, and $m_\mathrm{H}$ are the
radiation constant, the density, the temperature, Boltzmann's
constant, the mean molecular weight
\footnote{We assume complete ionization for all species, and note that
  the pressure is clearly dominated by the contribution from radiation
  inside the exploding star in our models.},
and the atomic mass unit, respectively.

After mapping we continued to take into account the effect of neutrino
heating near the proto-neutron star surface by means of a boundary
condition that describes a spherically symmetric neutrino-driven wind
injected through the inner grid boundary. Since the wind is
spherically symmetric it does not affect the development of
asymmetries in the ejecta.  The hydrodynamic and thermodynamic
properties of the wind are derived from the angle-averaged state
variables of the explosion models of \citet{Wongwathanaratetal13} at
$r = 500\,$km.

In most of our simulated models we imposed (for simplicity) a {\it
  constant wind} boundary condition, where all wind quantities are
time independent.  We also simulated two models with time-dependent
wind quantities (see Tab.\,\ref{tab:models}). In these latter models
we assumed a constant wind velocity $v_\mathrm{w}$. This leads to a
{\it power-law wind} boundary condition where the time-dependent
density $\rho_\mathrm{w}$, specific internal energy $e_\mathrm{w}$,
and pressure $p_\mathrm{w}$ are given by
\begin{eqnarray}
  \rho_\mathrm{w}(t)&=& \rho_{\mathrm{w}}(t_\mathrm{map})
                       \left(\frac{t}{t_\mathrm{map}}\right)^{-7/2}\,,
\label{eq:rhowind}\\
  e_\mathrm{w}(t)   &=& e_{\mathrm{w}}(t_\mathrm{map})
                       \left(\frac{t}{t_\mathrm{map}}\right)^{-7/6}\,,
\label{eq:ewind}
\end{eqnarray}
and
\begin{eqnarray}
  p_\mathrm{w}(t)   &=& \frac{1}{3} e_\mathrm{w}(t)\,\rho_\mathrm{w}(t)\,,
\label{eq:pwind}
\end{eqnarray}
with $t_\mathrm{map}$ from Tab.\,\ref{tab:expmod}.  The choice of the
power-law indices in Eqs.\,(\ref{eq:rhowind}) and (\ref{eq:ewind}) is
motivated by an extrapolation of the density and internal energy
evolution at $r = 500\,$km.  We refer to the two kinds of wind models
in the following by adding a suffix ``cw'' for the constant wind and
``pw'' for the power-law wind to the model names, respectively (see
Tab.\,\ref{tab:models}).

We applied both wind boundary conditions for a time period of 2\,s and
then switched to a reflecting boundary condition. At the outer radial
grid boundary we used a free outflow condition at all times. We tested
the influence of the wind boundary conditions with two simulations
(B15-1-cw and B15-1-pw; see Sect.\,\ref{subsec:overview} for details),
which represent quite an extreme case for the difference between
both. The test showed that the morphology of the metal-rich ejecta
does not depend on the choice of the wind boundary condition, although
in the power-law wind model the explosion energy saturates earlier at
a considerably lower value, and the maximum Ni velocities are almost a
factor of two smaller than in the constant wind model (see
Tab.\,\ref{tab:models}, and discussion in Sect.\,\ref{subsec:shock}
and \ref{subsubsec:eexp}).

To follow the evolution beyond shock breakout we embedded our stellar
models in a spherically symmetric circumstellar environment resembling
that of a stellar wind. In this environment, the density and
temperature distribution of the matter, which is assumed to be at
rest, is given for any grid cell $i$ with $r_i > R_\ast$ by
\begin{eqnarray}
  \rho_\mathrm{e}(r) &=& \rho_0\left(\frac{R_\ast}{r}\right)^2 \,,\\
     T_\mathrm{e}(r) &=& T_0\left(\frac{R_\ast}{r}\right)^2
\end{eqnarray}
with $\rho_0 = 3\times 10^{-10}$\,g\,cm$^{-3}$ and $T_0 = 10^4$\,K.
The stellar radius $R_\ast$ is given in Tab.\,\ref{tab:expmod}.


\section{Comparison with HJM10}
\label{sec:compareHJM10}

Before discussing the set of "standard" 3D simulations (see
Sect.\,\ref{subsec:overview}), we first consider two additional 3D
simulations that we performed specifically to compare the results with
those of the 3D simulation of HJM10.  The numerical setup and the
input physics differ slightly from the standard one used in all our
other simulations presented here, so that they closely resemble those
described in HJM10, except for the utilization of the Yin-Yang grid in
our simulations.

\subsection{Simulation setup}
\label{subsec:H15setup}

The simulations are initialized from the 3D explosion model of
\citet{Scheck07} that results from the core collapse of the BSG
progenitor model B15. \citet{Scheck07} simulated the evolution in 3D
from 15\,ms until 0.595\,s after core bounce using a spherical polar
grid with 2$^\circ$ angular resolution and 400 radial grid zones.  To
alleviate the CFL time step constraint he excised a cone of 5$^\circ$
half-opening angle around the polar axis from the computational
domain. The explosion energy was $0.6\,$B at the end of the
simulation, but had not yet saturated. \citet{Scheck07} neglected
nuclear burning and used the EoS of \citet{JankaMueller96} with four
nuclear species ($n$, $p$, $^4$He, and $^{54}$Mn), assumed to be in
nuclear statistical equilibrium.

We mapped the explosion model of \citet{Scheck07} onto the Yin-Yang
grid using two grid configurations with $1200(r) \times 92(\theta)
\times 272(\phi) \times 2$ and $1200(r) \times 47(\theta) \times
137(\phi) \times 2$ zones. This corresponds to an angular resolution
of 1$^\circ$ (model H15-1deg) and 2$^\circ$ (model H15-2deg),
respectively. Since a cone around the polar axis was excised in the
explosion model of \citet{Scheck07}, we supplemented the missing
initial data using tri-cubic spline interpolation.  The radial grid
extends from 200\,km to near the stellar surface, the fixed outer
boundary of the Eulerian grid being placed at $3.9\times 10^7\,$km. We
imposed a reflective boundary condition at the inner edge of the
radial grid, and a free-outflow boundary condition at the outer
one. During the simulations we repeatedly moved the inner boundary
outwards, as described in Sect.\,\ref{subsec:numerics}.

As in HJM10 we artificially boosted the explosion energy to a value of
1\,B by enhancing the thermal energy of the post-shock matter in the
mapped "initial" state (at 0.595\,s). We did neither take self-gravity
nor nuclear burning into account. We advected eight nuclear species
($n$, $p$, $^4$He, $^{12}$C, $^{16}$O, $^{20}$Ne, $^{24}$Mg, and
$^{56}$Ni) redefining the $^{54}$Mn in the explosion model of
\citet{Scheck07} as $^{56}$Ni in our simulations.

The setups employed for our two H15 simulations and the simulation of
HJM10 differ only with respect to the grid configuration. HJM10 used a
spherical polar grid excising a cone of 5$^\circ$ half-opening angle
around the polar axis as \citet{Scheck07}, while we performed our
present simulations with the Yin-Yang grid covering the full 4$\pi$
solid angle. Our model H15-1deg has the same angular resolution as the
3D simulation of HJM10. We note that in the simulation of HJM10 the
reflecting boundary condition imposed at the surface of the excised
cone might have affected the flow near this surface, while our
simulations based on the Yin-Yang grid avoid such a numerical problem.

%
\begin{figure}
\centering
\resizebox{0.495\hsize}{!}{\includegraphics*{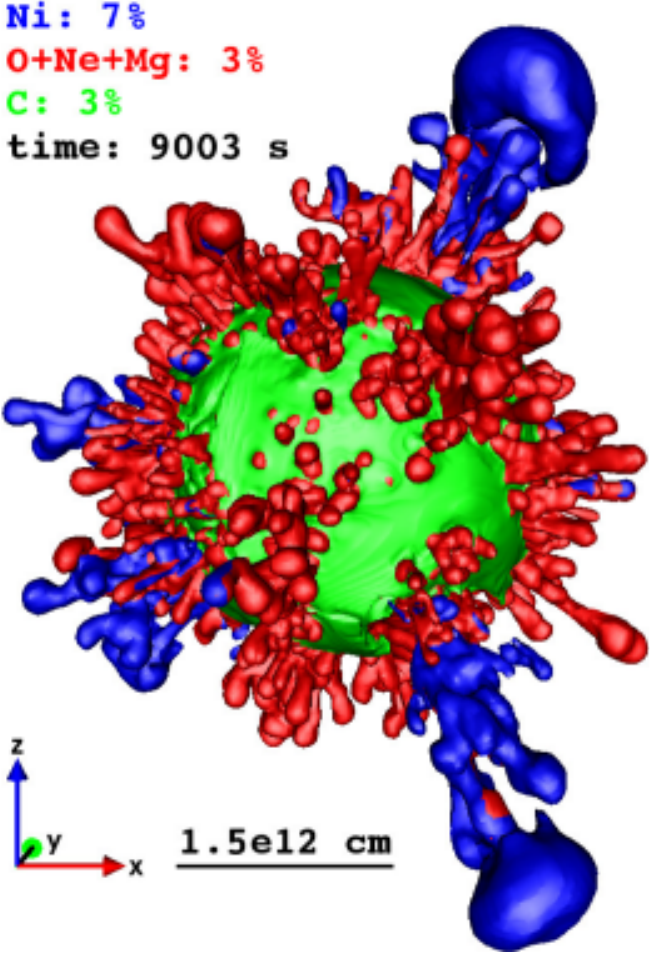}}
\resizebox{0.495\hsize}{!}{\includegraphics*{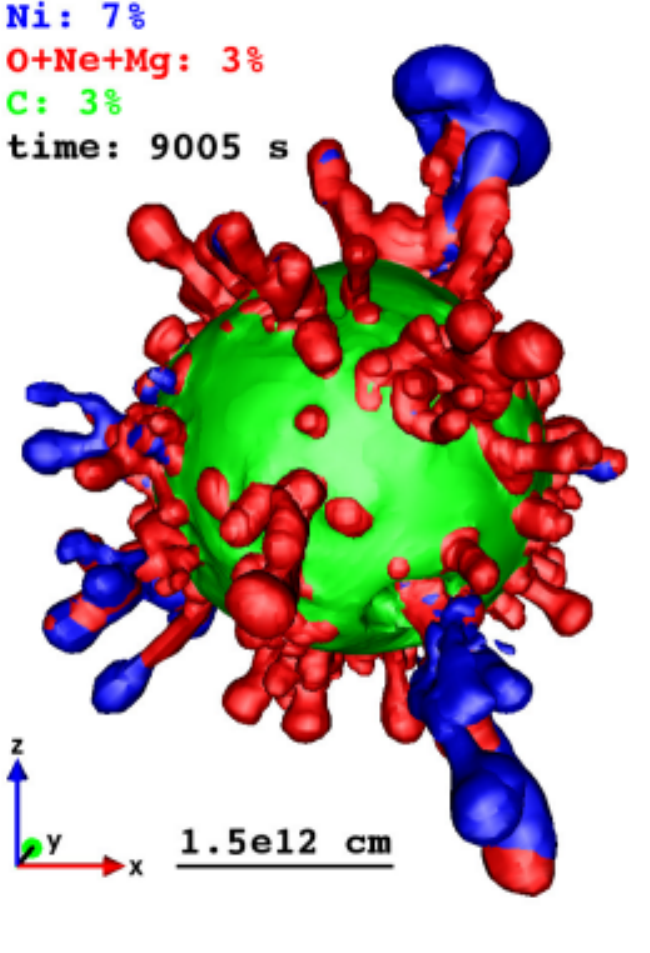}}\\
\caption{Isosurfaces of constant mass fractions at t$\approx$9000\,s
  for models H15-1deg (left) and H15-2deg (right), respectively. The
  mass fractions are 7\% for $^{56}$Ni (blue), and 3\% for
  $^{16}$O+$^{20}$Ne+$^{24}$Mg (red) and $^{12}$C (green). The
  morphology is almost identical to that shown in the lower left panel
  of Fig.\,2 in HJM10, except for some additional small-scale
  structures in the better resolved model. There are two pronounced
  nickel plumes (blue) visible on the right, which travel at
  velocities up to 3800\,km\,s$^{-1}$ and 4200\,km\,s$^{-1}$ in model
  H15-2deg and H15-1deg, respectively, and two smaller nickel fingers
  on the left.}
\label{fig:hammermodel}
\end{figure}
%

%
\begin{figure}
\resizebox{\hsize}{!}{\includegraphics*{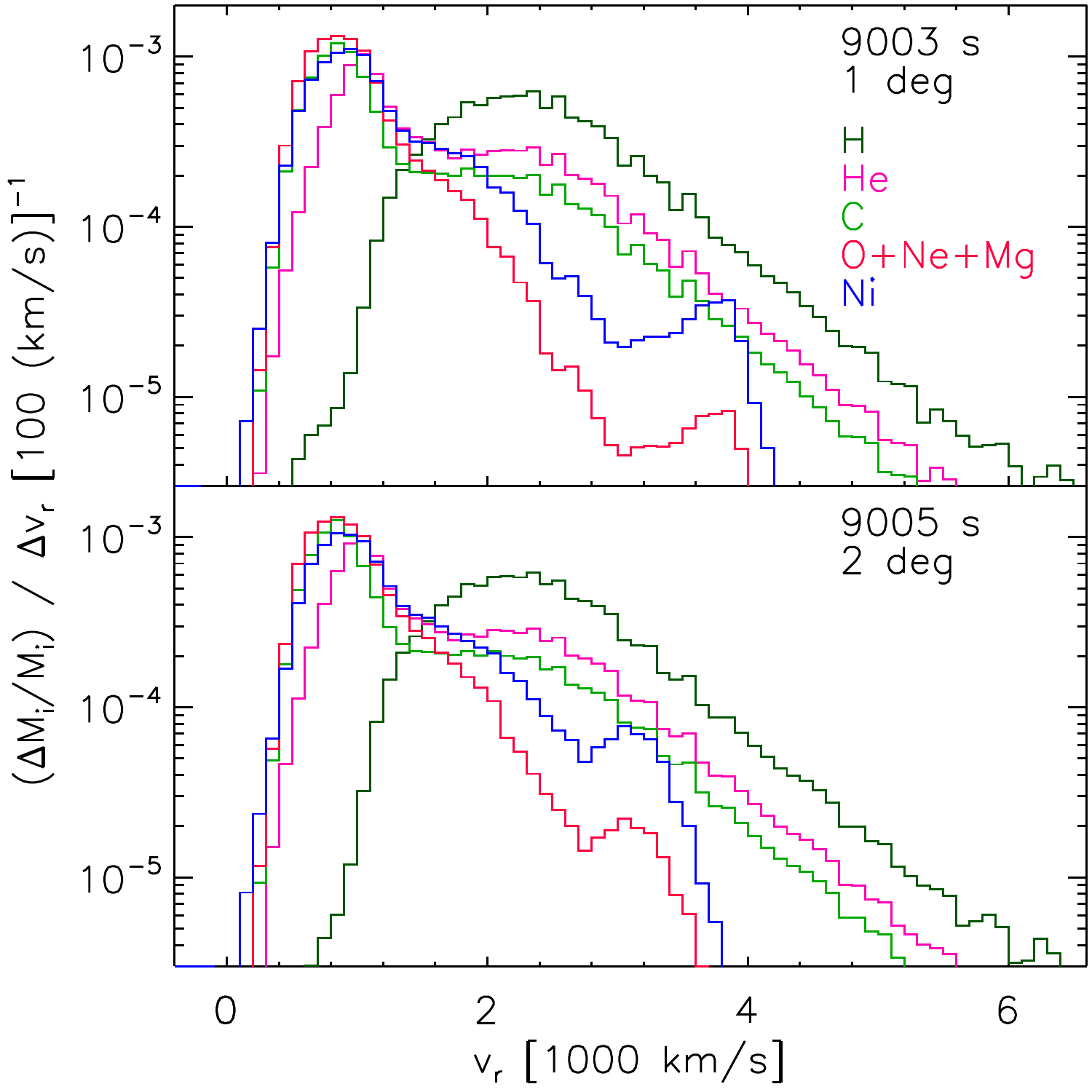}}\\
\caption{Normalized mass distributions of $^{1}$H (dark green),
  $^{4}$He (magenta), $^{12}$C (green), $^{16}$O+$^{20}$Ne+$^{24}$Mg
  (red), and $^{56}$Ni (blue) versus radial velocity $v_r$ for models
  H15-1deg (top) and H15-2deg (bottom). For all elements the
  distributions (especially of model H15-1deg) agree very well with
  those shown in the bottom right panel of Fig.\,6 in HJM10.}
\label{fig:hammermvsvr}
\end{figure}
%

\subsection{Results}
 
Fig.~\ref{fig:hammermodel} shows isosurfaces of constant mass
fractions of $^{56}$Ni, "oxygen", and $^{12}$C about 9000\,s after
core bounce for model H15-1deg (left) and H15-2deg (right),
respectively.  Note that as in HJM10, we denote in this section by
"oxygen" the sum of the mass fractions of $^{16}$O , $^{20}$Ne, and
$^{24}$Mg.  At first glance, both simulations exhibit similar RT
structures. Two pronounced nickel (blue) plumes, a few smaller nickel
fingers, and numerous "oxygen" (red) fingers burst out from a
quasi-spherical shell of carbon (green). The maximum radial velocity
of the pronounced nickel plumes is about 4200\,km\,s$^{-1}$ in model
H15-1deg and about 3800\,km\,s$^{-1}$ in model H15-2deg
(Fig.\,\ref{fig:hammermvsvr}). However, while at the tips of these
nickel plumes well-defined mushroom caps grow in model H15-1deg, they
are less developed in model H15-2deg, because the responsible
secondary Kelvin-Helmholtz (KH) instabilities are not captured very
well in the run with the lower angular resolution.

There are also more "oxygen" fingers in model H15-1deg than in model
H15-2deg. Nevertheless, these fingers grow along exactly the same
directions in both simulations.  Comparing the spatial distribution of
RT fingers in Fig.\,\ref{fig:hammermodel} and the lower left panel of
Fig.\,2 in HJM10 one also recognizes striking similarities.  Besides
the two pronounced nickel plumes, which grow into the same directions,
there are also two smaller nickel fingers pointing to the left side,
which are present in the 3D simulation of HJM10, too. Even the
distribution of the small "oxygen" fingers in HJM10's 3D simulation is
captured well by both of our H15 models.

Fig.\,\ref{fig:hammermvsvr} displays the normalized mass distributions
of $^{1}$H, $^{4}$He, $^{12}$C, "oxygen", and $^{56}$Ni versus radial
velocity for model H15-1deg (top) and H15-2deg (bottom). Although the
distributions from both simulations agree well for all nuclear
species, one can notice some differences at the high-velocity tail of
the nickel distribution and at the low-velocity tail of the hydrogen
distribution. The fastest nickel moves with a velocity of about
4200\,km\,s$^{-1}$ in model H15-1deg, while it is about 10\% slower in
model H15-2deg.  For the former model hydrogen is mixed down to radial
velocities of $~500\,$km\,s$^{-1}$, while the minimum velocity amounts
to $~700\,$km\,s$^{-1}$ in model H15-2deg.

These results are in excellent agreement with those of HJM10 (see the
bottom right panel of their Fig.\,6), except for some minor
differences. The peak radial velocity of nickel is about
4500\,km\,s$^{-1}$ in the model of HJM10, \ie roughly 10\% higher than
in model H15-1deg. The slowest hydrogen moves with about
300\,km\,s$^{-1}$ in their model, \ie it is slightly slower than in
ours.

We attribute these differences to the numerical resolution, which is
higher in the 3D model of HJM10 in the polar regions, and whose
influence can be seen by comparing our models H15-1deg and
H15-2deg. By construction the linear size of the angular zones is
almost uniform for the Yin-Yang grid, while it decreases considerably
near the poles for the spherical polar grid. Thus, although using the
same angular grid spacing of $1^\circ$ in both studies, the spatial
resolution is lower in our H15-1deg simulation in the polar regions
than in the simulation of HJM10. Consequently, the regions where the
two pronounced nickel plumes are located are better resolved in their
simulation than in ours.

The comparison of our results with those of HJM10 allows for two
conclusions:
\begin{itemize}
\item
 The overall similarity (qualitatively and quantitatively) of the
 results of models H15-1deg and H15-2deg suggests that an angular
 resolution of 2$^\circ$ is already sufficient for our current study,
 which aims at capturing neither details of small-scale structures nor
 determining the precise peak velocities of small fractions of heavy
 elements in the ejecta. Using an angular resolution of 2$^\circ$
 allows us to perform a parameter study changing both the explosion
 energy and the progenitor with a reasonable amount of computing time.
\item
 The similarity of the ejecta asymmetries and radial mixing confirms
 that the long-time evolution of the SN is determined by the initial
 asymmetries imprinted by the explosion mechanism rather than by
 stochastic effects of the secondary RT instabilities growing at the
 composition interfaces. Characteristic features of the earliest
 phases of the explosion therefore map into the ejecta morphology at
 later times.
\item
 The peak velocities of heavy elements and the minimal velocities of
 inward mixed hydrogen are more extreme for better resolved
 models. Numerical viscosity in less well resolved models reduces the
 extent of radial mixing.
\end{itemize}
In Sect.\,\ref{sec:3Dresults} we will investigate how this mapping
depends on the progenitor star and the progenitor-specific interaction
of initial explosion asymmetries with the RT instabilities that
develop at the composition interfaces after the passage of the
outgoing SN shock wave.


%
\begin{figure*}
\centering
\resizebox{\hsize}{!}{\includegraphics*{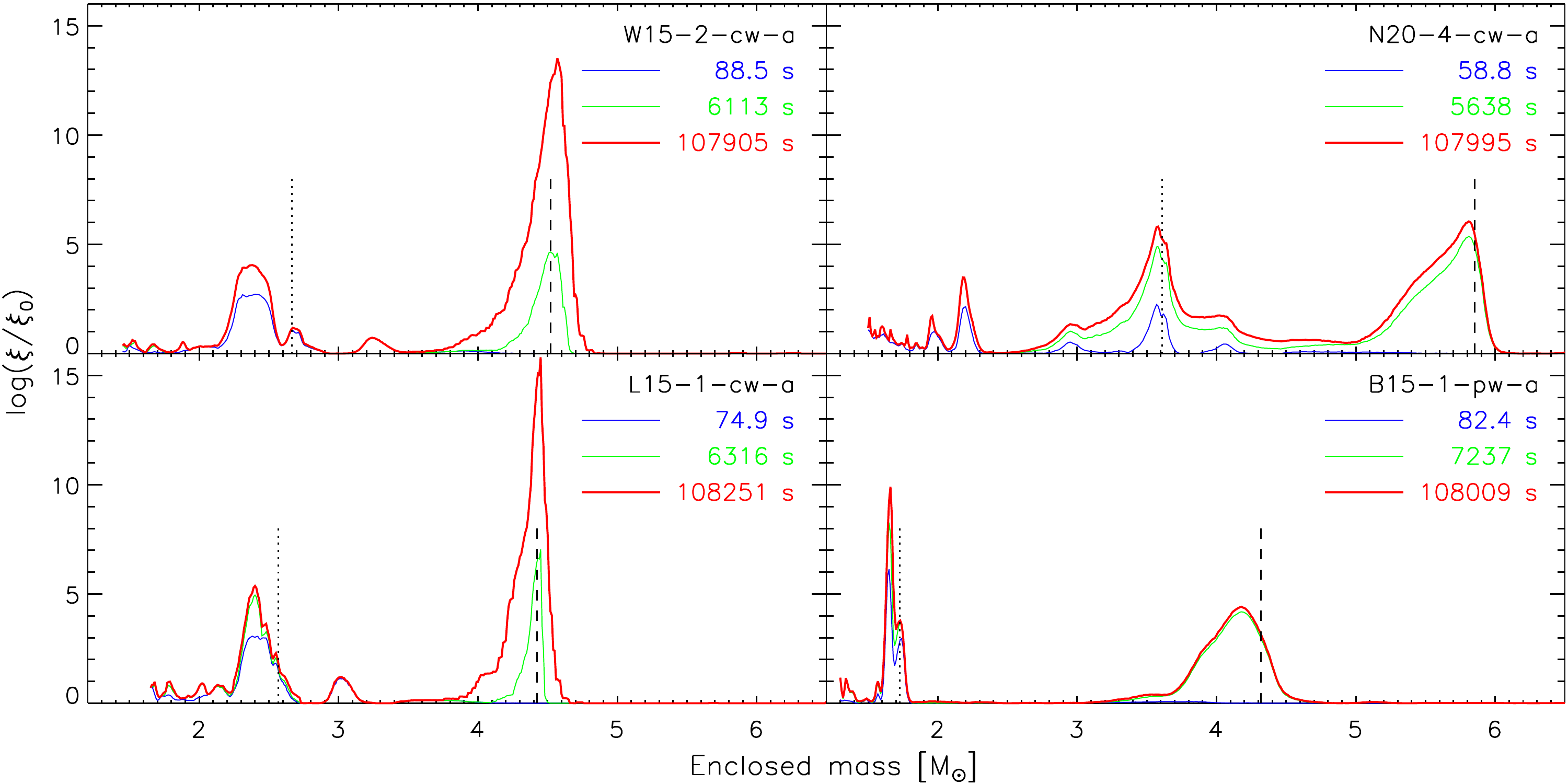}}
\caption{Time-integrated RT growth factors versus enclosed mass for
  the 1D models W15-2-cw-a, L15-1-cw-a, N20-4-cw-a, and B15-1-pw-a at
  the given times. The vertical dotted and dashed lines mark the mass
  coordinates of the C+O/He and He/H interfaces (defined in
  Sect.\,\ref{subsec:models}), respectively.}
\label{fig:lingrowth}
\end{figure*}
%

\section{One-dimensional models: Linear stability analysis}
\label{sec:1dmodels}

While propagating through the star the velocity of the SN shock
varies, because it accelerates or decelerates more or less strongly
depending on the density stratification it encounters, which in turn
depends on the progenitor star.  In particular, whenever the density
gradient is steeper than $r^{-3}$ ({\it i.e.}  $d (\rho r^3)/dr <0 $),
a shock wave accelerates according to the blast-wave solution
\citep{Sedov59}. This condition is fulfilled near composition
interfaces where the density varies most strongly, especially near the
C+O/He and He/H composition interfaces. An acceleration at the
interface follows a deceleration in the overlying stellar layers,
which causes a density inversion in the post-shock flow, \ie a dense
shell forms.  Such shells are prone to RT instabilities, because they
are associated with density and pressure gradients of opposite signs
\citep{Chevalier76}.

To aid us with the analysis of the 3D simulations, we have computed
the linear RT growth rates for each of our progenitor stars. For an
incompressible fluid
\footnote{The growth rate for an incompressible fluid provides a lower
  limit to the actual rate. This is sufficient here, because we are
  only interested in determining where in the star the growth rate is
  large, but do not mind the actual value.}
it is given by \citep{Bandiera84, BenzThielemann90, Muelleretal91}
\begin{eqnarray}
\sigma_\mathrm{RT} = \sqrt{-\frac{\partial p   }{\partial r}
                           \frac{\partial \rho}{\partial r}}.
\end{eqnarray}   
We monitor the growth of RT instabilities by calculating the
time-integrated growth factor
\begin{eqnarray}
\frac{\xi}{\xi_0}(t) =
\exp{\left(\int_0^t\,\sigma_\mathrm{RT}\,dt^\prime\right)} 
\end{eqnarray}   
at fixed Lagrangian mass coordinates, which tells by how much an
initial perturbation with an amplitude $\xi_0$ would grow until time
$t$.

Tracking a fixed mass coordinate is not easily feasible in our 3D
Eulerian simulations, especially when the flow shows strong
asymmetries.  Thus, we performed four 1D simulations, one for each of
the considered progenitor stars, based on the angle-averaged profiles
of the 3D models W15-2, L15-1, N20-4, and B15-1 at time
$t_\mathrm{map}$, which were also used for our 3D runs to late times
(see Sect.\,\ref{subsec:overview}). The numerical setup for these 1D
simulations was identical to that used in the 3D simulations described
in Sect.\,\ref{sec:methods}. The constant wind boundary condition (see
Sect.\,\ref{subsec:inputphysics}) was imposed in the simulations using
the explosion models W15-2, L15-1, and N20-4, while the power-law wind
boundary condition was applied in the run with the explosion model
B15-1. We denote the corresponding 1D simulations by W15-2-cw-a,
L15-1-cw-a, N20-4-cw-a, and B15-1-pw-a, which gave rise to explosions
with energies of $1.20\,$B, $1.53\,$B, $1.47\,$B, and $1.28\,$B,
respectively
\footnote{The explosion energies of the 1D models are about 10\% lower
  than those of the corresponding 3D ones (see
  Table\,\ref{tab:models}) because of an imperfect averaging of the 3D
  models. As we are only interested in determining where in the star
  the growth rate is large, we do not mind this discrepancy.}.

Figure\,\ref{fig:lingrowth} displays the time-integrated RT growth
factors as functions of enclosed mass for our 1D runs at three
different times. The blue lines show the growth factors at the time
when the SN shock crosses the C+O/He interface, which occurs between
60\,s and 90\,s depending on the model. The green lines give the
growth factors at the time of shock breakout for the BSG models
N20-4-cw-a and B15-1-pw-a, and at about 6000\,s (\ie roughly at the
time of shock breakout in the BSG models; see Table\,\ref{tab:models})
for the RSG models W15-2-cw-a and L15-1-cw-a. In these latter models
the SN shock reaches the stellar surface only after about 90\,000\,s
(see Table\,\ref{tab:models}).  The red lines, finally, display the
growth factors at the end of the simulations ($\approx 10^5\,$s).

The growth factors show a characteristic behavior near the C+O/He and
He/H interfaces, which depends on the density profile of the
respective progenitor.

In the RSG models W15 and L15 the growth factor is very large at the
He/H interface ($\log (\xi/\xi_0) \approx 14 \ldots 16$). It is largest
in model L15, since the density gradient at the He/H interface is
steepest in this model. In both progenitors the growth rate is much
smaller at the C+O/He interface ($\log (\xi/\xi_0) \approx 4\ldots
5$), because the density gradient is shallower there than at the He/H
interface.

In the BSG models N20 and B15 the RT growth factor at the He/H
interface is much smaller ($\log (\xi/\xi_0) \approx 5\ldots 6$) than
for the RSG progenitors, because the density decreases very little in
these progenitors at the He/H interface. Only for the B15 progenitor
the RT growth rate is higher at the C+O/He interface than at the He/H
interface, reaching already a value of $\log (\xi/\xi_0) \approx 6$ by
the time when the SN shock reaches the edge of the C+O core.

In all models, the RT growth factors continue to grow at the C+O/He
interface while the SN shock is propagating through the hydrogen
envelope, but they do not significantly increase in both BSG models at
the He/H interface after the SN shock has broken out of the stellar
surface.  The latter result holds despite of a much longer integration
time, implying saturation of the growth of RT instabilities at
relatively much earlier times than in the RSG models.

We point out that the results presented in Fig.\,\ref{fig:lingrowth}
can only provide a qualitative criterion for the relative strength of
the expected RT growth in different layers of the progenitor star. The
values of the linear growth factors depend on the flow structure in
the post-shock layers which, in turn, depends on the explosion
energy. The fact that the models shown in Fig.\,\ref{fig:lingrowth}
explode with comparable energies makes the comparison of the RT growth
factors among these models meaningful. More importantly, we have
calculated the RT growth rates by means of linear perturbation theory
of 1D models, while in multi-D simulations the instabilities will
quickly enter the non-linear regime. Nevertheless, we find that this
linear stability analysis is particularly useful in understanding the
results of our 3D runs, which we will discuss in the next section.


\begin{table*}
\caption{Some properties of the simulated 3D models at the given times
  (see text for details).}
\centering
\renewcommand{\arraystretch}{1.5}
\begin{tabular}{ccccccccc}
\hline \hline
   progenitor & 3D & $t_\mathrm{surf}$ & $E_\mathrm{exp}$ &
   avg$^\mathrm{(max)}_\mathrm{(min)}$ $R_\mathrm{s}$ & 
   $M_\mathrm{Ni}$ $(M_{\mathrm{Ni}+X})$ &
   $v_\textrm{max}(\textrm{Ni})$ & $\langle v \rangle_{1\%}(\textrm{Ni})$ 
   \\ 
   type & model & [s] & [B] & [$10^6\,$km] & [$M_\odot$] & 
   [$10^3\,$km\,s$^{-1}$] & [$10^3\,$km\,s$^{-1}$] \\
\hline
 \multirow{4}{*}{RSG}
 & W15-1-cw & 84974 & 1.48 & $389^{(443)}_{(355)}$ & 0.05 (0.13) & 5.29 & 3.72 \\ 
 & W15-2-cw & 85408 & 1.47 & $393^{(458)}_{(349)}$ & 0.05 (0.14) & 4.20 & 3.47 \\
 & L15-1-cw & 95659 & 1.75 & $478^{(530)}_{(448)}$ & 0.03 (0.15) & 4.78 & 3.90 \\
 & L15-2-cw & 76915 & 2.75 & $475^{(500)}_{(458)}$ & 0.04 (0.21) & 5.01 & 4.51 \\
\hline
 \multirow{4}{*}{BSG} 
 & N20-4-cw & 5589  & 1.65 & $39.7^{(43.6)}_{(35.6)}$ & 0.04 (0.12) & 2.23 & 1.95 \\ 
 & B15-1-cw & 5372  & 2.56 & $41.5^{(43.6)}_{(39.5)}$ & 0.05 (0.11) & 6.25 & 5.01 \\
 & B15-1-pw & 7258  & 1.39 & $42.7^{(45.7)}_{(40.0)}$ & 0.03 (0.09) & 3.34 & 3.17 \\
 & B15-3-pw & 8202  & 1.14 & $48.1^{(51.1)}_{(44.7)}$ & 0.03 (0.08) & 3.18 & 2.95 \\
\hline
\end{tabular}
\renewcommand{\arraystretch}{1}
\label{tab:models}
\end{table*}

\section{Three-dimensional models}
\label{sec:3Dresults}

\subsection{Model overview}
\label{subsec:overview}

We have simulated eight 3D models (see Table\,\ref{tab:models}) using
our four progenitor stars, two of which are RSG and two of which are
BSG (see Sect.\,\ref{subsec:models}). The first part of the model name
refers to the model for the initial neutrino-driven onset of the
explosion (\eg W15-1), while the second part denotes the kind of
neutrino-driven wind boundary condition imposed in the respective
long-time simulation (\ie either {\it cw} or {\it pw}).

The RSG models W15-1-cw and W15-2-cw are initialized from two
explosion models, which differ only in the initial perturbation
pattern that was imposed to break the spherical symmetry of the 1D
collapse model at 15\,ms after core bounce (for details, see
\citet{Wongwathanaratetal13}).  Although both models have nearly
identical explosion energies, the ejecta asymmetries developed
differently in these models during the shock revival phase because of
the chaotic nature of the non-radial hydrodynamic instabilities.
Hence, comparing models W15-1-cw and W15-2-cw, we can infer how
asymmetries developing during the revival of the SN shock influence
the ejecta distribution at late times when all other conditions are
the same.

The RSG models L15-1-cw and L15-2-cw represent two cases of
significantly different explosion energies of the L15 progenitor, the
explosion of model L15-2-cw being approximately 60\% more energetic
and also developing faster than the one of model L15-1-cw. Moreover,
the shock surface is less deformed, appearing almost spherical, in
explosion model L15-2 than in L15-1, because low-mode instabilities
have less time to grow during the shock revival phase in the former
explosion model.

The BSG model N20-4-cw is the only model in our set which does not
have a 15 but a 20 solar mass progenitor. Comparing its evolution with
those of our other models we gain some insight into the influence of
variations of BSG models for SN\,1987A progenitors on the ejecta
morphology.

To test the influence of the neutrino-driven wind boundary condition
we simulated the BSG models B15-1-cw and B15-1-pw, which are both
based on the explosion model B15-1. We chose this explosion model for
the test, because the neutrino luminosities radiated by the neutron
star at the time of mapping (1.1\,s after bounce; see
Table\,\ref{tab:expmod}) are high in this model. This results in a
denser and faster neutrino-driven wind at 500\,km, where we placed the
(initial) inner grid boundary in the long-time runs. The mapping time
(1.1\,s) was chosen to be earlier than in all other models, because
the explosion timescales of the B15 models ($\approx 170$\,ms) are
smaller than in all other models (see Table\,\ref{tab:expmod}).
Finally, with model B15-3-pw we also simulated a lower-energy
explosion of the B15 progenitor.

In Table\,\ref{tab:models} we summarize some properties of our 3D
simulations at the shock breakout time $t_\mathrm{surf}$ (column 3),
defined as the time when the minimum shock radius becomes larger than
the initial stellar radius $R_\ast$. This time, which depends on the
progenitor and the explosion energy (and to a lesser degree on the
shock deformation), is less than two hours for the compact BSG
progenitors, while it reaches and even exceeds 20 hours for the RSG
progenitors. The additional quantities listed in the table are
(columns 4 to 8):
\begin{itemize}
\item $E_\mathrm{exp}$ is the explosion energy defined as the sum of
  the total (internal plus kinetic plus gravitational) energy over all
  grid zones where the total energy is positive,
\item $R_\mathrm{s}$ is the angle-averaged, maximum, and minimum value
  of the SN shock radius, respectively,
\item $M$(Ni) is the total nickel mass, the number in the brackets
  giving an upper limit of the total mass of $^{56}$Ni plus tracer
  $X$,
\item $v_\textrm{max}$(Ni) is the maximum radial velocity on the
  outermost surface where the mass fraction of $^{56}$Ni plus the
  tracer $X$ equals 3\%, and
\item $\langle v \rangle_{1\%}$(Ni) is the mass-weighted average
  radial velocity of the fastest 1\% (by mass) of nickel.
\end{itemize}
\noindent The average SN shock radius, $\langle R_\mathrm{s} \rangle$,
is given by
\begin{eqnarray}
  \langle R_\mathrm{s} \rangle = \frac{1}{4\pi} 
                                \int R_\mathrm{s}(\theta,\phi)\,d\Omega
\end{eqnarray}
with $d\Omega=\sin{\theta}\,d\theta\,d\phi$. 

%
\begin{figure*}
\centering
\resizebox{\hsize}{!}{\includegraphics*{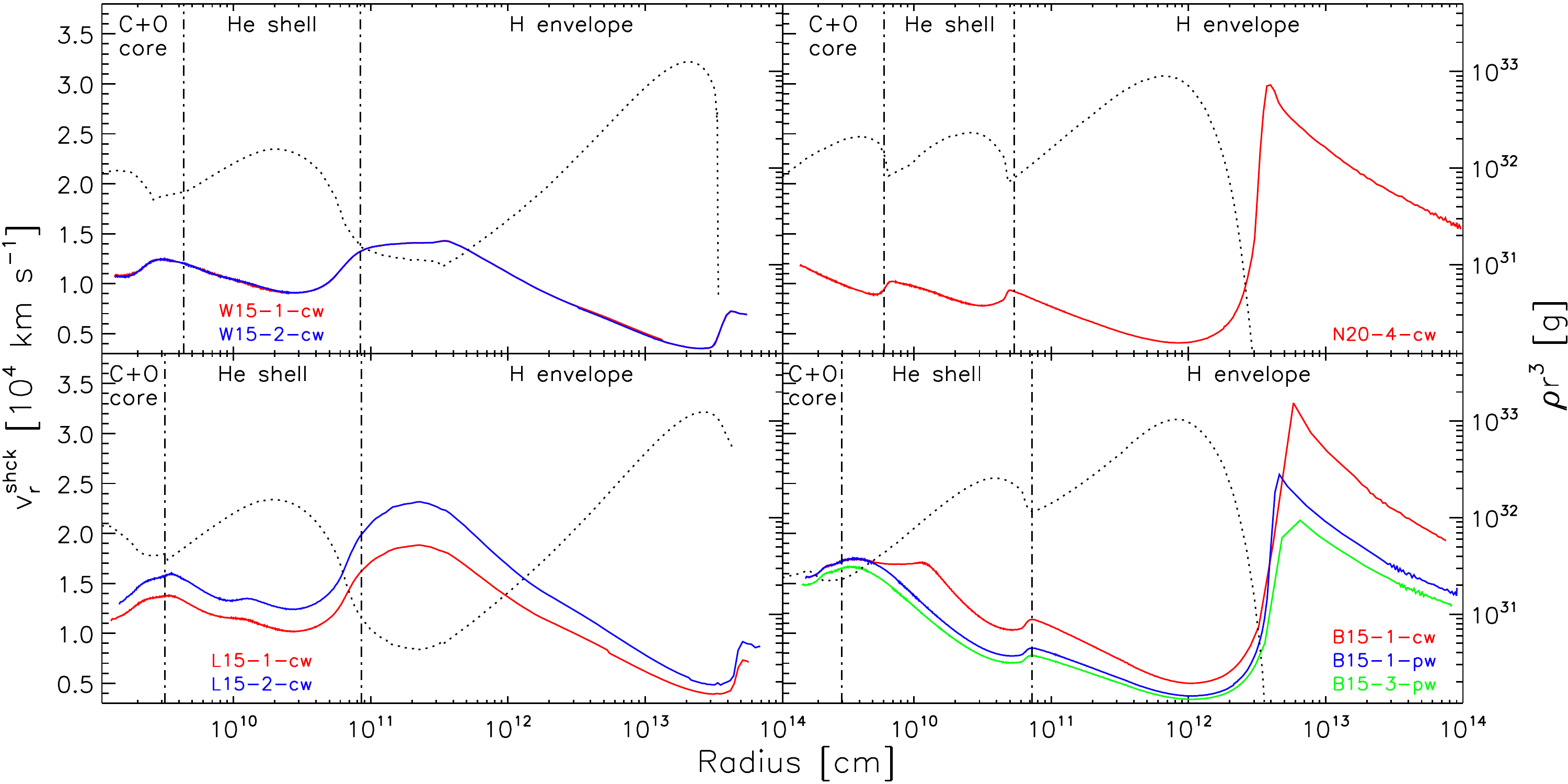}}
\caption{Radial velocity of the angle-averaged shock radius,
  $v_r^\textrm{shck}$, as a function of radius for our eight 3D models
  grouped in panels according to the respective progenitor model: W15
  (top left), L15 (bottom left), N20 (top right), and B15 (bottom
  right). In each panel, we also display the corresponding
  distribution of $\rho r^3$ inside the progenitor star (black dotted
  line), the scale being given at the right hand side of the plot. The
  radii of the C+O/He and He/H composition interfaces in the
  progenitor star (defined in Sect.\,\ref{subsec:models}) are
  indicated by the vertical dash-dotted lines.}
\label{fig:vrho}
\end{figure*}
%

\subsection{Dynamics of the supernova shock}
\label{subsec:shock}

Fig.\,\ref{fig:vrho} displays the radial velocity of the
angle-averaged SN shock, $v_r^\textrm{shck}$, as a function of radius
for our eight 3D models. To obtain this velocity we compute a backward
finite difference in time of the angle-averaged shock radius $\langle
R_\mathrm{s} \rangle$ using two subsequent outputs (typically 50 time
steps apart). The figure also shows the $\rho r^3$ profile of each
progenitor star (dotted lines), and the radial locations of the C+O/He
and He/H composition interfaces (vertical dash-dotted lines).

Models W15-1-cw and W15-2-cw (Fig.\,\ref{fig:vrho}, top left panel)
have almost the same explosion energy and differ only in the initial
seed perturbations. Hence, the propagation of the SN shock is similar
in both models. It experiences two episodes of acceleration inside the
star, one just before it crosses the C+O/He interface and another one
at the He/H interface, followed by a strong deceleration in both
cases. A third period of acceleration takes place right at the stellar
surface.

The density structure of progenitor model L15 is overall quite similar
to that of model W15, but the density decreases more rapidly with
radius at the He/H interface in the former model. Therefore, the SN
shock is accelerated more strongly at the He/H interface in model L15
than in model W15. According to Fig.\,\ref{fig:vrho} the shock speed
increases nearly by a factor of two in model L15 and by about 50\% in
model W15.  A comparison of models L15-1-cw and L15-2-cw
(Fig.\,\ref{fig:vrho}, lower right panel) shows that an increase of
the explosion energy leads to the expected overall increase of the
shock velocity, because the latter roughly scales as
$\sqrt{E_\mathrm{exp}}$ (from which one expects about 25\% higher
values in model L15-2-cw).

Otherwise, the duration and the amount of
acceleration/deceleration are very similar in both models.

The results obtained for the BSG progenitors N20 and B15 are displayed
in the right panels of Fig.\,\ref{fig:vrho}. In these models the SN
shock experiences only a modest amount of acceleration at the He/H
interface compared to models W15 and L15, because the density drops at
that interface much less steeply in the BSG models than in the RSG
models (see Fig.\, \ref{fig:rhor-progenitor}). The SN shock
decelerates strongly while propagating through the helium shell of the
B15 models, whereas it experiences only a modest deceleration in model
N20-4-cw.

Comparing the dynamics of the SN shock in models B15-1-cw and B15-1-pw
the effect of the neutrino-driven wind imposed at the inner radial
grid boundary becomes obvious, in contrast to all other models with
constant but much weaker wind power.  Initially, the shock velocity is
identical in both models. However, the constant supply of energy by
the constant wind condition delays the saturation of the explosion
energy in model B15-1-cw, \ie the SN shock enters the blast-wave phase
in model B15-1-cw later than in model B15-1-pw. The shock velocity
remains almost constant in model B15-1-cw for $4\times 10^{9} \la r
\la 11\times 10^{9}$\,cm, although $\rho r^3$ increases in these
layers. After the inflow of neutrino-driven wind has ceased and the
explosion energy has saturated, the shock velocity behaves as in all
other models and decreases (increases) with increasing (decreasing)
$\rho r^3$ according to the blast-wave solution \citep{Sedov59}.

When the SN shocks reach the surfaces of the progenitor stars they
encounter the steep density gradients prevailing there. They
accelerate briefly, the acceleration being stronger in the BSG models
than in the RSG models since the density gradients are steeper in the
former models. Subsequently, the shocks decelerate again, because we
assume a $r^{-2}$ density profile outside of the progenitor stars.

%
\begin{figure*}
\centering
\resizebox{0.252\hsize}{!}{\includegraphics*{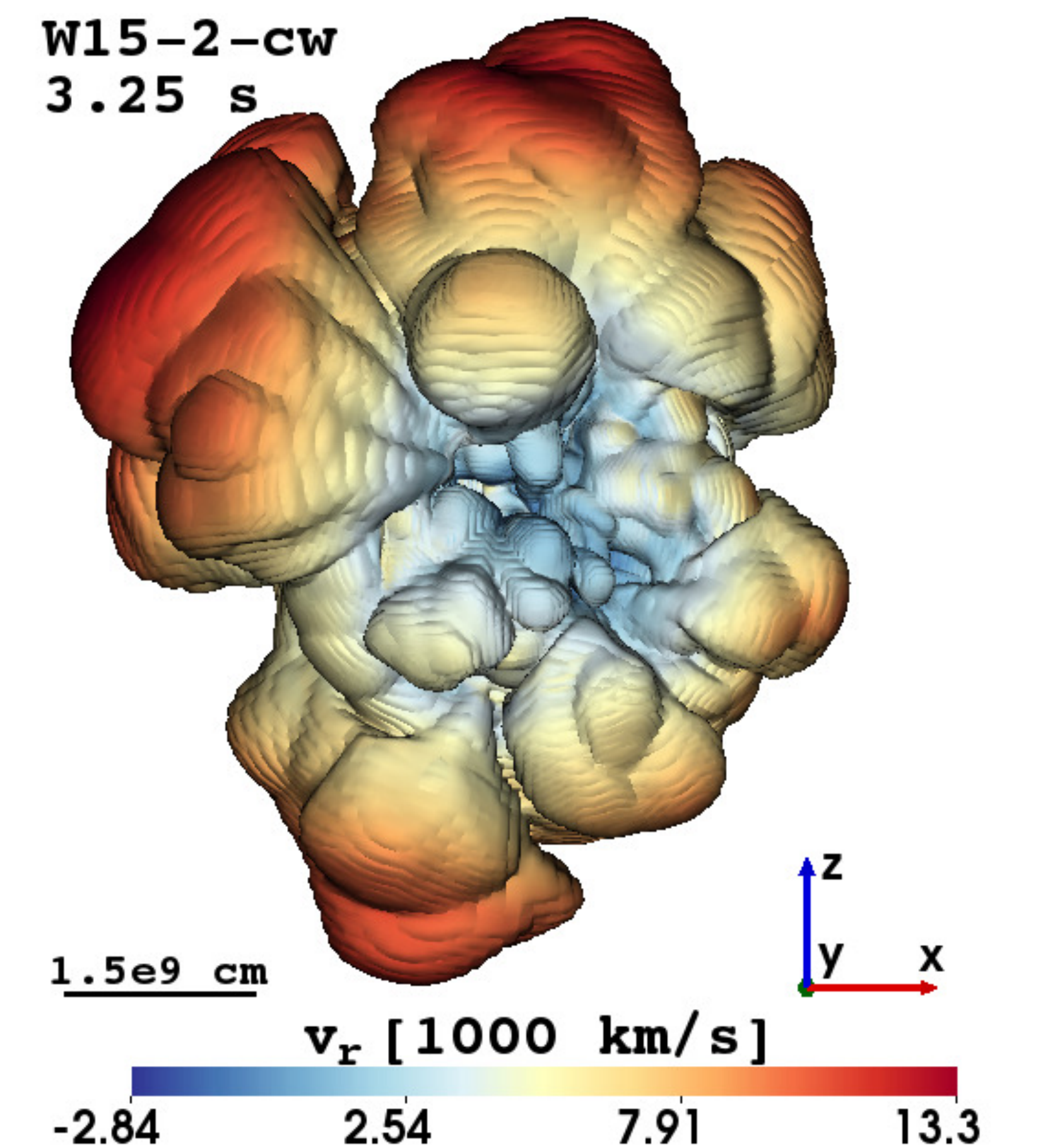}}
\hspace{-0.21cm}
\resizebox{0.252\hsize}{!}{\includegraphics*{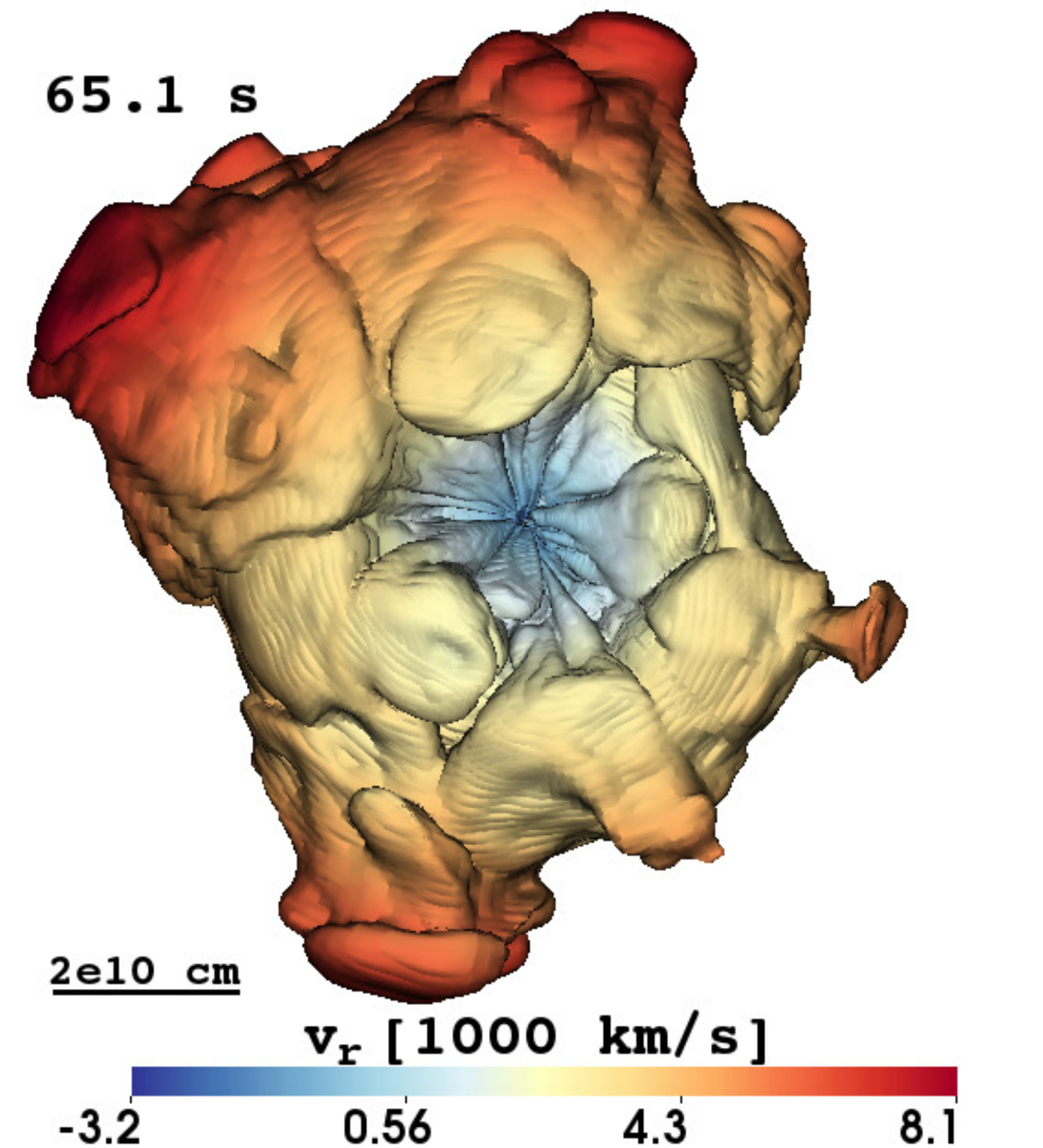}}
\hspace{-0.21cm}
\resizebox{0.252\hsize}{!}{\includegraphics*{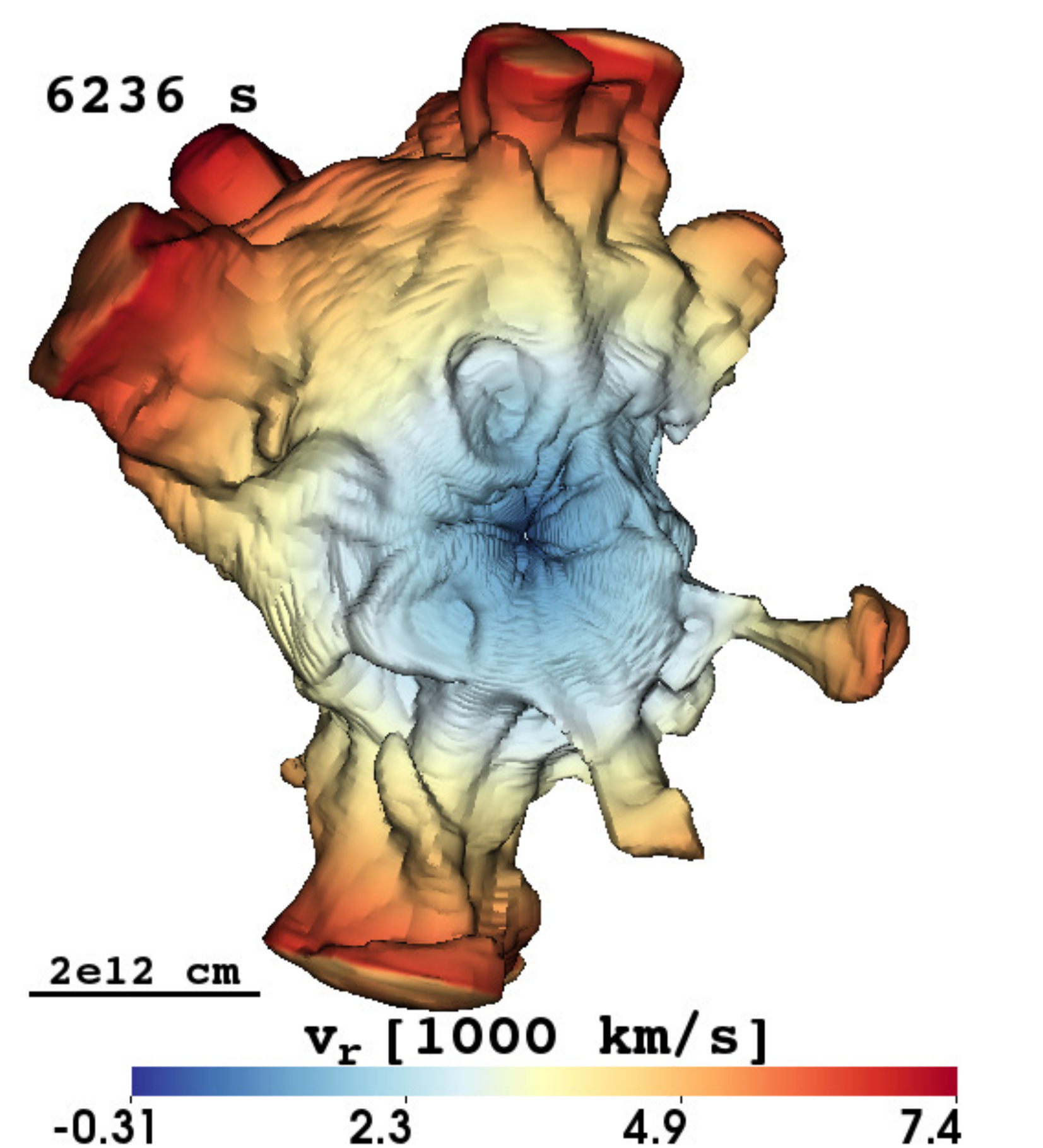}}
\hspace{-0.21cm}
\resizebox{0.252\hsize}{!}{\includegraphics*{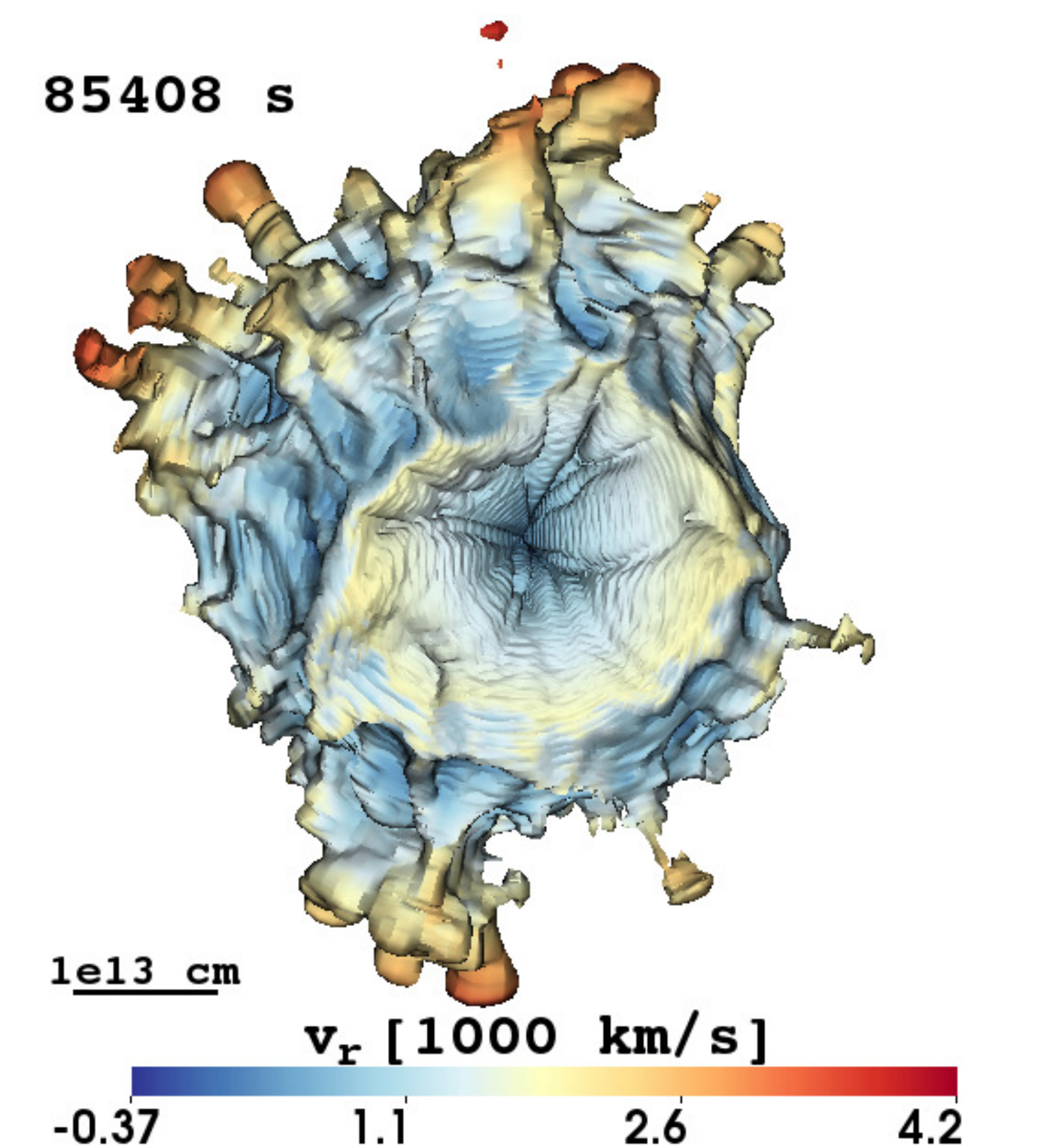}}\\
\vspace{0.3cm}
\resizebox{0.252\hsize}{!}{\includegraphics*{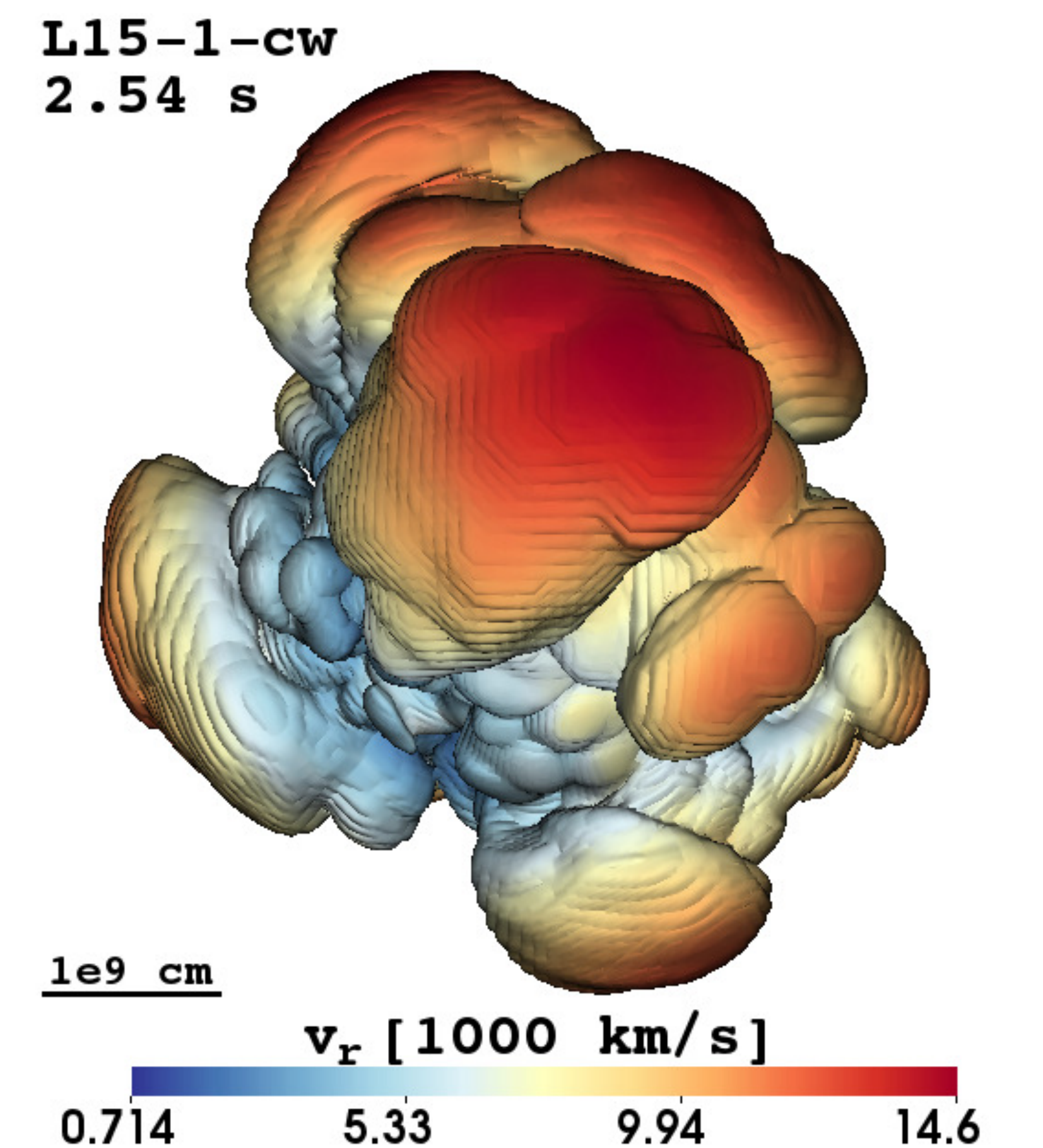}}
\hspace{-0.21cm}
\resizebox{0.252\hsize}{!}{\includegraphics*{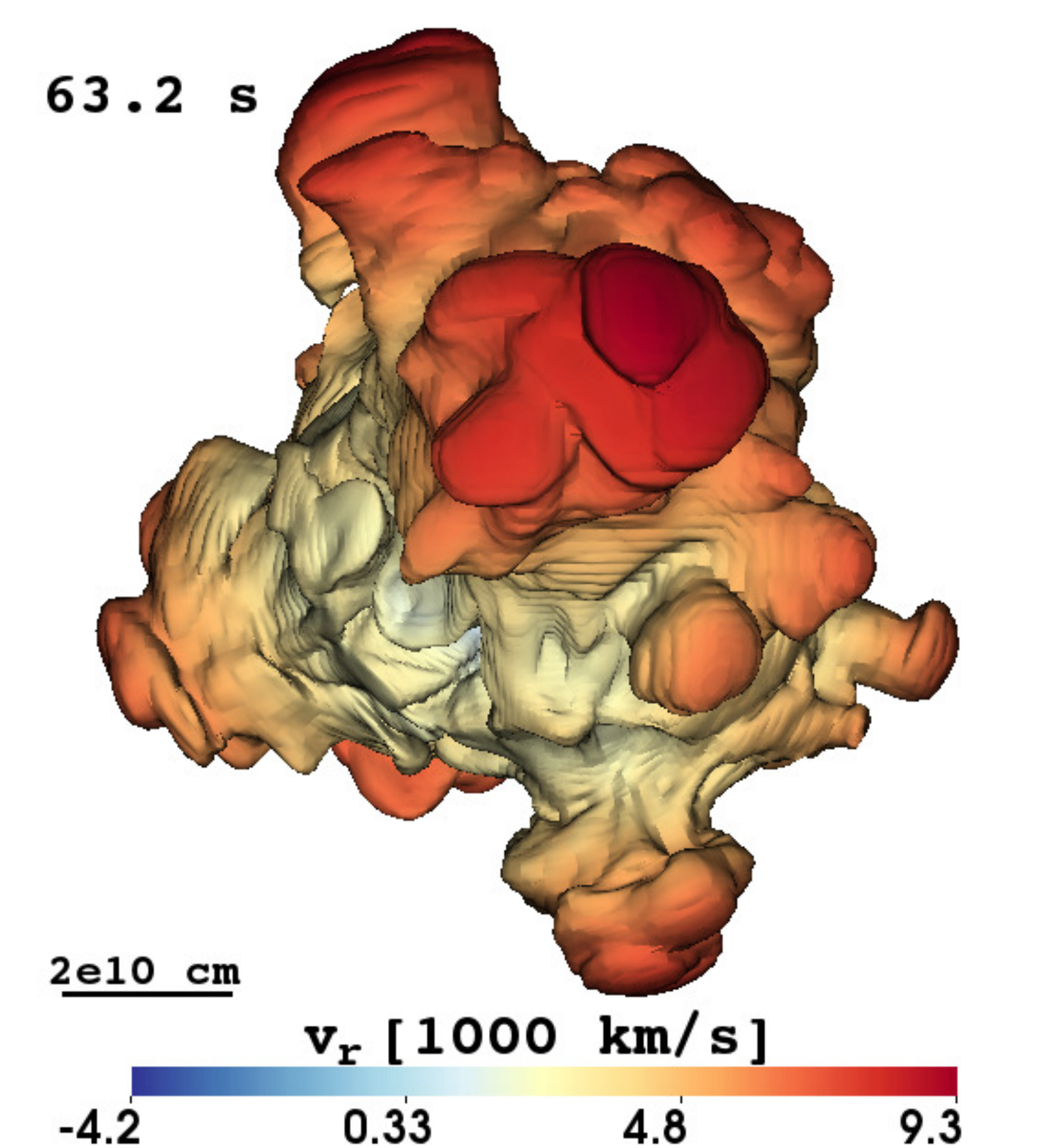}}
\hspace{-0.21cm}
\resizebox{0.252\hsize}{!}{\includegraphics*{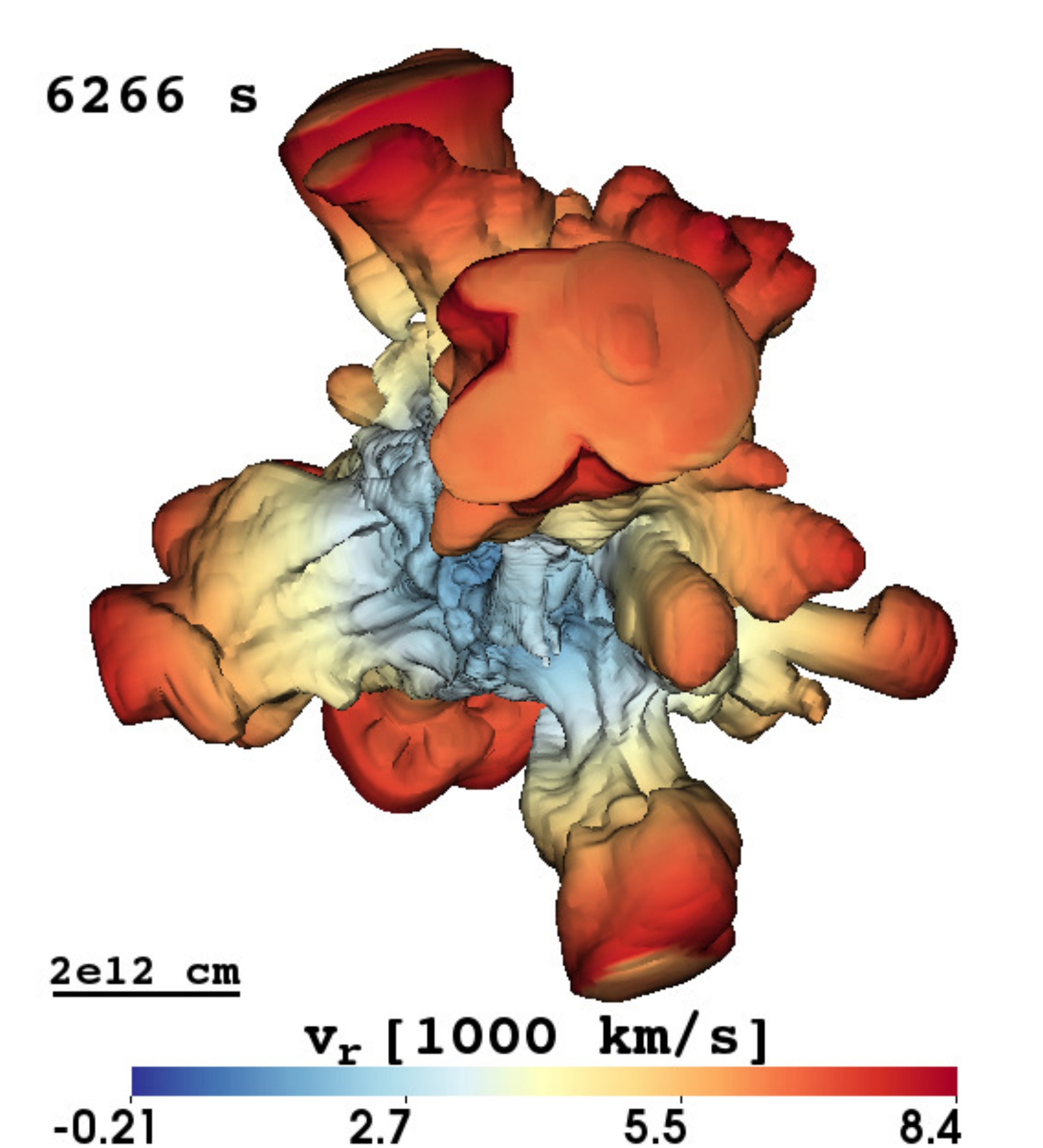}}
\hspace{-0.21cm}
\resizebox{0.252\hsize}{!}{\includegraphics*{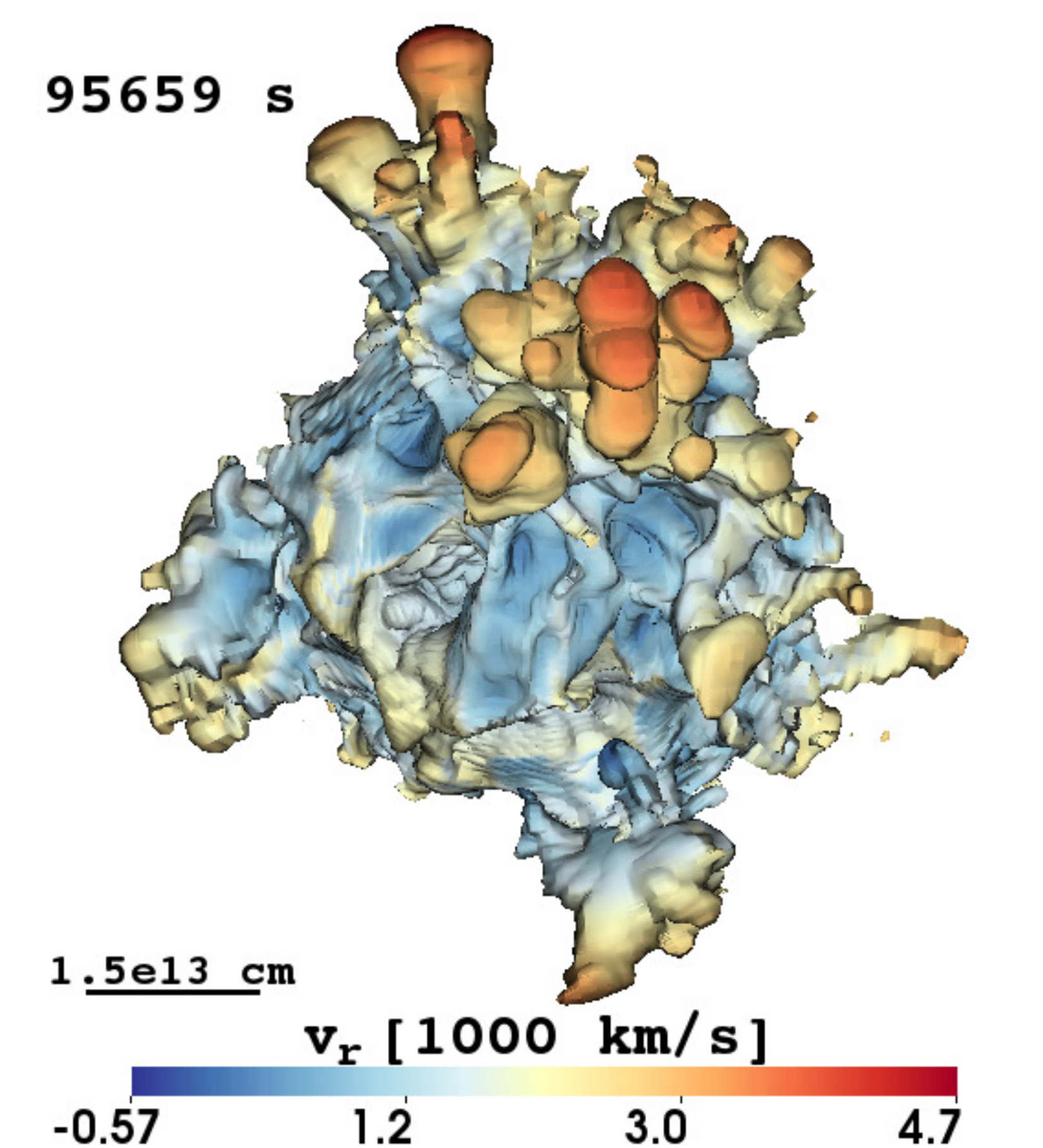}}\\
\vspace{0.3cm}
\resizebox{0.252\hsize}{!}{\includegraphics*{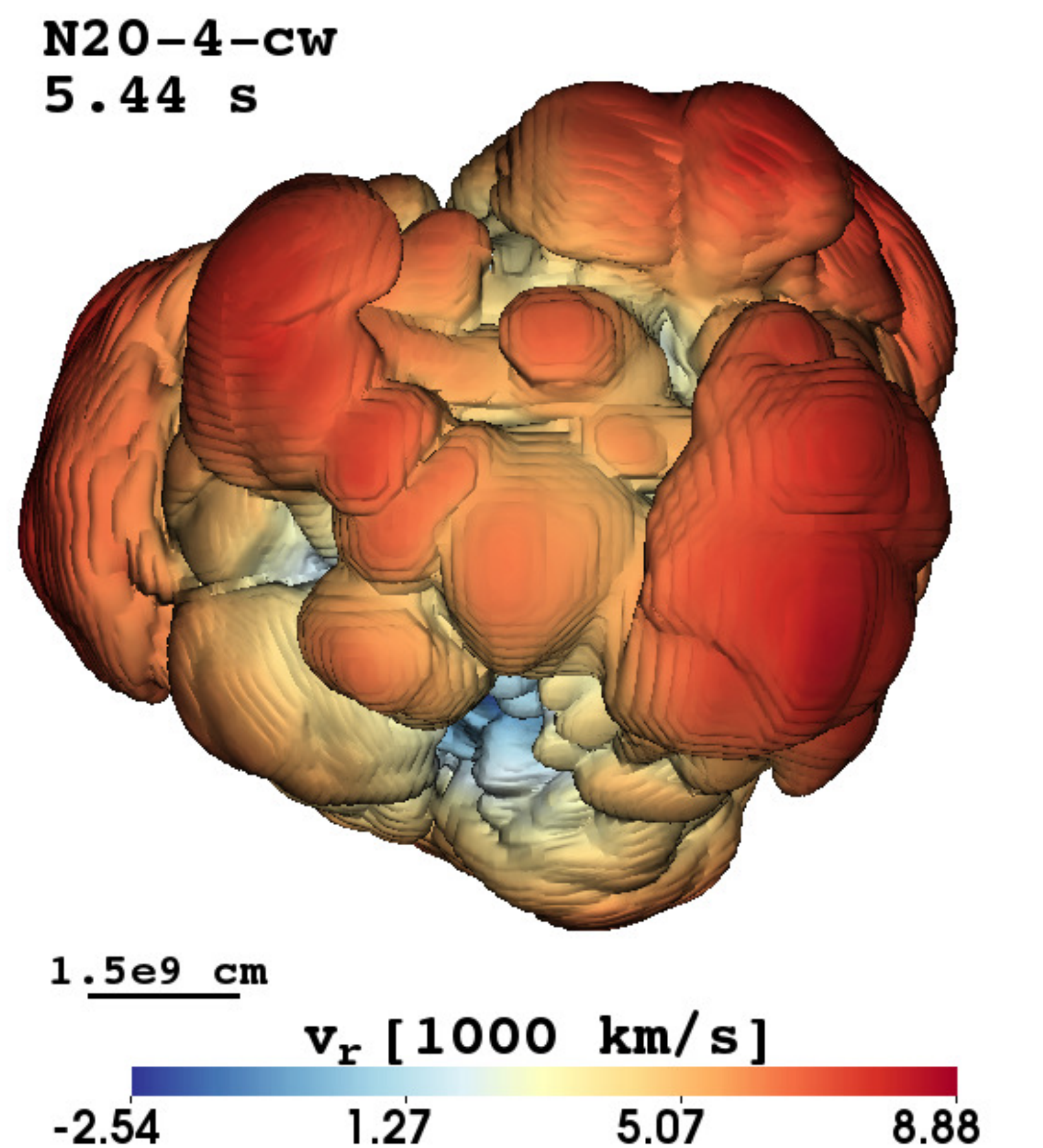}}
\hspace{-0.21cm}
\resizebox{0.252\hsize}{!}{\includegraphics*{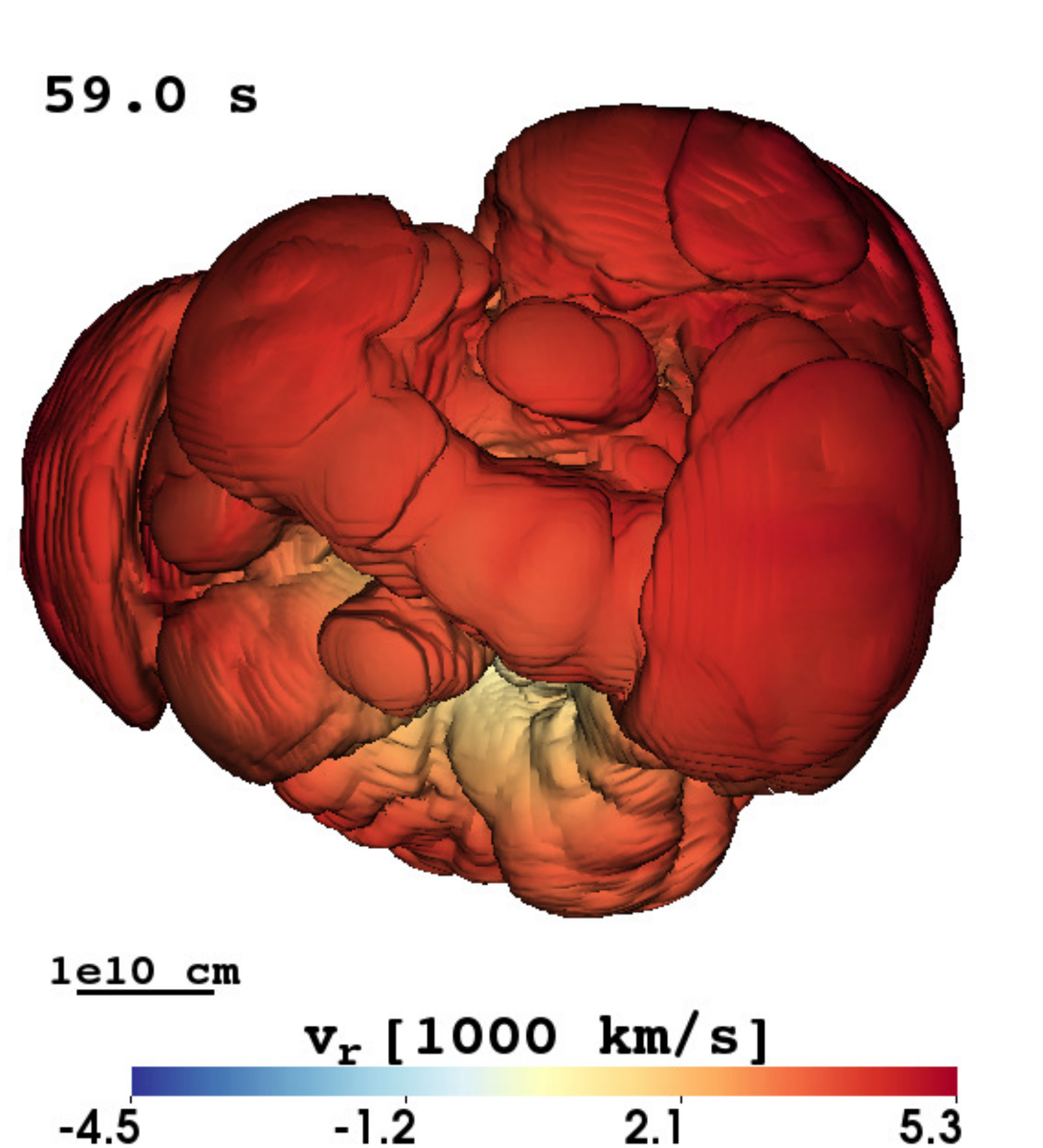}}
\hspace{-0.21cm}
\resizebox{0.252\hsize}{!}{\includegraphics*{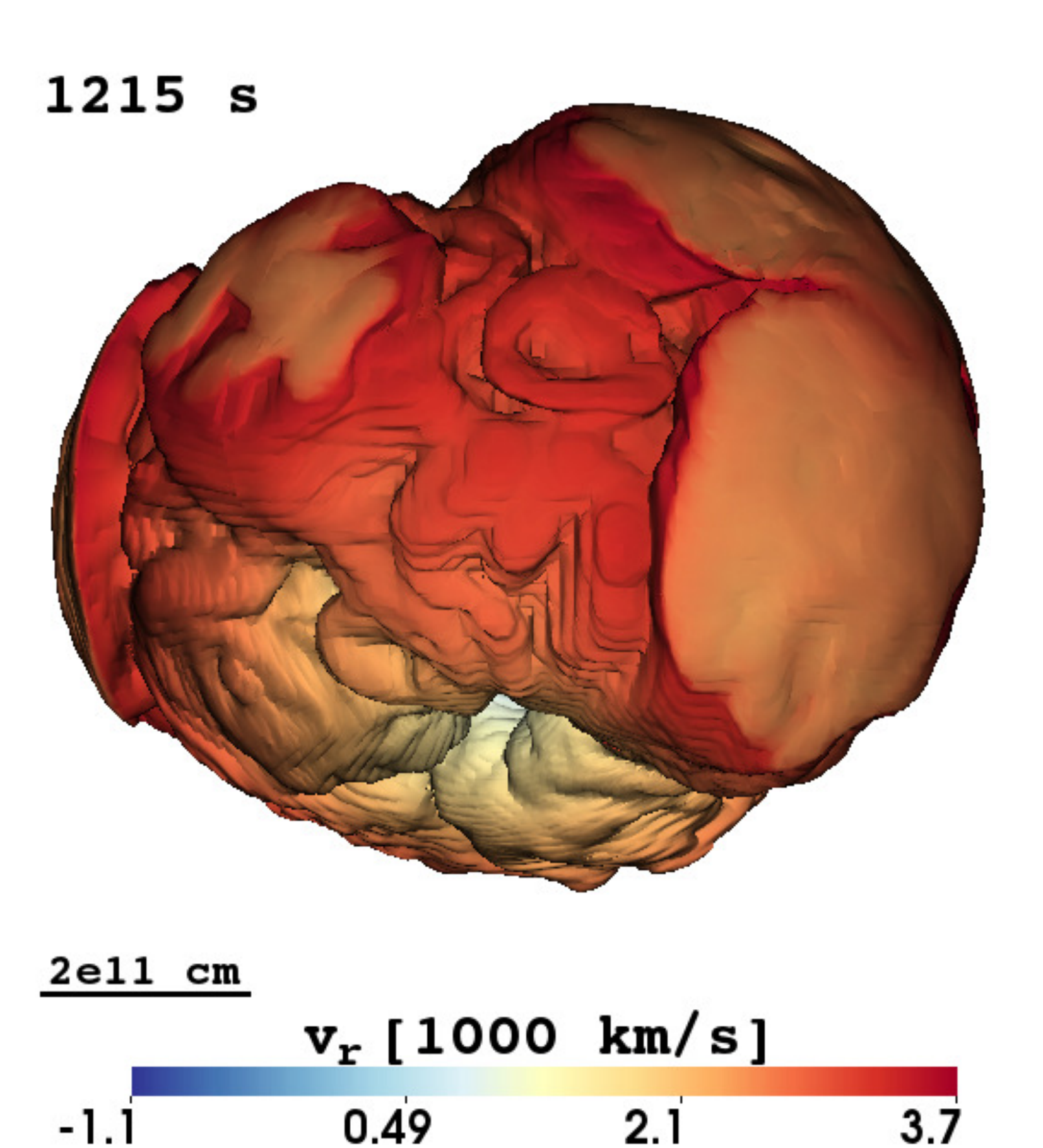}}
\hspace{-0.21cm}
\resizebox{0.252\hsize}{!}{\includegraphics*{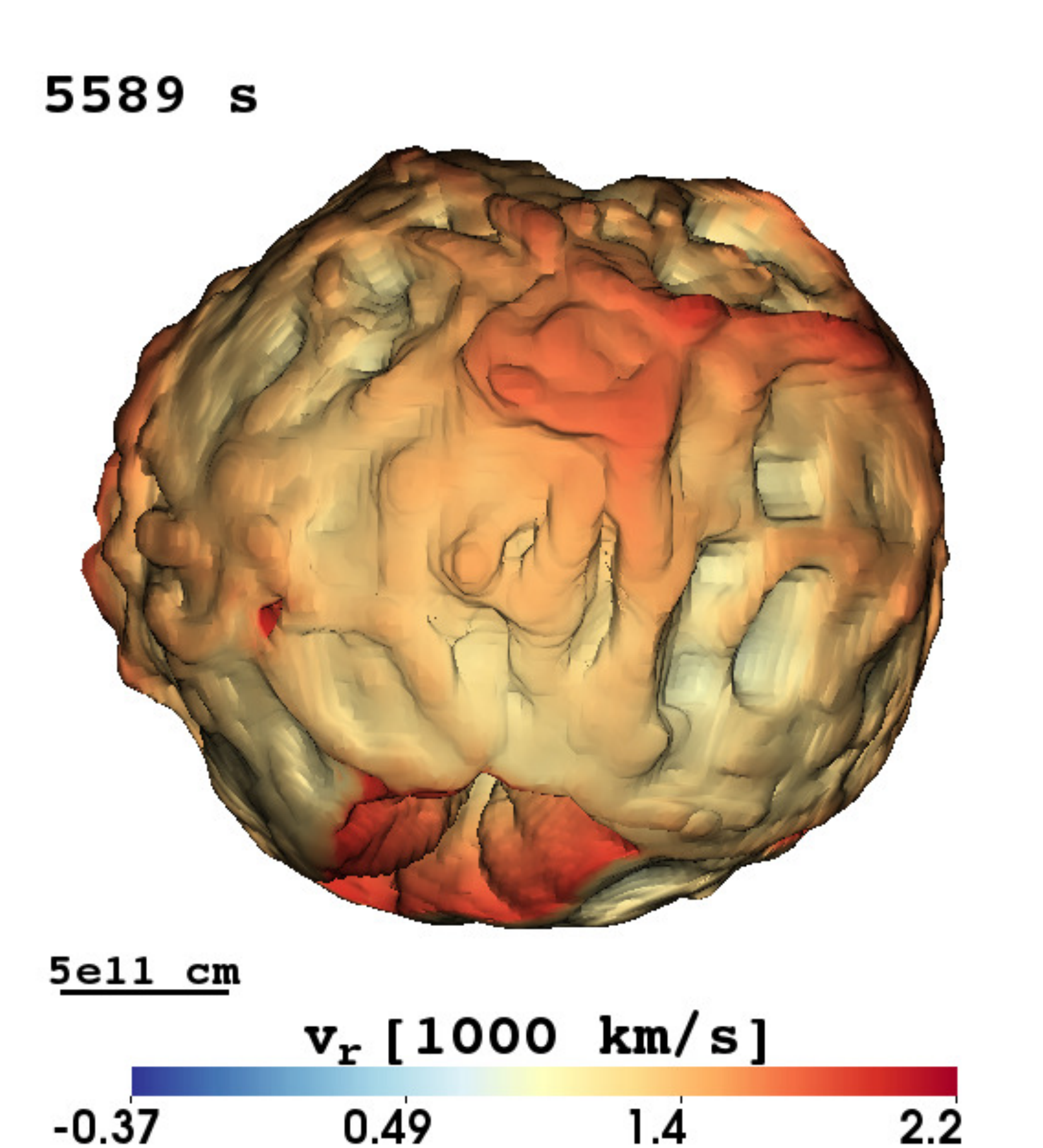}}\\
\vspace{0.3cm}
\resizebox{0.252\hsize}{!}{\includegraphics*{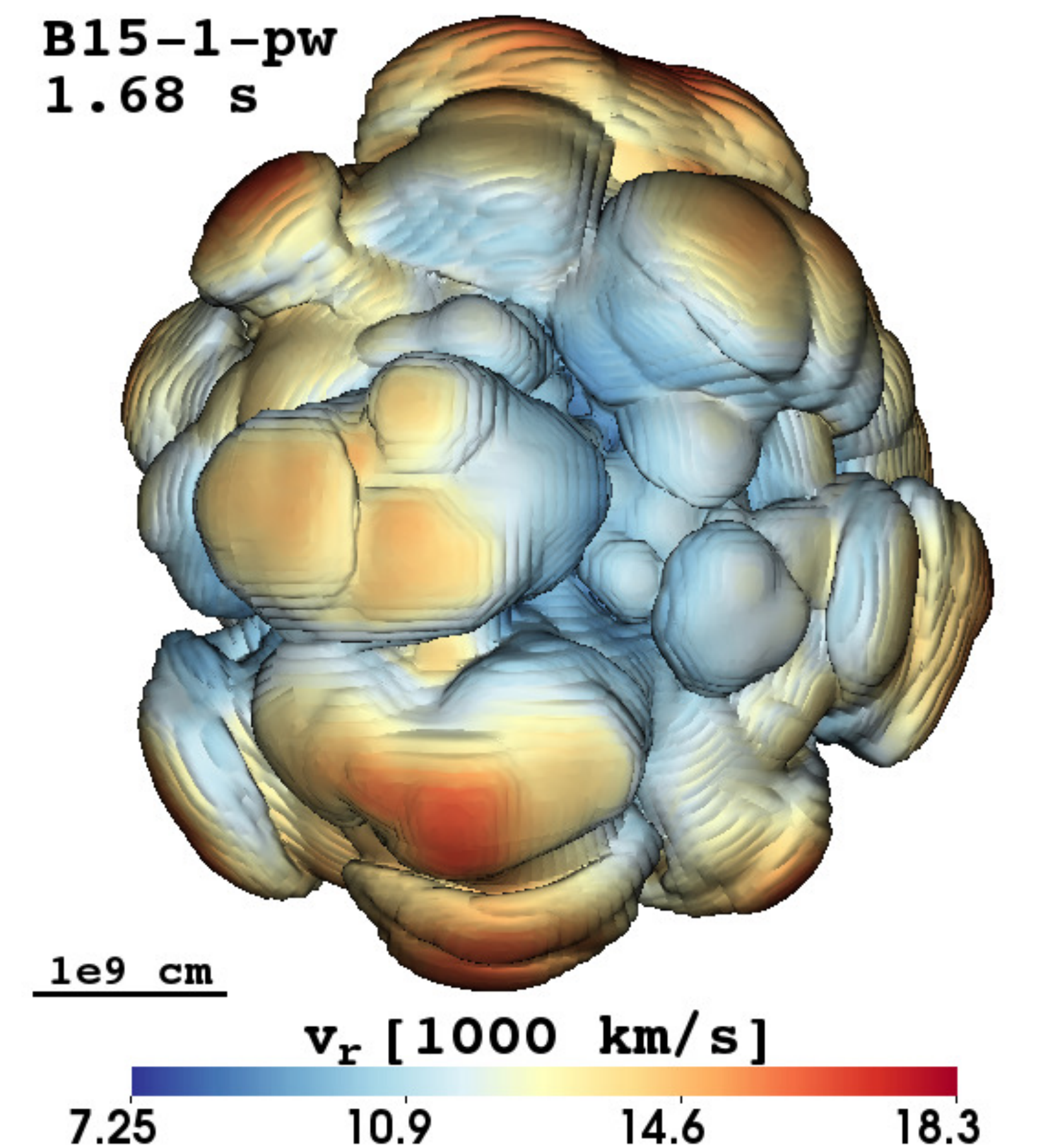}}
\hspace{-0.21cm}
\resizebox{0.252\hsize}{!}{\includegraphics*{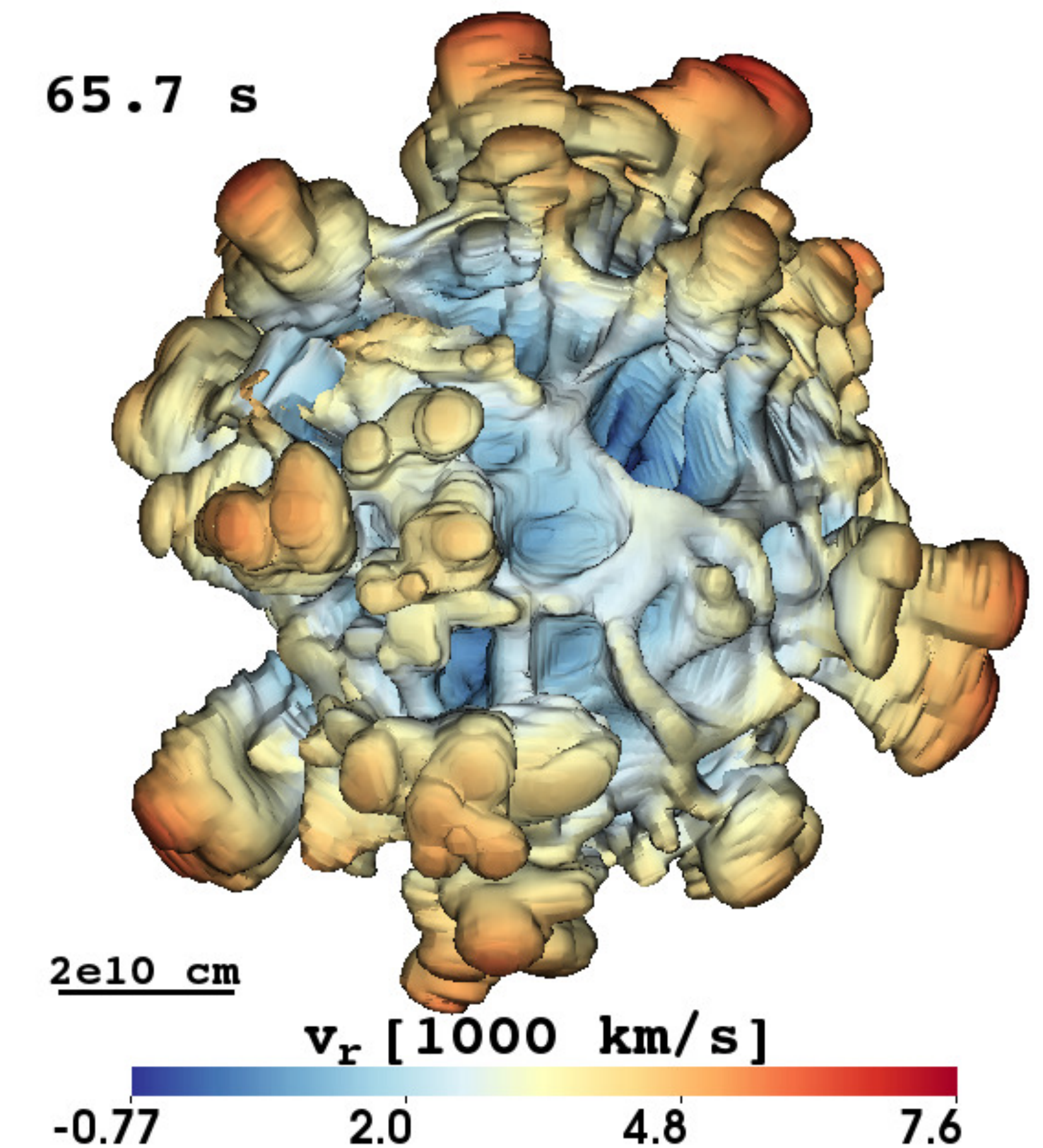}}
\hspace{-0.21cm}
\resizebox{0.252\hsize}{!}{\includegraphics*{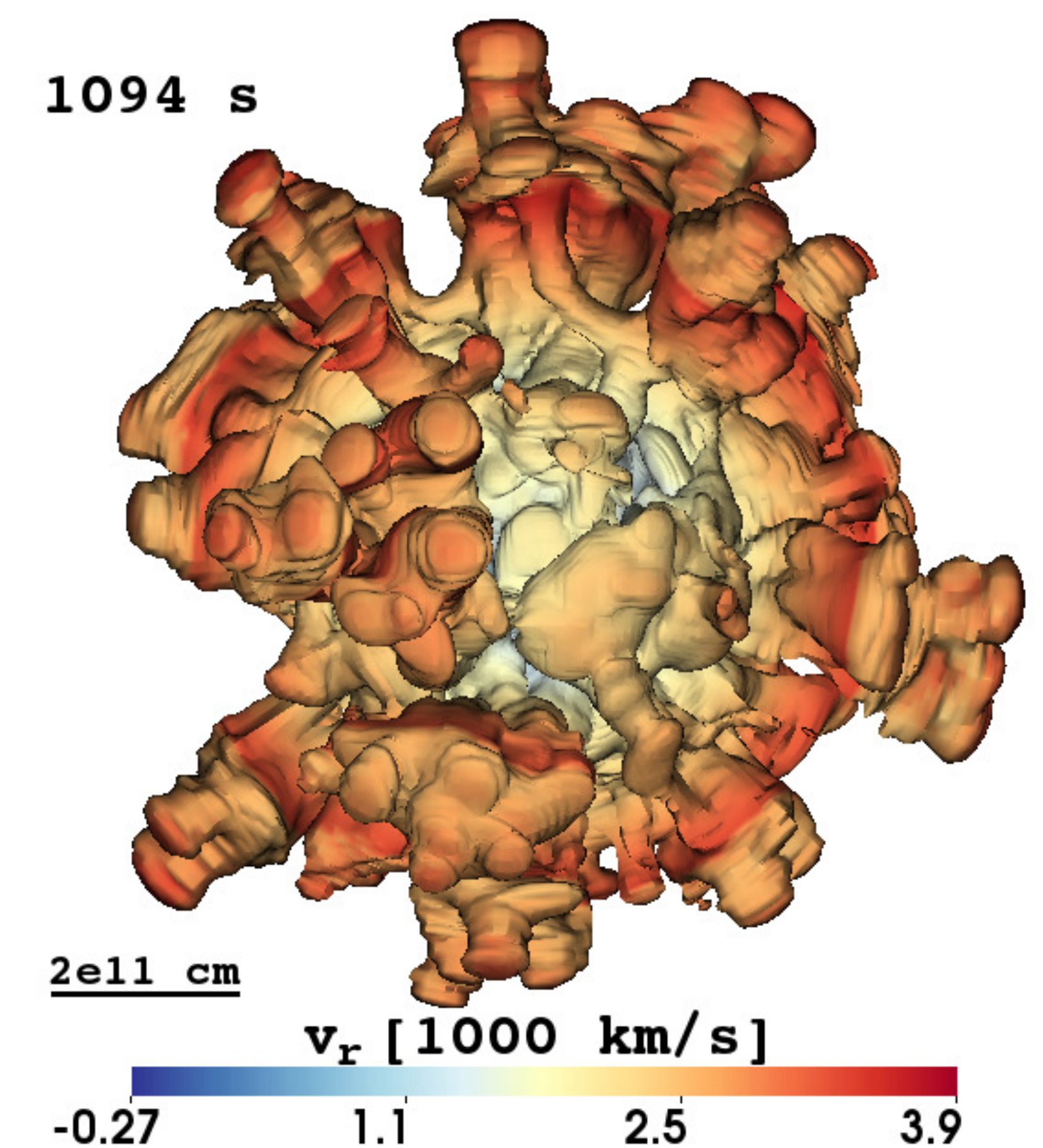}}
\hspace{-0.21cm}
\resizebox{0.252\hsize}{!}{\includegraphics*{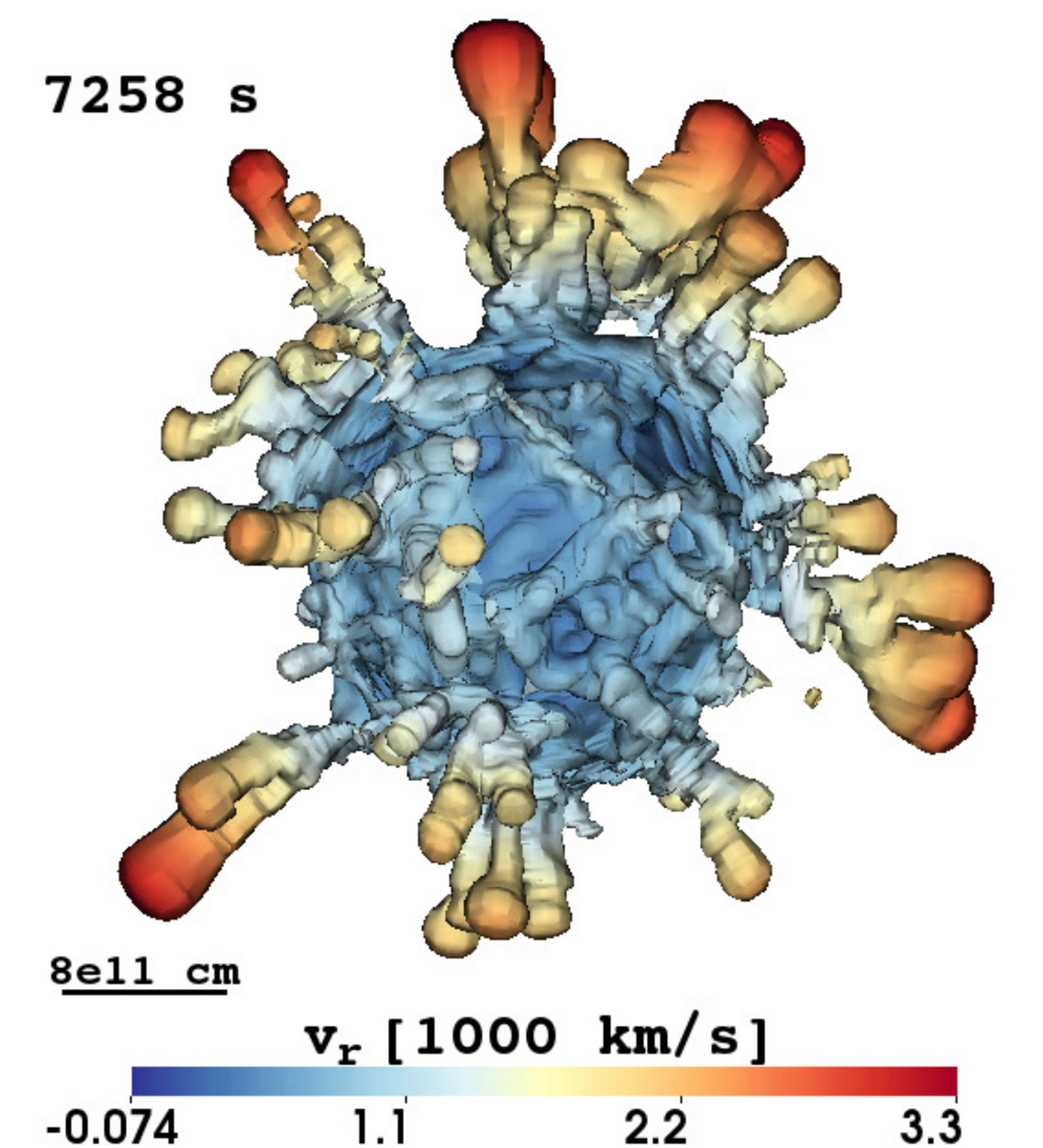}}\\
\vspace{0.3cm}
\caption{Snapshots displaying isosurfaces where the mass fraction of
  $^{56}$Ni plus n-rich tracer $X$ equals 3\% for model W15-2-cw (top
  row), L15-1-cw (second row), N20-4-cw (third row), and B15-1-pw
  (bottom row). The isosurfaces, which roughly coincide with the
  outermost edge of the neutrino-heated ejecta, are shown at four
  different epochs starting from shortly before the SN shock crosses
  the C+O/He composition interface in the progenitor star until the
  shock breakout time. The colors give the radial velocity (in units
  of km\,s$^{-1}$) on the isosurface, the color coding being defined
  at the bottom of each panel. In the top left corner of each panel we
  give the post-bounce time of the snapshot and in the bottom left
  corner a yardstick indicating the length scale. The negative y-axis
  is pointing towards the reader. One notices distinct differences in
  the final morphology of the nickel-rich ejecta of all models, which
  arise from their specific progenitor structures.}
\label{fig:time-nickel}
\end{figure*}
%

%
\begin{figure*}
\centering
\resizebox{\hsize}{!}{\includegraphics*{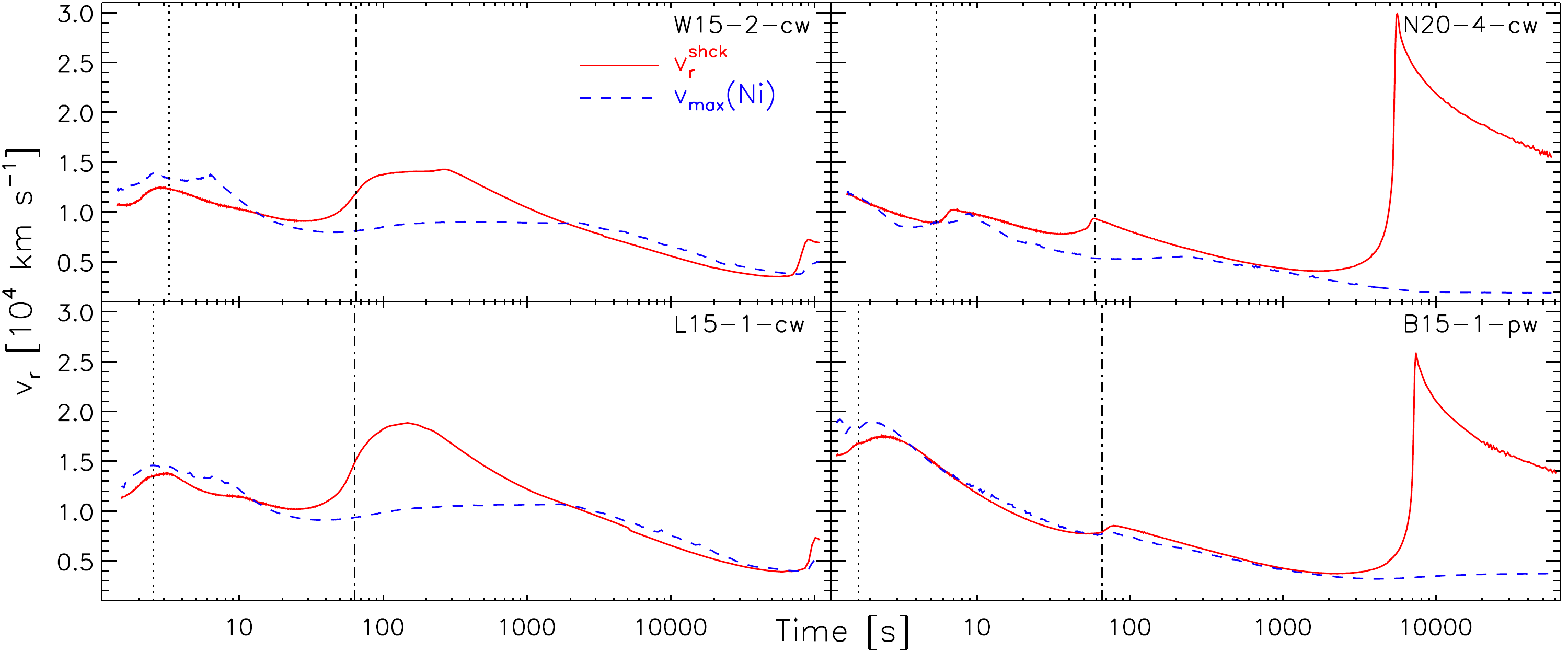}}
\caption{Time evolution of the radial velocity of the SN shock,
  $v_r^\textrm{shck}$ (solid red), and of the maximum radial velocity,
  $v_\textrm{max}(\textrm{Ni})$ (dashed blue), on the surface where
  the mass fraction of $^{56}$Ni plus the n-rich tracer X equals 3\%,
  for models W15-2-cw, N20-4-cw, L15-1-cw, and B15-1-pw,
  respectively. The vertical dotted and dashed-dotted lines mark the
  times when the shock crosses the C+O/He and He/H interfaces,
  respectively. These times are identical to those of the snapshots
  shown in the first and second column of
  Fig.\,\ref{fig:time-nickel}.}
\label{fig:v_ejecta}
\end{figure*}
%

\subsection{Propagation of the neutrino-heated ejecta}
\label{subsec:ejecta}

We find interesting differences between the results obtained for
different progenitor stars concerning the development of the
early-time asymmetries which we trace by comparing the propagation
history of the neutrino-heated ejecta bubbles. We illustrate these
differences in Fig.\,\ref{fig:time-nickel} by means of snapshots
displaying isosurfaces where the mass fraction of $^{56}$Ni plus the
n-rich tracer $X$ equals 3\% for models W15-2-cw, L15-1-cw, N20-4-cw,
and B15-1-pw (from top to bottom), respectively.  We stress that
\citet{Kifonidisetal03} showed that $^{56}$Ni is explosively produced
in "pockets" between the high-entropy bubbles of neutrino-heated
matter, \ie $^{56}$Ni reflects the asymmetries of the first second of
the explosion.  All of these models yield comparable explosion
energies with $1.39\,\mathrm{B} \leq E_\mathrm{exp} \leq 1.75\,$B. The
isosurfaces, which roughly coincide with the outermost edge of the
neutrino-heated ejecta, are shown at four epochs for each model.

Snapshots in column\,1 and 2 of Fig.\,\ref{fig:time-nickel} display
the isosurfaces at the moments when the SN shock is about to cross the
C+O/He and He/H interfaces, \ie when the maximum $R_\mathrm{s}$
becomes greater than $R_\mathrm{C+O/He}$ and $R_\mathrm{He/H}$,
respectively. Column\,3 depicts the isosurfaces at a time when the
neutrino-heated ejecta are strongly slowed down by the reverse shock
forming as a consequence of the deceleration of the SN shock in the
hydrogen envelope, and column\,4 shows the situation at the shock
breakout time. The isosurfaces are color-coded according to the value
of the radial velocity to depict the angle-dependent ejecta speed.  In
Fig.\,\ref{fig:v_ejecta}, we further show $v_\textrm{max}
(\textrm{Ni})$, the maximum radial velocity on the isosurface, as a
function of time together with $v_r^\textrm{shck}$.

At the time when the SN shock reaches the C+O/He interface the
morphology of the neutrino-heated ejecta still resembles that imposed
upon them by neutrino-driven convection and SASI mass motions during
the launch of the explosion (Fig.\,\ref{fig:time-nickel}, column
1). The ejecta morphology features fast plumes accelerated by the
buoyant rise of neutrino-heated postshock matter, the number and
angular sizes of the plumes being similar in all models. The plumes
are somewhat compressed in model N20-4-cw since they are decelerated
while propagating through the C+O shell (Fig.\,\ref{fig:v_ejecta}, top
right panel). This contrasts with the situation in the other three
models, where the plumes of neutrino-heated ejecta accelerate inside
the C+O layer.

During the later evolution (\ie after the SN shock has entered the
helium layer of the star) differences develop in the morphology of the
iron/nickel-rich ejecta between models from different progenitor
stars. They are the result of the interrelation between the
propagation of the iron/nickel-rich ejecta and the SN shock, which
depends on the density structure of the progenitor star and the
explosion energy.

\subsubsection{Fragmentation in the helium shell} 
After the SN shock has crossed the C+O/He interface it decelerates
while propagating through the helium layer of the star. Dense shells
form behind the decelerating shock and become RT unstable, the
strength of the RT instability being determined by the amount of shock
deceleration \citep{HerantBenz91}. The RT instability changes the
initial morphology of the neutrino-heated ejecta. 

Among all models we observe the largest change of the early-time
morphology in model B15-1-pw (Fig.\,\ref{fig:time-nickel}, 2nd
column), where the buoyant fast plumes of neutrino-heated ejecta
fragment into numerous smaller fingers. This behavior agrees well with
the results of the 1D linear analysis presented in
Sect.\,\ref{sec:1dmodels}, which yielded the largest time-integrated
RT growth factor at the C+O/He interface in the B15-1-pw-a model
(Fig.\,\ref{fig:lingrowth}, bottom right panel), fully compatible with
the strongest deceleration in the He-layer of the B15 models. Note
that the largest secondary RT fingers grow from the biggest initial
buoyant bubbles.

The shock decelerates in the helium layer of the RSG models, W15-2-cw
and L15-1-cw, less than in model B15-1-pw, \ie the RT instabilities
grow slower at the C+O/He interface in these two models. Consequently,
the neutrino-heated ejecta in models W15-2-cw and L15-1-cw show only
minor fragmentation at the tips of the rising fast plumes.

Contrary to all other models, the neutrino-heated ejecta do not
exhibit any significant morphological change in model N20-4-cw,
implying no or very little radial mixing of iron/nickel-rich matter
with lighter elements. Why does model N20-4-cw show no vigorous mixing
at this stage of the evolution, although the SN shock decelerates
quite similarly in the helium shell of this model as in the RSG
models?  The answer can be inferred from the relative velocity between
the neutrino-heated ejecta and the SN shock during the period of time
when they are approaching the C+O/He interface. As discussed above
(see Sect.\,\ref{subsec:shock}) the shock accelerates when crossing
the interface and then decelerates inside the helium layer, which
causes the post-shock C+O rich matter to decelerate, too.  In
addition, a closer inspection of Fig.\,\ref{fig:v_ejecta} reveals that
when approaching the C+O/He interface the ejecta propagate slower than
the shock in model N20-4-cw, but faster in all other models.  This
means that the neutrino-heated ejecta of model N20-4-cw impact the
decelerating post-shock layer with a smaller relative velocity than in
all other models, which results in a smaller perturbation of the RT
unstable post-shock layer, and less radial mixing.

\subsubsection{Encounter with the reverse shock} 
The morphology of the iron/nickel-rich ejecta changes considerably
after the SN shock has been decelerated in the hydrogen envelope of
the star. The deceleration of the SN shock causes a pileup of
postshock He-rich matter into a thick, dense shell, called the
``helium wall'' \citep{Kifonidisetal03}. A pressure wave propagates
upstream and eventually steepens into a strong reverse shock, which
decelerates and compresses the neutrino-heated ejecta when
encountering them.

In the RSG models W15-2-cw and L15-1-cw a few broad plumes of
iron/nickel-rich ejecta that dominate the ejecta morphology at a time
of $\sim 60\,$s (Fig.\,\ref{fig:time-nickel}, 2nd column) stretch into
finger-like structures (Fig.\,\ref{fig:time-nickel}, 3rd column),
because the plumes possess larger radial velocities than the bulk of
the iron/nickel-rich ejecta. The tips of the fingers are compressed
and flattened when hit by the reverse shock.  

The ejecta morphology evolves quite differently in model N20-4-cw.
The iron/nickel-rich plumes do not have enough time in this model to
grow into finger-like structures before they are slowed down by the
reverse shock, because of the earlier onset of SN shock deceleration
and reverse shock formation in the more compact BSG stars (see
Fig.\,\ref{fig:v_ejecta}).  Hence, the iron/nickel-rich ejecta consist
of broader and roundish structures in this model.  The snapshots of
model B15-1-pw show no imprint of the reverse shock on the morphology
of the neutrino-heated ejecta. Differently from the other three models
the reverse shock forms {\it behind} the fast-moving iron/nickel-rich
plumes, but ahead of the bulk of the ejecta in this model. It
manifests itself as a color discontinuity (from red to orange) in the
plot depicting the radial velocity of the iron/nickel-rich ejecta
(Fig.\,\ref{fig:time-nickel}, 3rd column, bottom row).

How can the fast iron/nickel-rich fingers escape the deceleration by
the reverse shock in model B15-1-pw, while this is not the case in the
other three models? Again, the answer to this question can be inferred
by comparing the time evolution of $v_\mathrm{max}$(Ni) and
$v_r^\mathrm{shck}$ (Fig.\,\ref{fig:v_ejecta}). In model B15-1-pw
matter in the fingers has very high velocities (nearly
20\,000\,km\,s$^{-1}$) at the time when the SN shock crosses the
C+O/He interface. Both the fast iron/nickel-rich fingers and the SN
shock are decelerated afterwards while propagating through the helium
layer of the exploding star, their radial velocities being almost the
same during this period ($3\,\mathrm{s} \la t \la 70\,$s). Hence, the
fast iron/nickel-rich fingers stay close to the SN shock and propagate
ahead of the mass shell where the reverse shock will form at a later
time. This situation was also found in the 2D simulations of
\citet{Kifonidisetal06} and the 3D simulation of HJM10, who both used
the B15 progenitor model, too.

In contrast, the fastest iron/nickel-rich ejecta of model N20-4-cw
have a lower radial velocity than the SN shock when they enter the
helium shell. They also experience a stronger deceleration than the
shock while propagating through the helium layer, \ie the difference
between their radial velocity and that of the SN shock increases
(Fig.\,\ref{fig:v_ejecta}). Thus, the iron/nickel-rich ejecta stay
further and further behind the SN shock, and even their fastest parts
(the fingers) will be compressed later by the reverse shock. The
situation seems to differ for the RSG models, W15-2-cw and L15-1-cw,
because the radial velocities of their iron/nickel-rich ejecta are
higher than the velocity of the SN shock at the time when the ejecta
cross the C+O/He interface. However, as the SN shock strongly
accelerates at the He/H interface owing to the steep density gradient
there, it propagates well ahead of the iron/nickel-rich ejecta before
it begins to decelerate again. Thus, the reverse shock forms ahead of
the extended fast iron/nickel-rich fingers, which then have time to
grow in the RSG stars where the reverse shock develops much later than
in the BSG progenitors.

\subsubsection{Morphology at shock breakout} 
After the iron/nickel-rich ejecta have been compressed by the reverse
shock below the He/H interface, subsequent fragmentation by RT
instabilities can lead to further significant changes in their
morphology.  This is illustrated in Fig.\,\ref{fig:time-nickel} (4th
column), where we display the morphology of the iron/nickel-rich
ejecta at the shock breakout time. The RT instabilities result from
the strong deceleration of the SN shock in the hydrogen
envelope. According to the time-integrated RT growth factors obtained
with our 1D models in Sect.\,\ref{sec:1dmodels} we expect the RT
instabilities at the He/H interface to be stronger in the RSG models
than in the BSG models (see Fig.\,\ref{fig:lingrowth}).  Indeed, the
RSG models W15-2-cw and L15-1-cw show strong fragmentation of the
iron/nickel-rich ejecta, especially at the tips of the extended fast
fingers, which are hit by the reverse shock first. On the other hand,
the morphology of the iron/nickel-rich ejecta surface of the BSG model
N20-4-cw remains overall roundish except for some small amount of
additional fine structure.

%
\begin{figure}
\centering
\resizebox{0.498\hsize}{!}{\includegraphics*{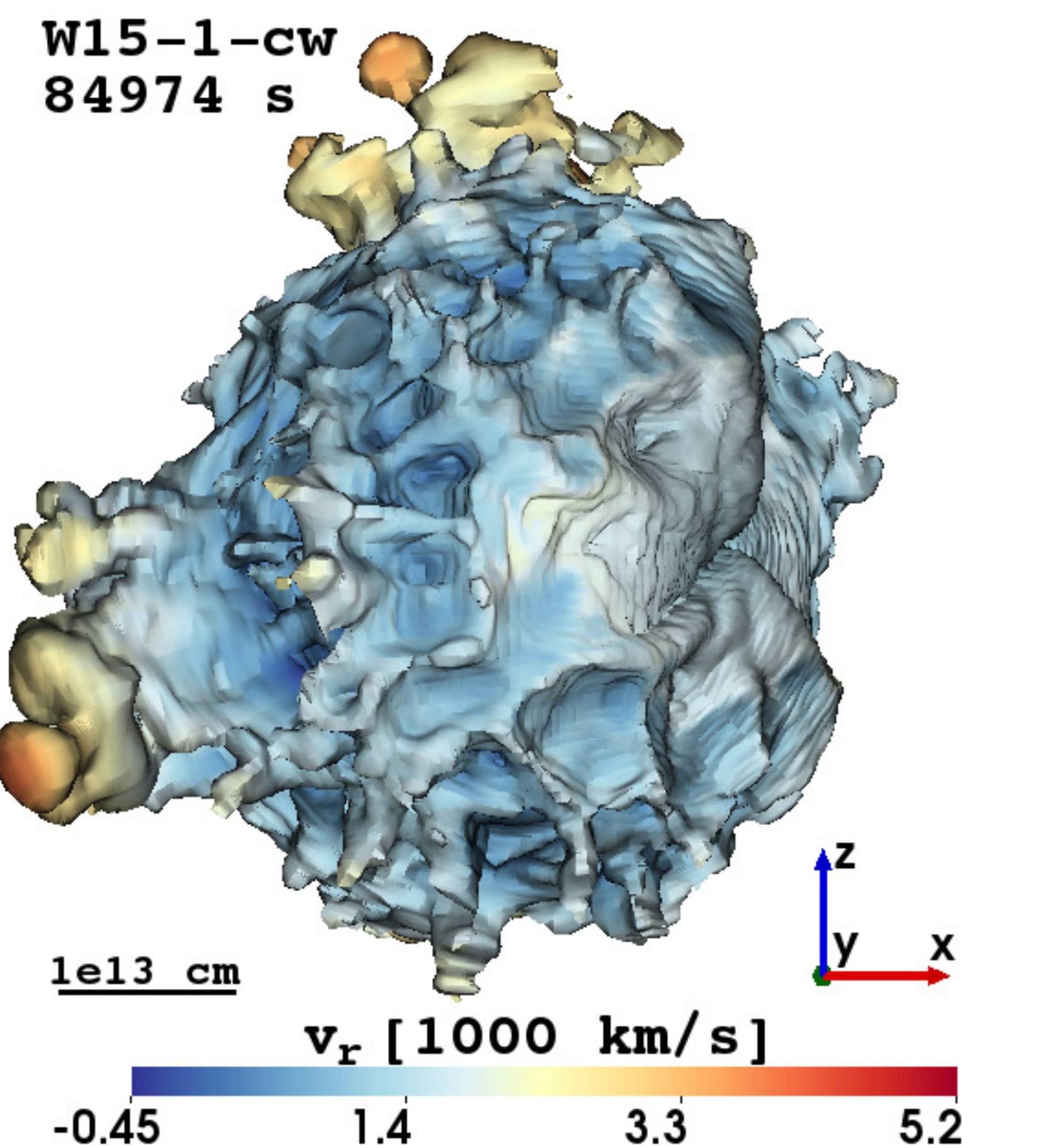}}
\hspace{-0.22cm}
\resizebox{0.498\hsize}{!}{\includegraphics*{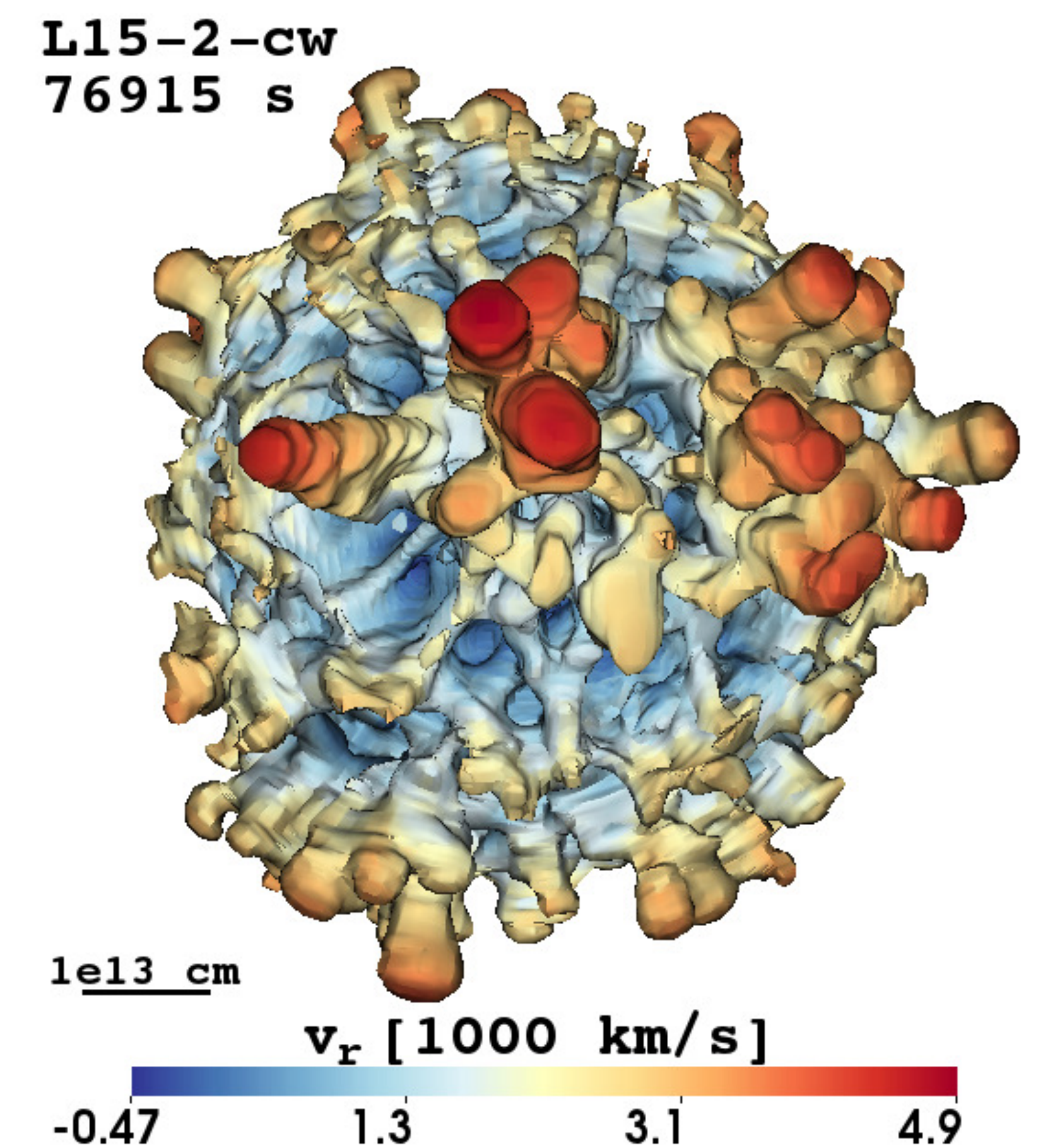}}
\resizebox{0.498\hsize}{!}{\includegraphics*{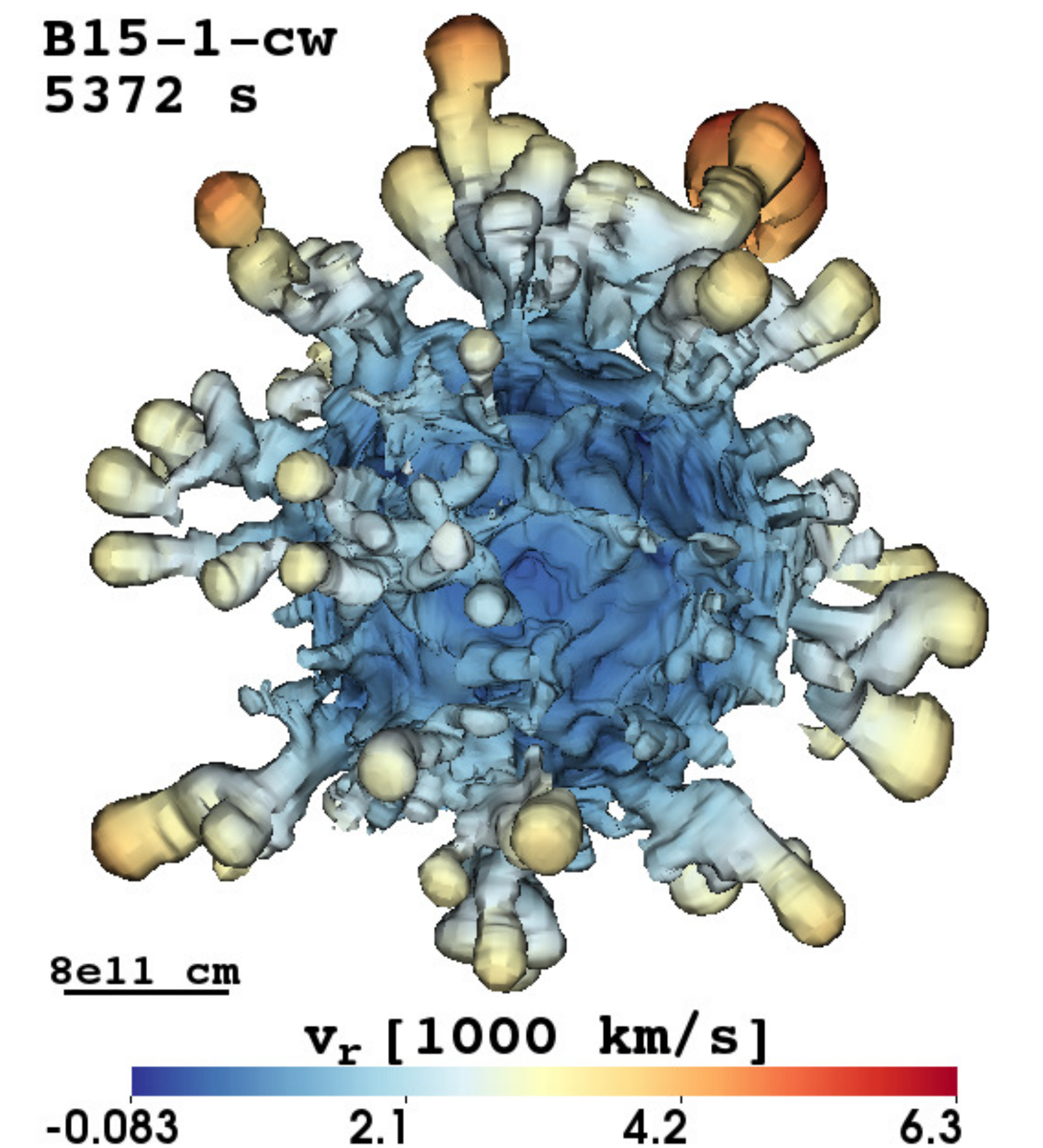}}
\hspace{-0.22cm}
\resizebox{0.498\hsize}{!}{\includegraphics*{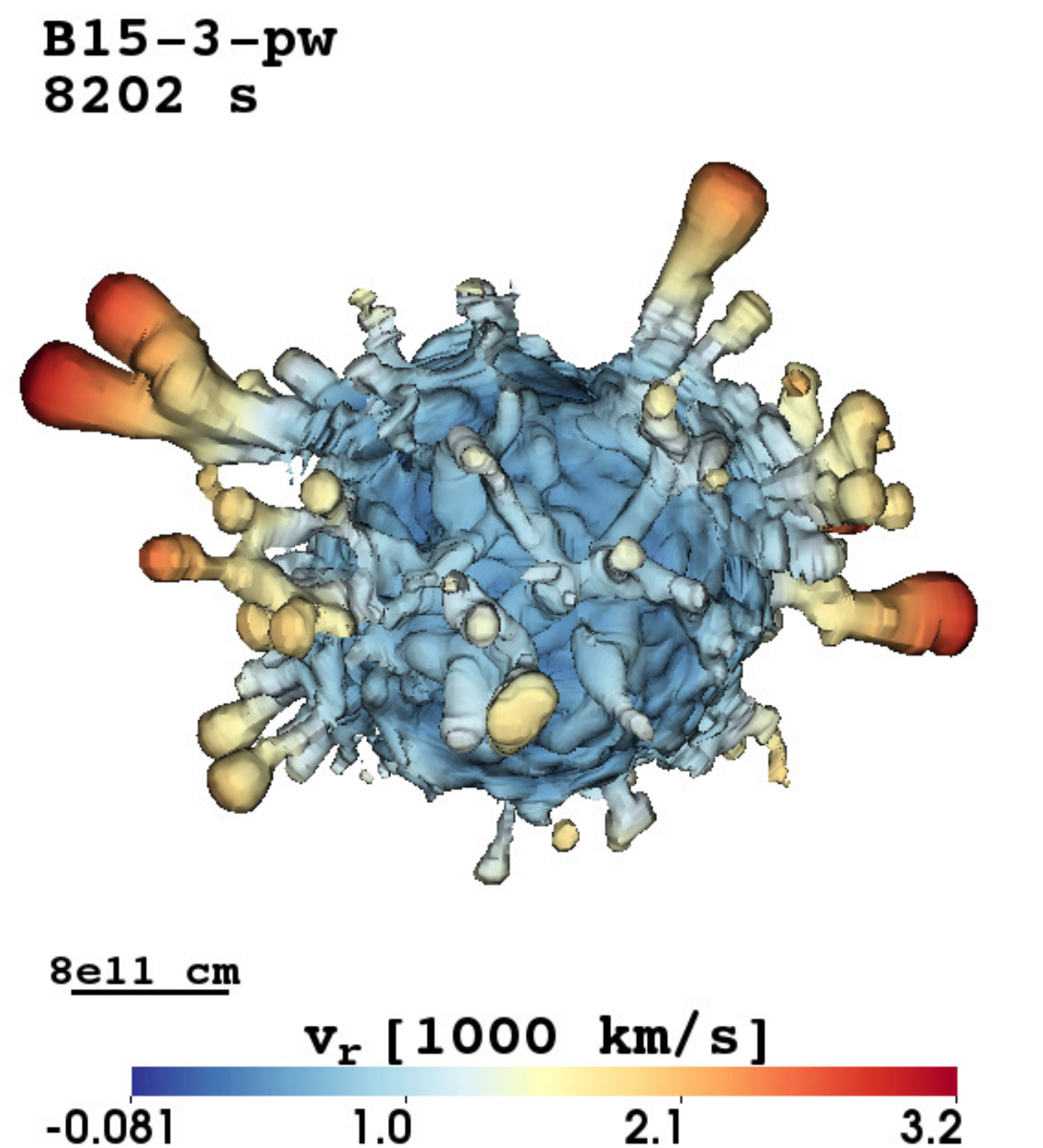}}

\caption{Same as Fig.\,\ref{fig:time-nickel}, but for models W15-1-cw,
  L15-2-cw, B15-1-cw, and B15-3-pw at the shock breakout time.}
\label{fig:time-nickel2}

\end{figure}
%

%
\begin{figure}
\centering
\resizebox{0.498\hsize}{!}{\includegraphics*{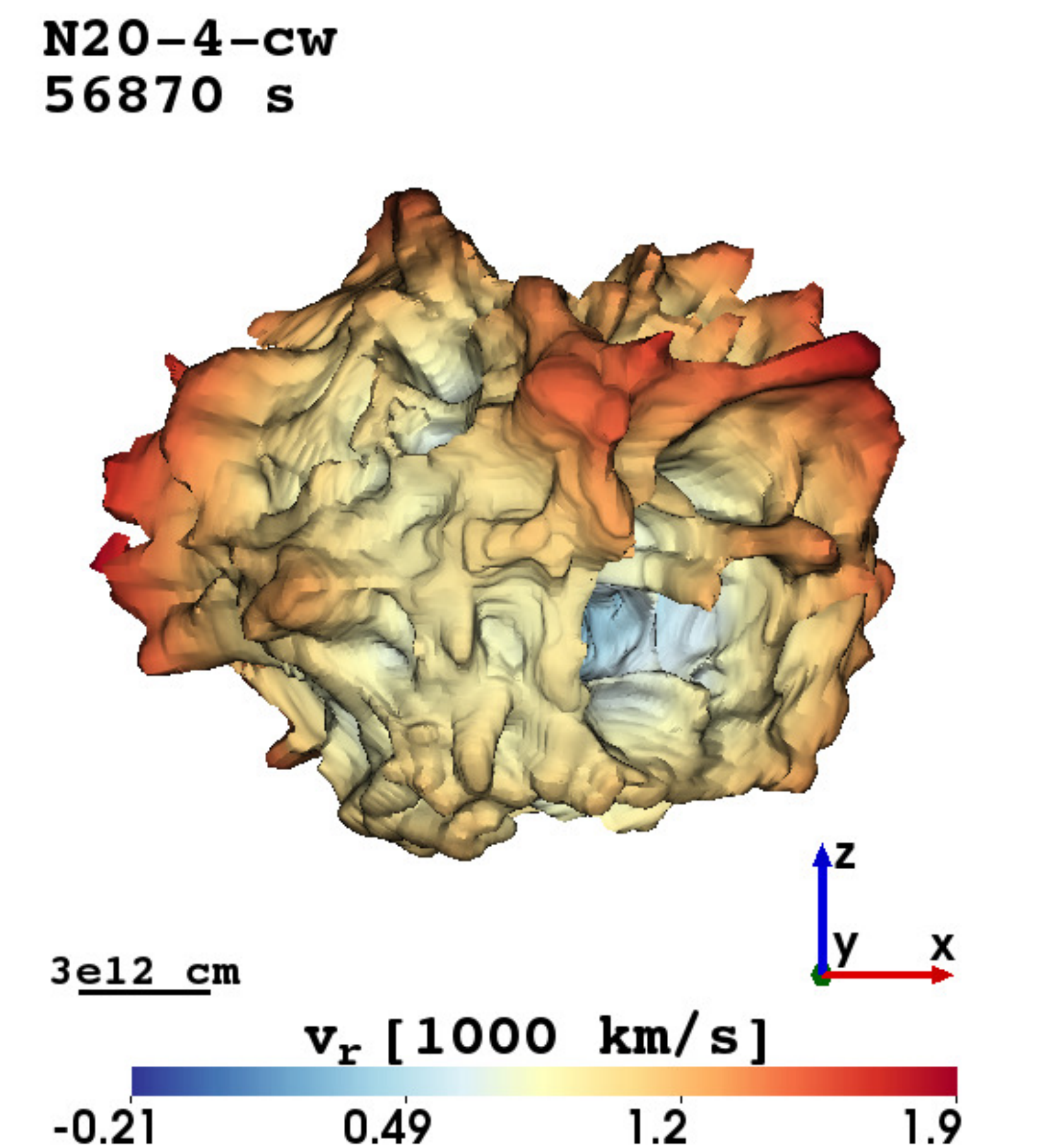}}
\hspace{-0.22cm}
\resizebox{0.498\hsize}{!}{\includegraphics*{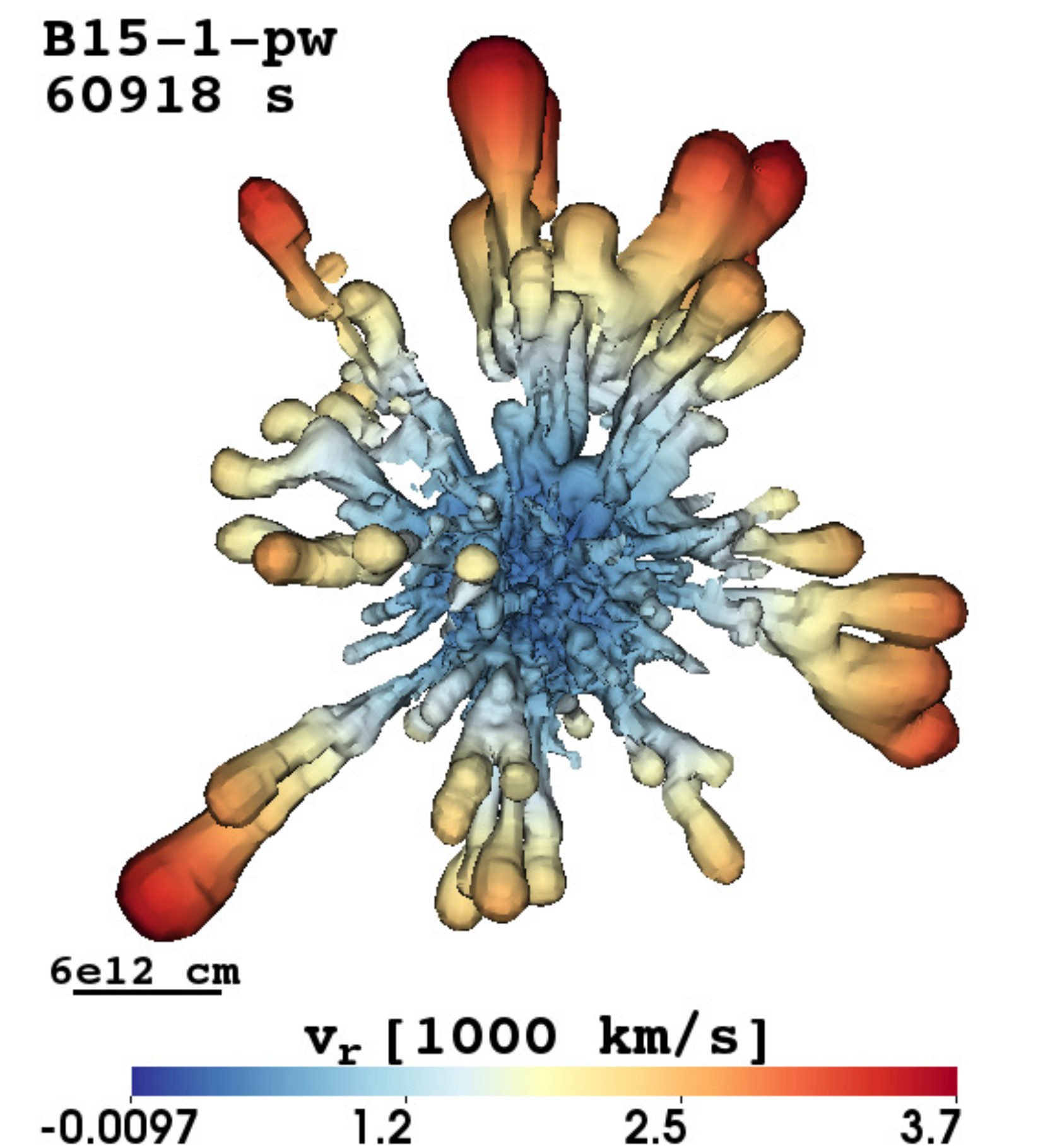}}

\caption{Same as Fig.\,\ref{fig:time-nickel}, but for models N20-4-cw
  and B15-1-pw at 56870\,s and 60918\,s, respectively.}
\label{fig:bsg-ext}

\end{figure}
%

%
\begin{figure}
\centering
\resizebox{0.34\hsize}{!}{\includegraphics*{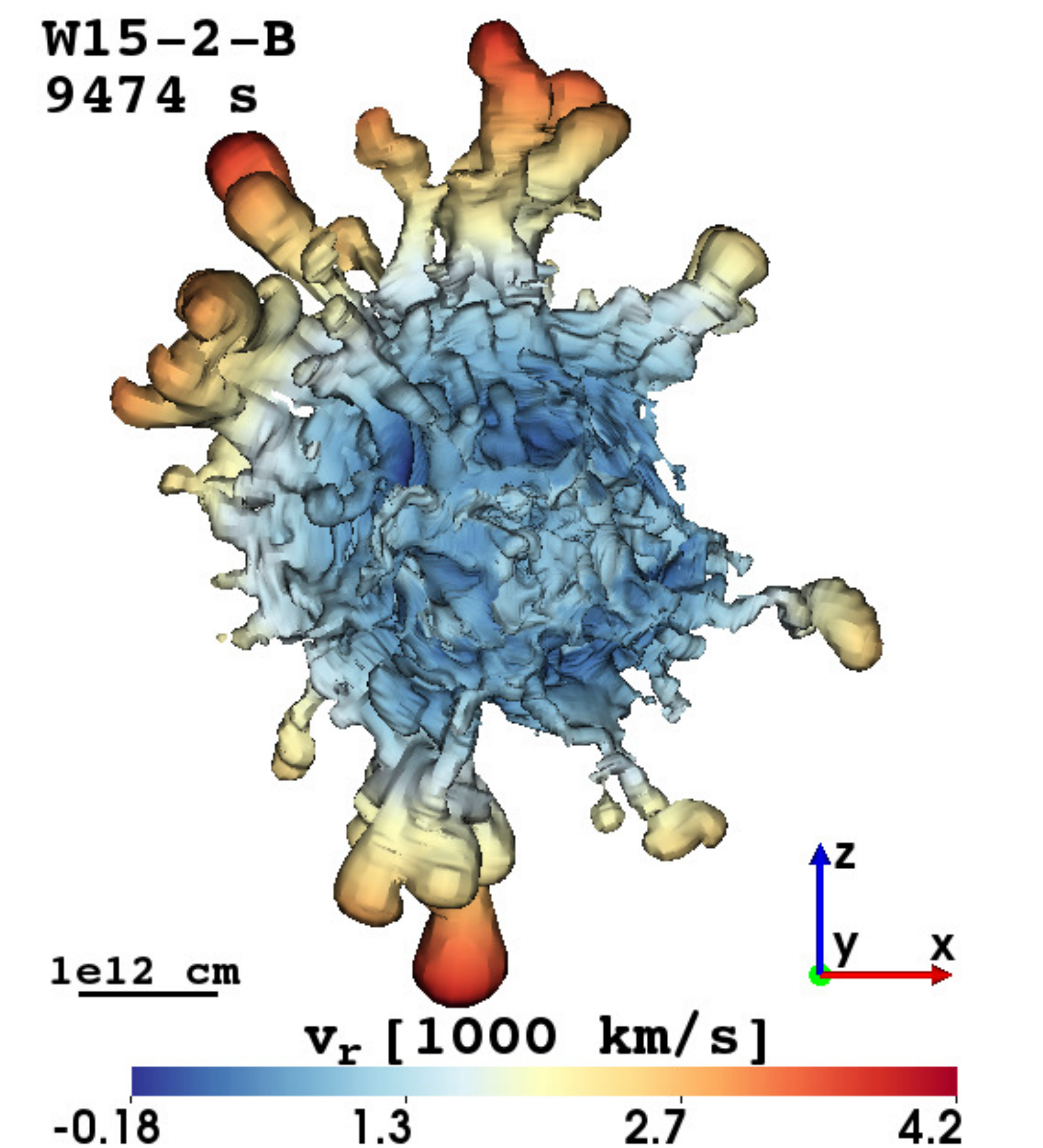}}
\hspace{-0.3cm}
\resizebox{0.34\hsize}{!}{\includegraphics*{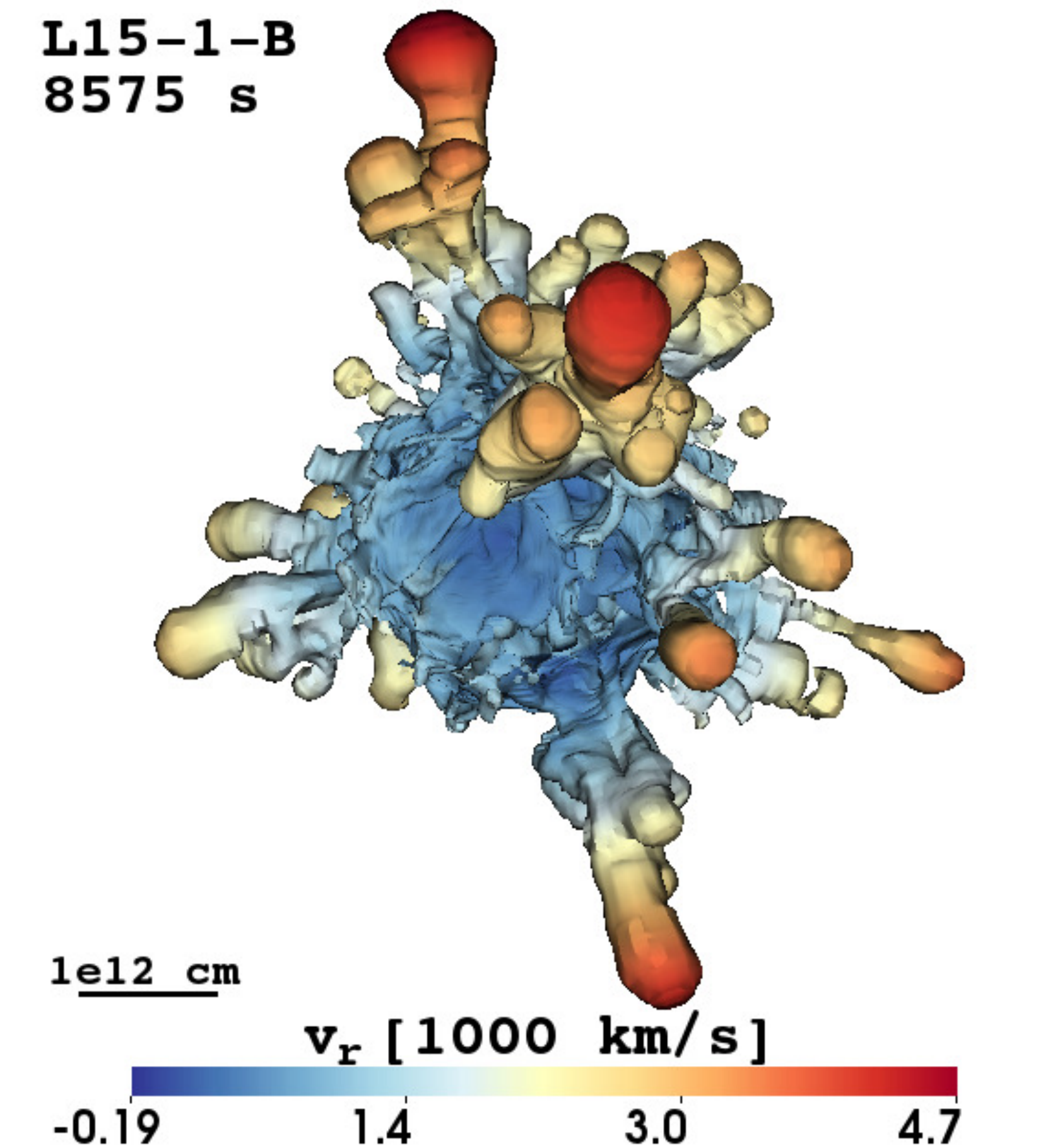}}
\hspace{-0.3cm}
\resizebox{0.34\hsize}{!}{\includegraphics*{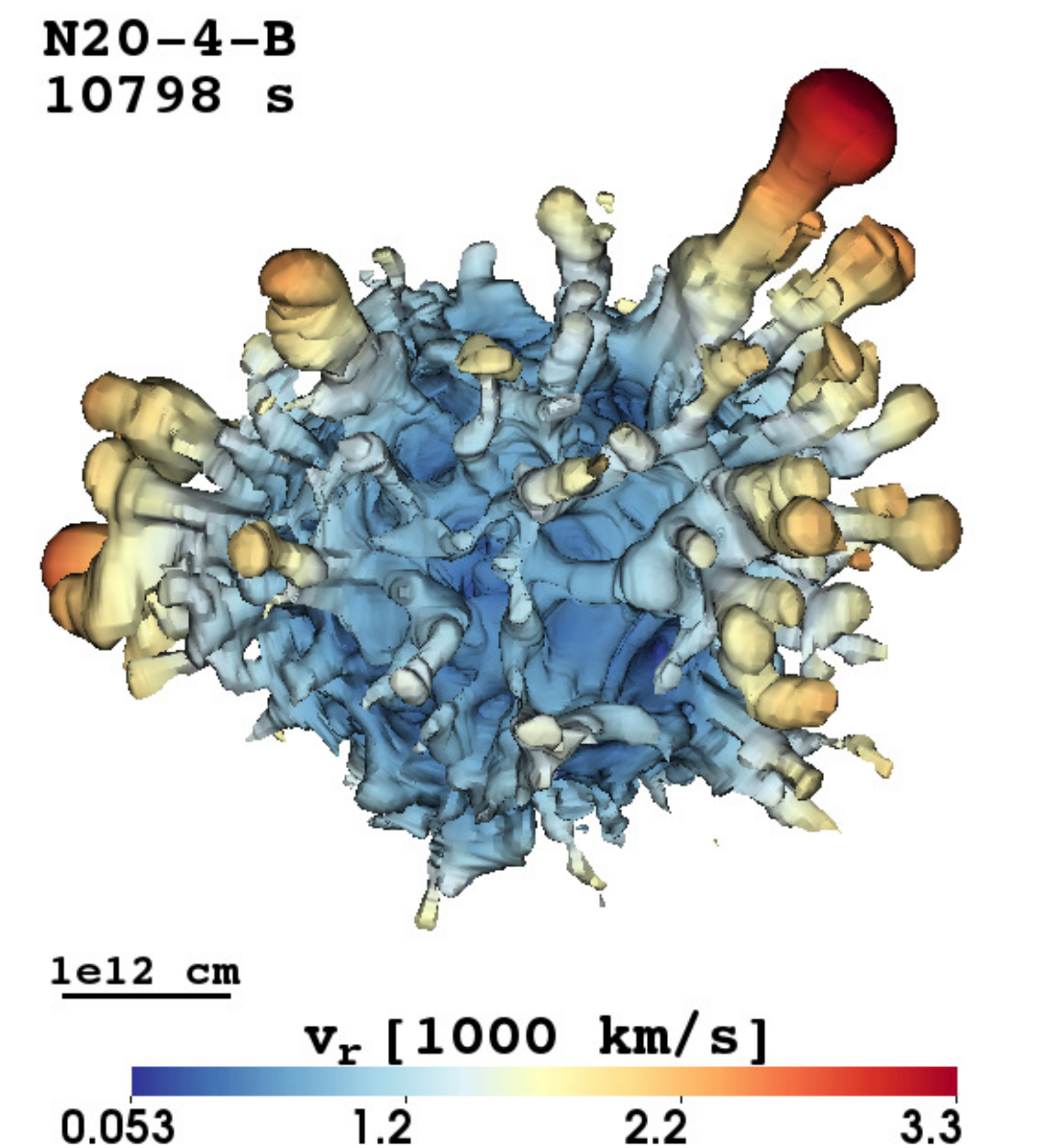}}

\caption{Same as Fig.\,\ref{fig:time-nickel}, but for the additional
  3D models W15-2-B, L15-1-B, and N20-4-B, discussed in
  Sect.\,\ref{subsec:additional3D}.}
\label{fig:additional3D}

\end{figure}
%

The morphology of the iron/nickel-rich ejecta of model B15-1-pw
differs distinctly and visibly from that of the other three models at
shock breakout time. Because a fraction of the iron/nickel-rich ejecta
can avoid deceleration by the reverse shock the ejecta separate into
two components: a fast component that was able to avoid the reverse
shock, and a second, slower one that was drastically slowed down by
it. The fast component evolves into elongated RT fingers stretching
very far into the hydrogen envelope and experiencing almost no further
deceleration. The slower component constitutes the spherical central
part of the ejecta expanding at much lower velocities than the fast
component.

\subsubsection{Dependence on explosion energy} 
\label{subsubsec:eexp}
The models shown in Fig.\,\ref{fig:time-nickel}, one for each
considered progenitor model, have almost the same explosion
energy. This fact does not imply, however, a restriction concerning
our discussion of the iron/nickel-rich ejecta morphology, because the
explosion energy is not responsible for the differences we described
above. To prove this statement, we show in
Fig.\,\ref{fig:time-nickel2} the morphology of the iron/nickel-rich
ejecta of models W15-1-cw, L15-2-cw, B15-1-cw, and B15-3-pw at the
time of shock breakout. Despite having the highest explosion energy of
all calculated models the iron/nickel-rich ejecta in model L15-2-cw
are still trapped completely by the reverse shock. Moreover, model
B15-3-pw, which possesses the lowest explosion energy of all models,
yields the same type of morphology of the iron/nickel-rich ejecta as
model B15-1-pw, because the fast iron/nickel-rich RT filaments avoid a
strong deceleration by the reverse shock in both models.

Increasing the explosion energy results in a faster propagation of the
iron/nickel-rich ejecta, but also gives rise to a faster SN shock, \ie
the relative velocity between the shock and the iron/nickel-rich
ejecta, which is the most crucial factor determining whether or not
the iron/nickel-rich ejecta will be strongly decelerated by the
reverse shock, does not change. We further note that although the
iron/nickel-rich ejecta show a lesser degree of global asymmetry in
model L15-2-cw than in model L15-1-cw, they evolve in a very similar
manner. The asymmetry is less in model L15-2-cw because it explodes
faster, which results in a more spherically symmetric distribution of
iron/nickel-rich hot bubbles. We also tested whether the fundamental
features of radial mixing depend on the seed perturbation pattern
imposed during the explosion phase. A comparison of the results of
models W15-1-cw and W15-2-cw shows that this is not the case (see
Figs.\,\ref{fig:time-nickel} and \ref{fig:time-nickel2}). However, we
find interesting differences: $v_\textrm{max}(\textrm{Ni})$ varies by
up to 20\% among the models, while $\langle v
\rangle_{1\%}(\textrm{Ni})$ is similar for all models.

\subsubsection{Dependence on progenitor star} 
To be able to discuss possible differences between the morphology of
the iron/nickel-rich ejecta of the RSG and BSG models at the time of
shock breakout in the RSG progenitors, we had to evolve the BSG models
until approximately 60000\,s, which is about a factor of ten beyond
the time of shock breakout in the BSG models.  Fig.\,\ref{fig:bsg-ext}
displays the morphology of the iron/nickel-rich ejecta of the BSG
models N20-4-cw and B15-1-pw at 56870\,s and 60918\,s, respectively.
After the SN shock has crossed the surface of the progenitor star,
instabilities continue to grow in model N20-4-cw and the morphology of
the iron/nickel-rich ejecta remains no longer smooth and roundish.
Narrow spikes/walls of iron/nickel-rich matter stretch in radial
direction and further fragmentation of the overall ejecta structure
occurs. On the other hand, in our second BSG model B15-1-pw the
elongated fingers containing iron/nickel-rich matter grow
significantly in length. They even reaccelerate from $~
3300\,$km\,s$^{-1}$ at shock breakout to $~3700\,$km\,s$^{-1}$ at
61\,000\,s, but the angular extent and the orientation of these
fingers remain unchanged.

%
\begin{figure*}
\centering
\resizebox{\hsize}{!}{\includegraphics*{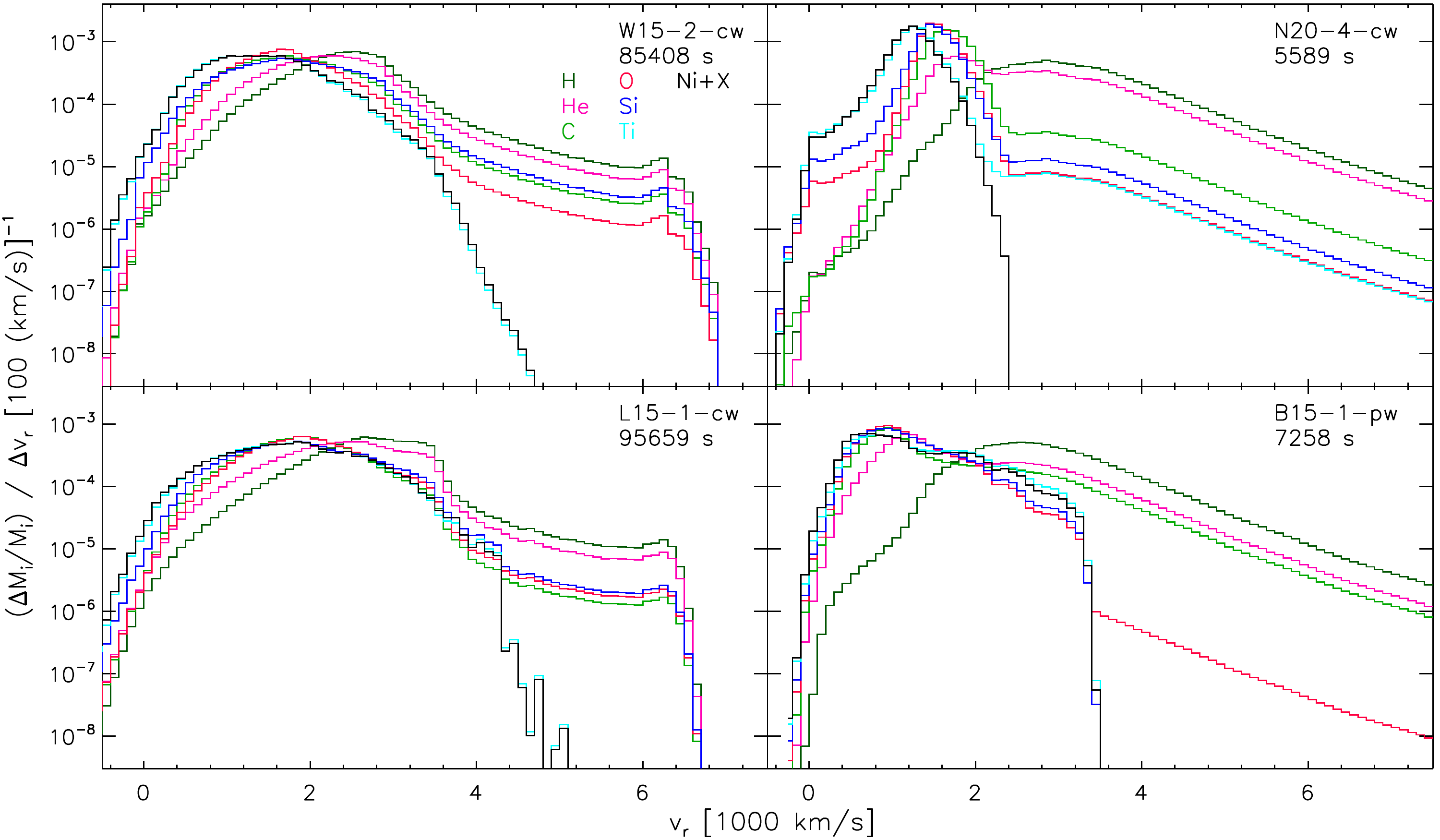}}
\caption{Normalized mass distributions of hydrogen (dark green),
  helium (magenta), carbon (green), oxygen (red), silicon (blue),
  titanium (cyan), and ``nickel'' (black) versus radial velocity for
  models W15-2-cw (top left), L15-1-cw (bottom left), N20-4-cw (top
  right), and B15-1-pw (bottom right) at the shock breakout time. The
  radial velocity bins $\Delta v_r$ are 100\,km\,s$^{-1}$ wide, and
  the normalized mass is given per velocity bin.}
\label{fig:mvsvr} 

\end{figure*}
%

\subsection{Additional 3D models}
\label{subsec:additional3D}

We have seen that the structure of the B15 progenitor provides
favorable conditions for the iron/nickel-rich ejecta to escape a
deceleration by the reverse shock at the He/H interface. The shallow
density gradient (steep rise of $\rho r^3$) inside the He layer and
the large width of the He layer of the B15 progenitor cause a strong
deceleration of the SN shock while it propagates through the
layer. This reduces the velocity difference between the shock and the
iron-group ejecta allowing the latter to stay close behind the
shock. Moreover, the shallow density gradient encountered at the He/H
interface results only in brief and slight acceleration of the SN
shock, \ie the relative velocity between shock and ejecta does not
increase by much.

To elaborate on the importance of the progenitor structure we
performed three additional 3D simulations, W15-2-B, L15-1-B, and
N20-4-B, which were initialized using the 3D explosion models W15-2,
L15-1, and N20-4, respectively. The numerical setup of these models
was identical to that of models W15-2-cw, L15-1-cw, and N20-4-cw,
except for one important difference.  Instead of extending the 3D
explosion models to the stellar surface using the data from the
corresponding progenitor model, we initialized the pre-shock state of
models W15-2-B, L15-1-B, and N20-4-B using the data from the B15
progenitor model. Therefore, these models are hybrid models of W15-2,
L15-1, and N20-4 explosions inside the B15 progenitor envelope, which
allow us to demonstrate the effect of the envelope structure outside
of the C+O core on the ejecta morphology.  We mapped the hybrid models
W15-2-B and N20-4-B at $t_\mathrm{map} = 1.3\,$s and model L15-1-B at
$t_\mathrm{map} = 1.4\,$s, \ie at the same times as the corresponding
models W15-2-cw, N20-4-cw, and L15-1-cw, respectively (see
Table\,\ref{tab:expmod}). At the time of mapping the SN shock still
resides in the C+O core, but is already close to the C+O/He interface
of the progenitor star.

Fig.\,\ref{fig:additional3D} shows the iron/nickel-rich ejecta
morphologies of the hybrid models W15-2-B, L15-1-B, and N20-4-B at the
end of the simulations using the same rendering technique as in
Fig.\,\ref{fig:time-nickel}. The evolution of these three models
resembles that of the other B15 models qualitatively (see
Sect.\,\ref{subsec:overview}, \ref{subsec:shock}, and
\ref{subsec:ejecta}). The SN shock speeds up and slows down depending
on the $\rho r^3$ profile as discussed in Section\,\ref{subsec:shock},
and the morphology of the ejecta evolves similarly to that of model
B15-1-pw described in Section\,\ref{subsec:ejecta}. Thus, we refrain
from repeating the details here.

It is quite evident from the morphology of the iron/nickel-rich ejecta
displayed in Fig.\,\ref{fig:additional3D} that basic features of the
explosion asymmetries seen in model B15-1-pw are also found in all
other models exploding inside the B15 progenitor envelope. Fast
iron/nickel-rich filaments emerging from the largest bubble of
neutrino-heated matter at the time of mapping can escape strong
deceleration by the reverse shock at the base of the helium wall and
move unhamperedly through the hydrogen envelope. More importantly, the
morphologies of the iron/nickel-rich ejecta in the hybrid models are
very different from those of models W15-2-cw, L15-1-cw, and N20-4-cw,
respectively.  This difference goes back to the substitution of the
outer structure (\ie the He and H envelope) of the progenitor models
W15, L15, and N20 by that of the progenitor model B15.  It is
remarkable that in all cases the biggest and strongest plumes created
by the neutrino-driven mechanism are the seeds of the longest and most
prominent Ni fingers at late stages.

%
\begin{figure}
\centering
\resizebox{\hsize}{!}{\includegraphics*{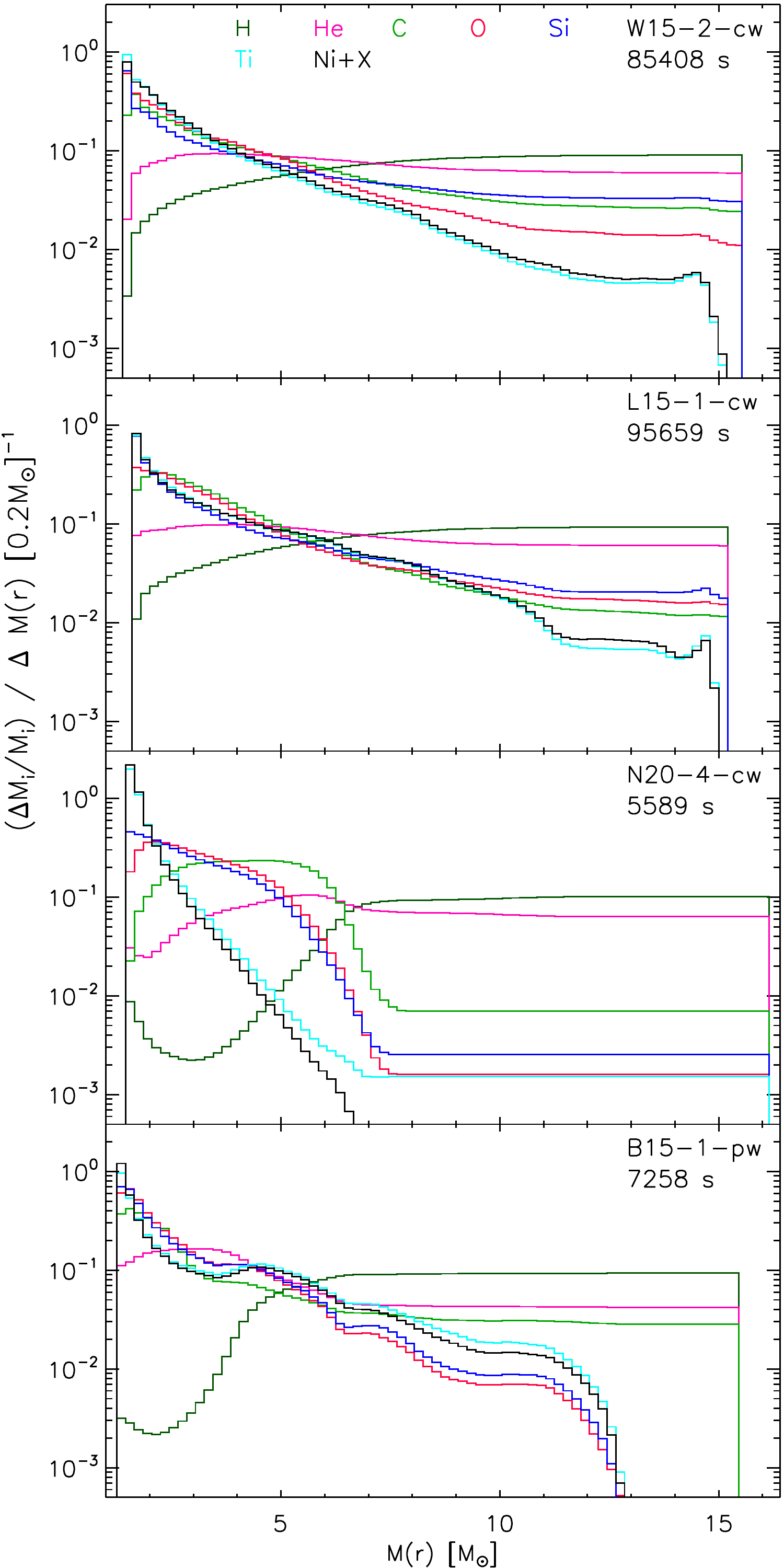}}
\caption{Same as Fig.\,\ref{fig:mvsvr} but versus mass coordinate. The
  normalized mass is given per mass bin of 0.2\,$M_\odot$.}
\label{fig:mvsm} 

\end{figure}
%

\subsection{Radial mixing}

Figures\,\ref{fig:mvsvr} and \ref{fig:mvsm} display the normalized
mass distributions of hydrogen, helium, carbon, oxygen, silicon,
titanium, and nickel plus the ``tracer'' nucleus for models W15-2-cw,
L15-1-cw, N20-4-cw, and B15-1-pw as functions of radial velocity and
mass coordinate
\footnote{Because of interpolation errors when mapping the 1D data
  from the fine Lagrangian grids of the progenitor models to the
  coarser radial (Eulerian) grids of the 3D hydrodynamic models, the
  masses of the simulated 3D SN explosion models differ from those of
  the corresponding 1D progenitor models, \eg in case of the
  progenitor B15 the 3D models have a mass of $15.5\,M_\odot$, like
  the model in \citet{Hammeretal10}.},
respectively.  The figures show the distributions at
the time of shock breakout, \ie at the time when the minimum
(angle-dependent) shock radius exceeds $R_\ast$.  When interpreting
these figures one should keep in mind that the distributions at high
velocities and large mass coordinates just reflect the initial
composition of the progenitor star.

The heavy ejecta (\ie the metals) of our RSG models (W15 and L15) are
at first compressed into a thin shell once the SN shock begins to
decelerate in the hydrogen envelope, and then they are mixed in radial
direction due to the growth of RT instabilities at the He/H interface.
The resulting distributions of the metals versus radial velocity are
of similar shape at the time of shock breakout. They are characterized
by a broad maximum centered around a radial velocity of $\approx
2000$\,km\,s$^{-1}$ and a high velocity wing extending up to $\approx
6500$\,km\,s$^{-1}$ (Fig.\,\ref{fig:mvsvr}, left panels). The maximum
velocity of "nickel" is $4200\,$km\,s$^{-1}$ in model W15-2-cw, and
$4780\,$km\,s$^{-1}$ in model L15-1-cw (see $v_\mathrm{max}$ in
Table\,\ref{tab:models}).  The matter (mostly H and He, and some C, O,
and Si) having even higher velocities represents the rapidly expanding
unmixed outer part of the hydrogen envelope. The velocity distribution
of the heavy ejecta corresponds to a radial mixing in mass coordinate
that comprises almost the whole hydrogen envelope
(Fig.\,\ref{fig:mvsm}, top two panels). Figures\,\ref{fig:mvsvr} and
\ref{fig:mvsm} further show that some hydrogen (and helium) is mixed
down to velocities below $1000$\,km\,s$^{-1}$, \ie into the innermost
2\,$M_\odot$ of the ejecta.

The amount of both the outward mixing of metals into the hydrogen
envelope and the inward mixing of hydrogen into innermost mass layers
is much less in the BSG models than in the RSG models.  The amount of
mixing also differs strongly among the simulated BSG models.  The
metal distributions are narrower in radial velocity space compared to
the corresponding ones of the RSG models. In model N20-4-cw the metal
distributions peak at 1500\,km\,s$^{-1}$, while the maxima of the
metal distributions are located at around 1000\,km\,s$^{-1}$ in model
B15-1-pw (Fig.\,\ref{fig:mvsvr}, right panels).  The distributions of
the latter model also show a pronounced high velocity shoulder
extending to 3400\,km\,s$^{-1}$ (Fig.\,\ref{fig:mvsvr}, bottom right),
while no metal ejecta from the deep core move at velocities higher
than 2400\,km\,s$^{-1}$ in model N20-4-cw (Fig.\,\ref{fig:mvsvr}, top
right). We note again that the matter having higher velocities
represents the rapidly expanding unmixed outer part of the hydrogen
envelope.  The velocity distributions of model N20-4-cw correspond to
a radial mixing of the lighter metals C, O, and Si out to a mass
coordinate of about 7.5\,$M_\odot$, and of Ti and Ni+X to about
6.7\,$M_\odot$ (Fig.\,\ref{fig:mvsm}, 3rd panel from top).  In model
B15-1-pw, all metal ejecta are mixed out to a mass coordinate of
$\approx 13\,M_\odot$ (Fig.\,\ref{fig:mvsm}, bottom panel).
Hydrogen is mixed less deeply into the metal core in the BSG models
than in the RSG ones, so that there is no significant amount of
hydrogen at velocities lower than 1000\,km\,s$^{-1}$, except for a
low-velocity tail down to 500\,km\,s$^{-1}$ in model B15-1-pw
(Fig.\,\ref{fig:mvsvr}, bottom right).

The particular shape of the velocity distribution of the metal ejecta
in model B15-1-pw (Fig.\,\ref{fig:mvsvr}, bottom right) results from
the two-component ejecta morphology that is imprinted by the B15
progenitor structure (see discussion in Sect.\,\ref{subsec:ejecta}).
The peak in the metal distributions at low velocities is associated
with the ejecta that are decelerated by the reverse shock, while the
broad high-velocity shoulder results from the elongated fingers that
escaped the reverse shock. These elongated fingers are responsible for
the mixing of metals up to an enclosed mass coordinate of about
13\,$M_\odot$.

\subsection{Element distributions}

Figure\,\ref{fig:mvsvr}, which shows the element distributions as
functions of radial velocity, provides no (direct) spatial information
and Fig.\,\ref{fig:mvsm}, which gives the element distributions as
functions of enclosed mass, contains only 1D information about the
spatial (radial) distribution. In addition, using the enclosed mass at
a given radius as a radial coordinate becomes questionable when the
departure of the mass distribution from spherical symmetry is
large. Hence, we sought for a visualization method that incorporates
mass, velocity, and spatial information into a single plot. We opted
for a method using a set of particles for each chemical element, where
the particles are distributed according to the mass distribution of
the considered element, and colored each particle by the radial
velocity at the particle's position, which is obtained from the 3D
simulation data.

%
\begin{figure*}[p]
\centering
\resizebox{0.245\hsize}{!}{\includegraphics*{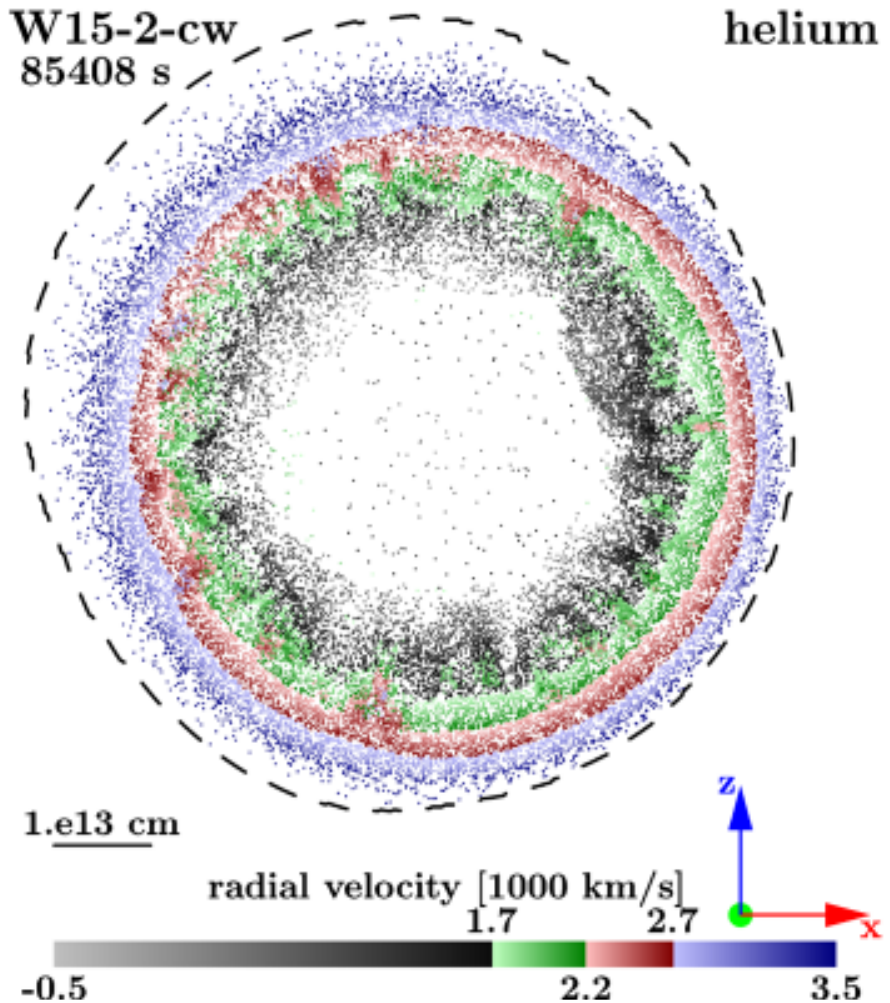}}
\resizebox{0.245\hsize}{!}{\includegraphics*{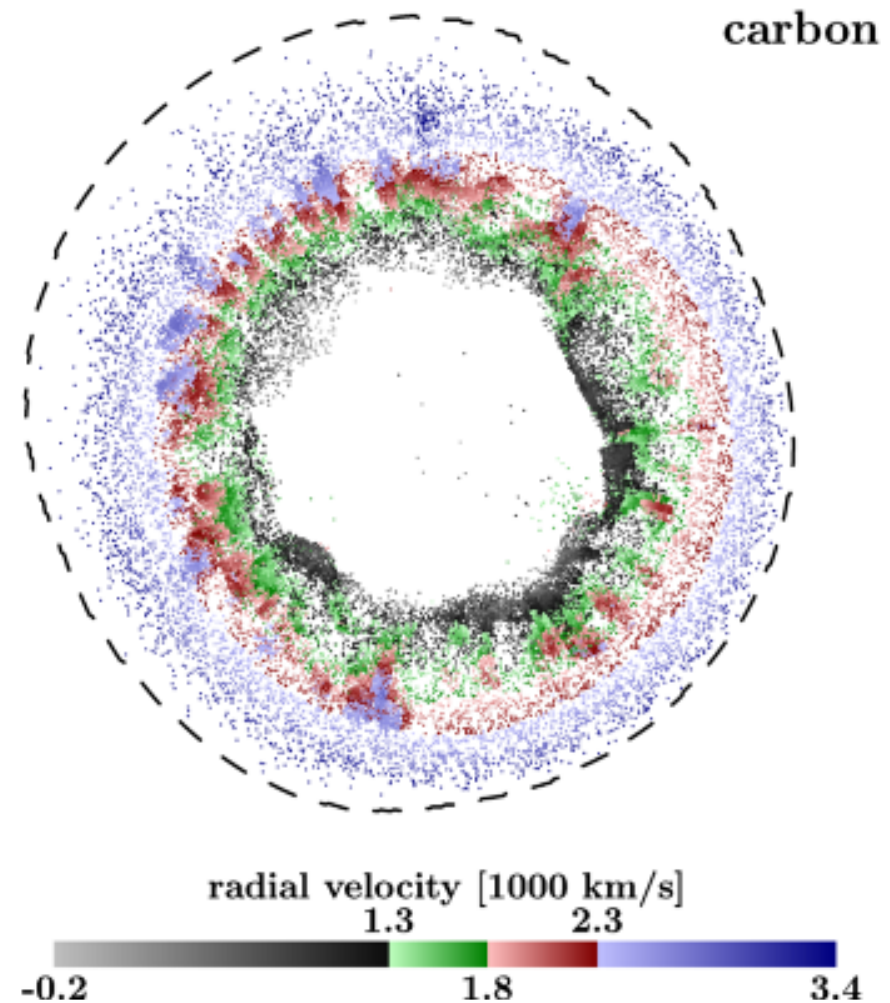}}
\resizebox{0.245\hsize}{!}{\includegraphics*{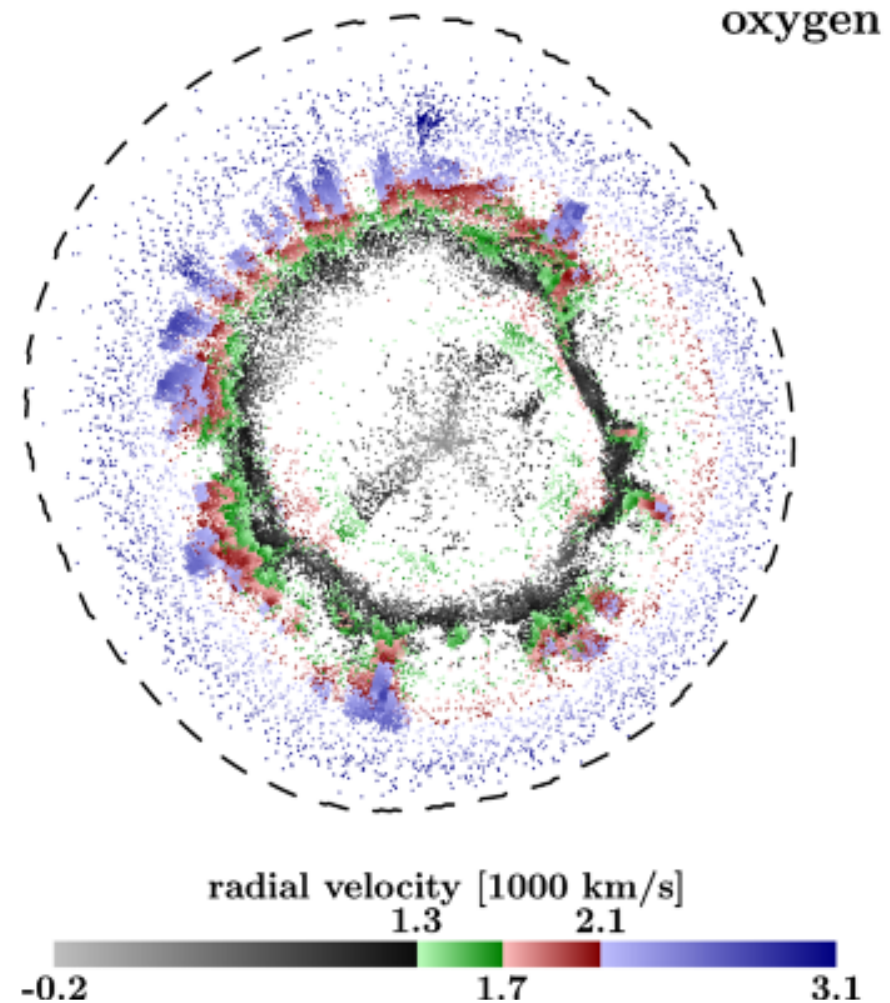}} 
\resizebox{0.245\hsize}{!}{\includegraphics*{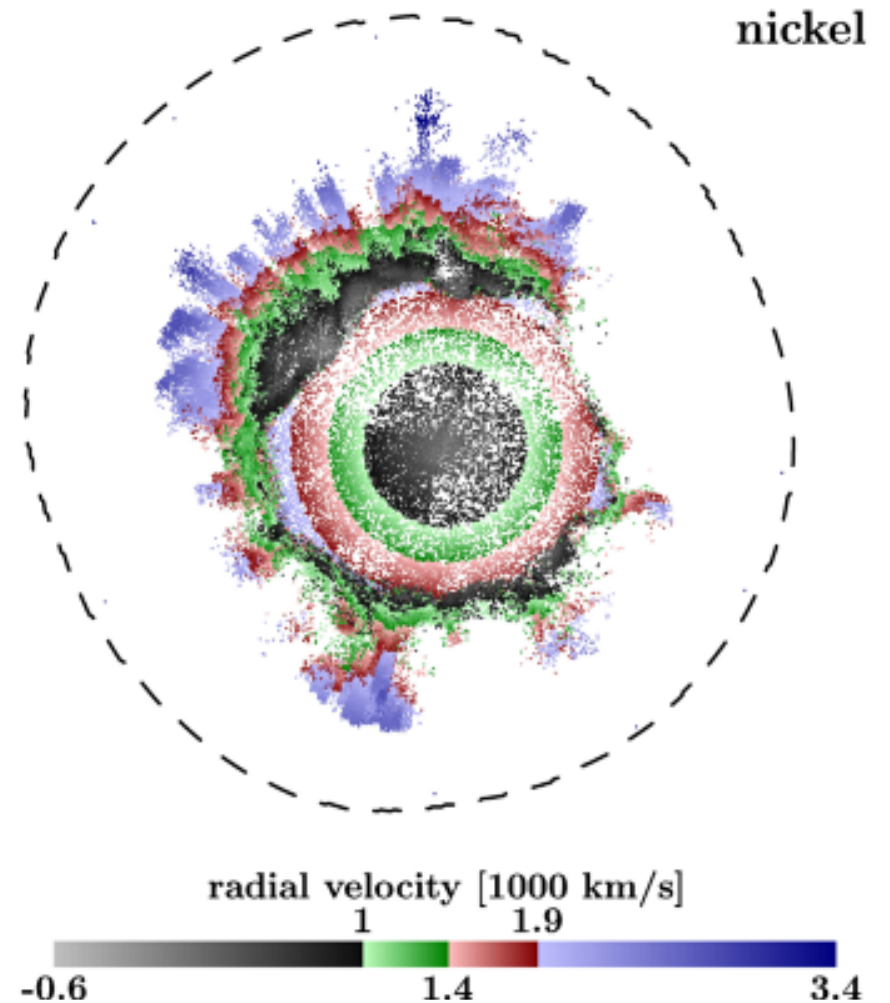}}\\
\bigskip\medskip
\resizebox{0.245\hsize}{!}{\includegraphics*{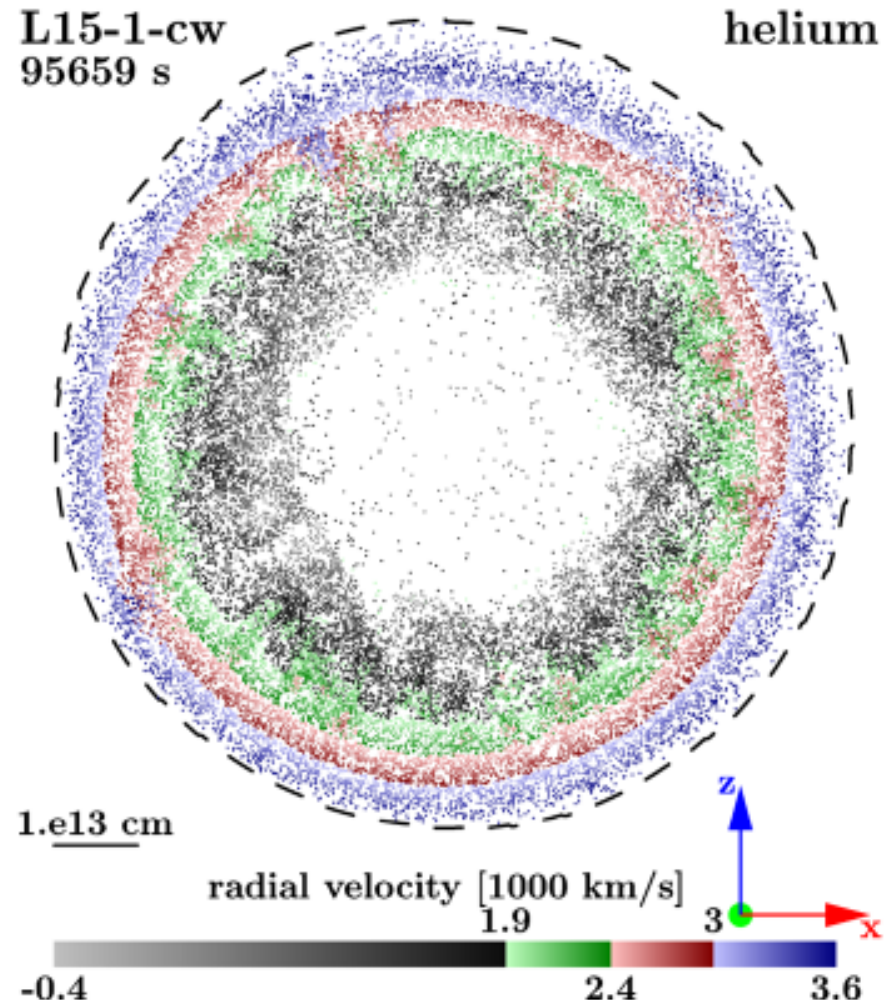}}
\resizebox{0.245\hsize}{!}{\includegraphics*{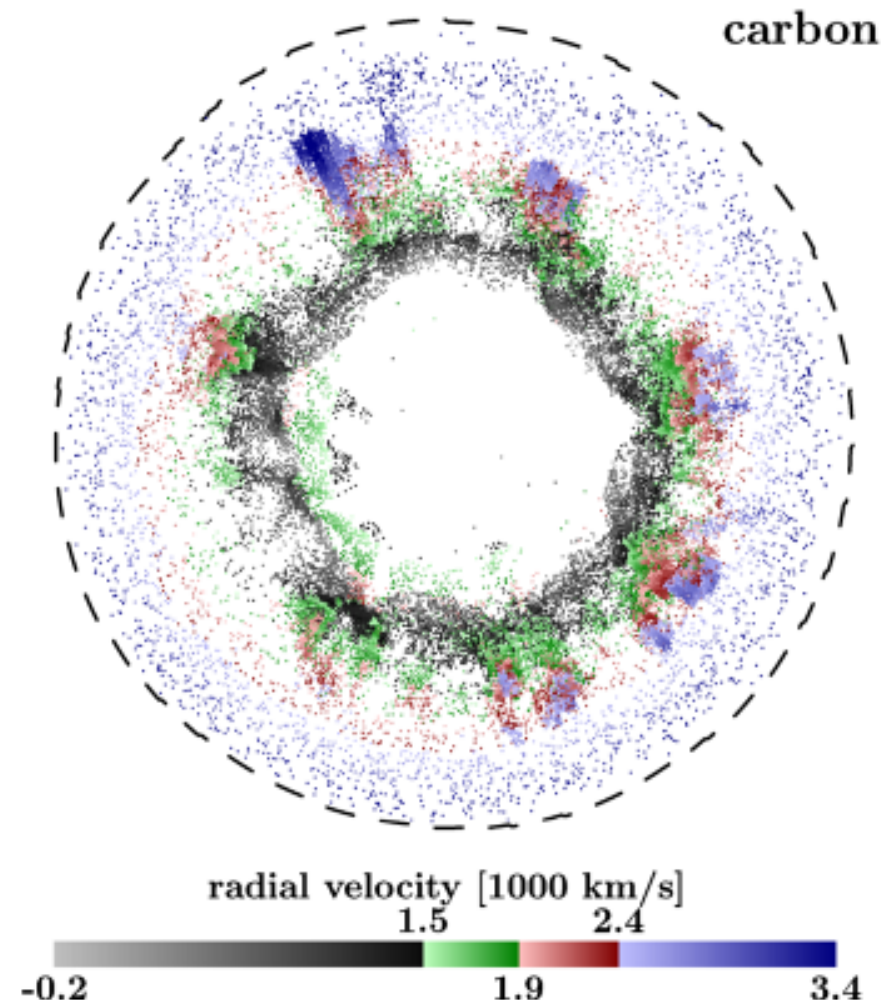}}
\resizebox{0.245\hsize}{!}{\includegraphics*{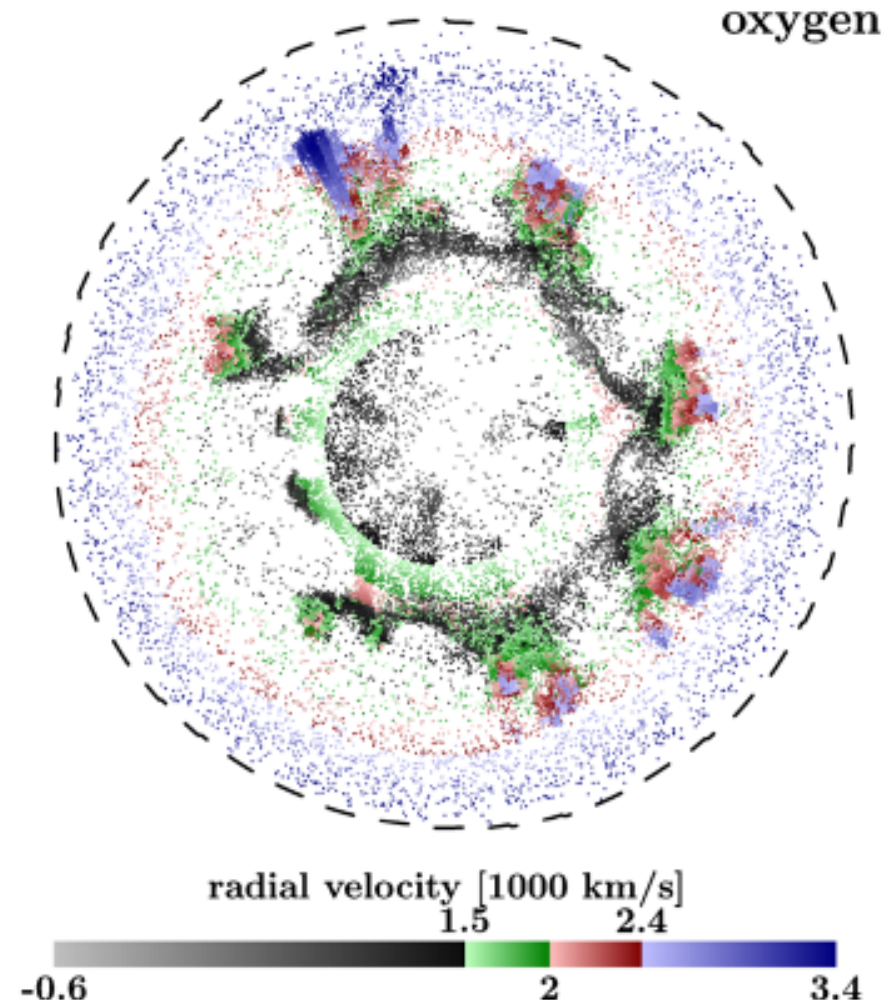}} 
\resizebox{0.245\hsize}{!}{\includegraphics*{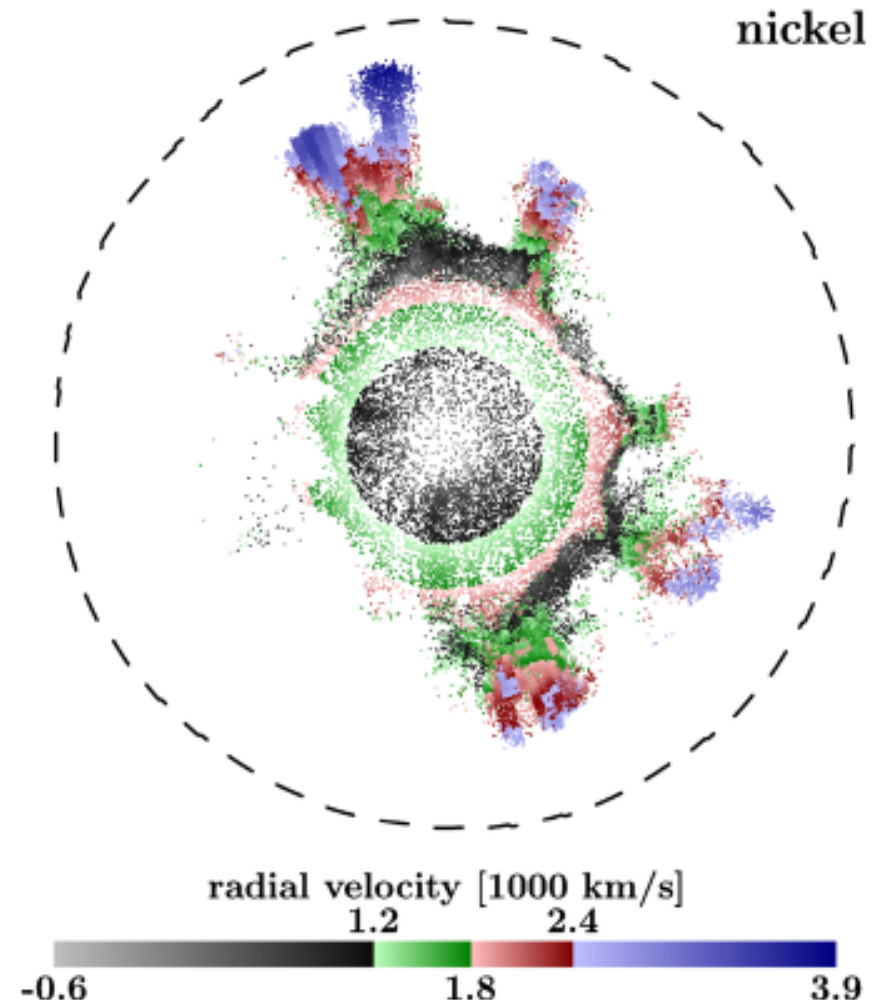}}\\
\bigskip\medskip
\resizebox{0.245\hsize}{!}{\includegraphics*{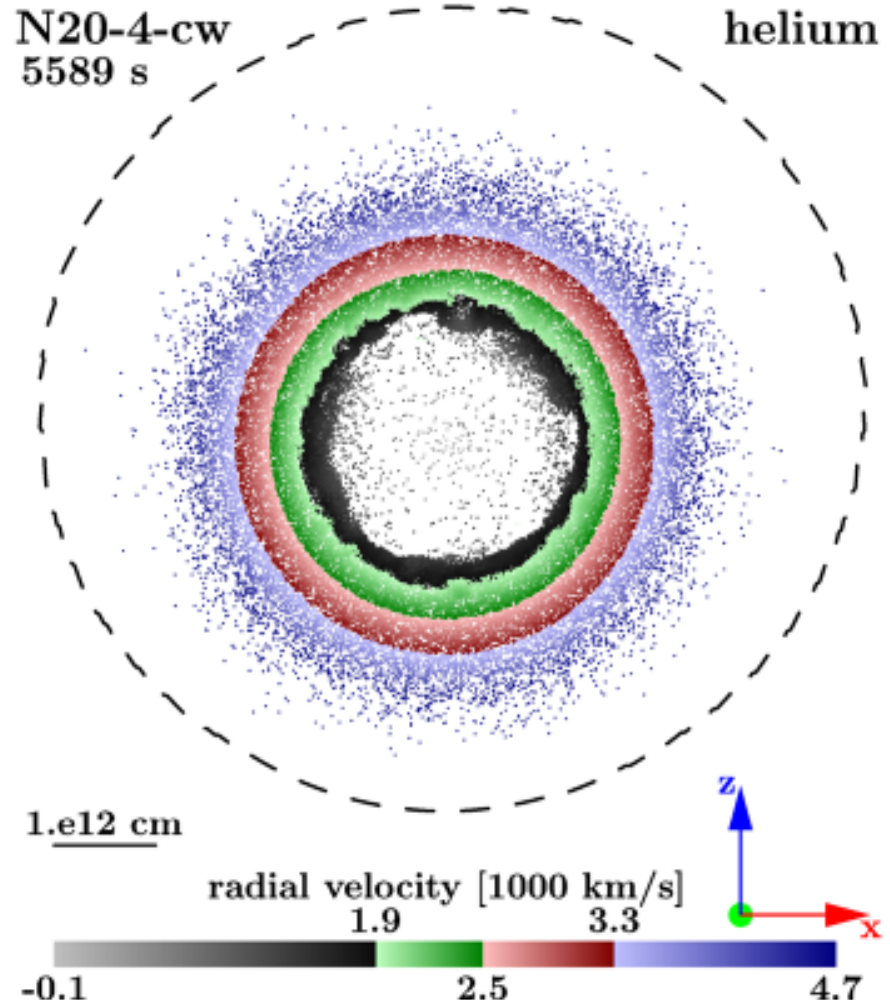}}
\resizebox{0.245\hsize}{!}{\includegraphics*{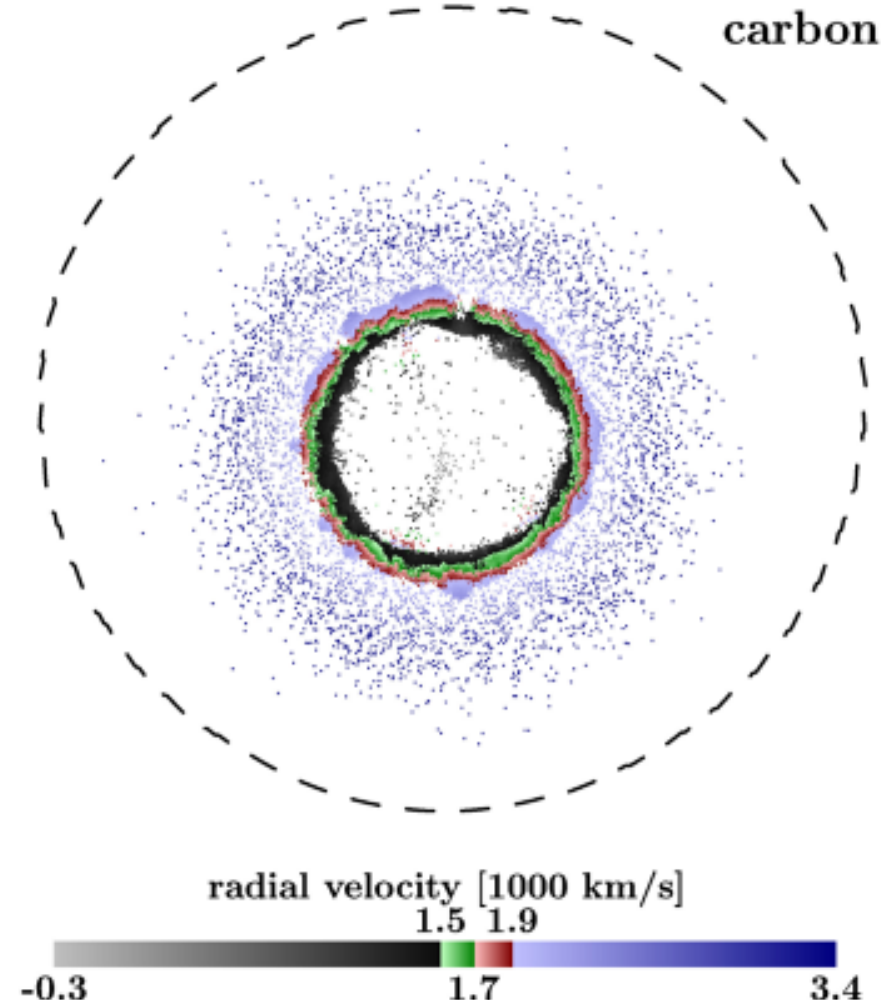}}
\resizebox{0.245\hsize}{!}{\includegraphics*{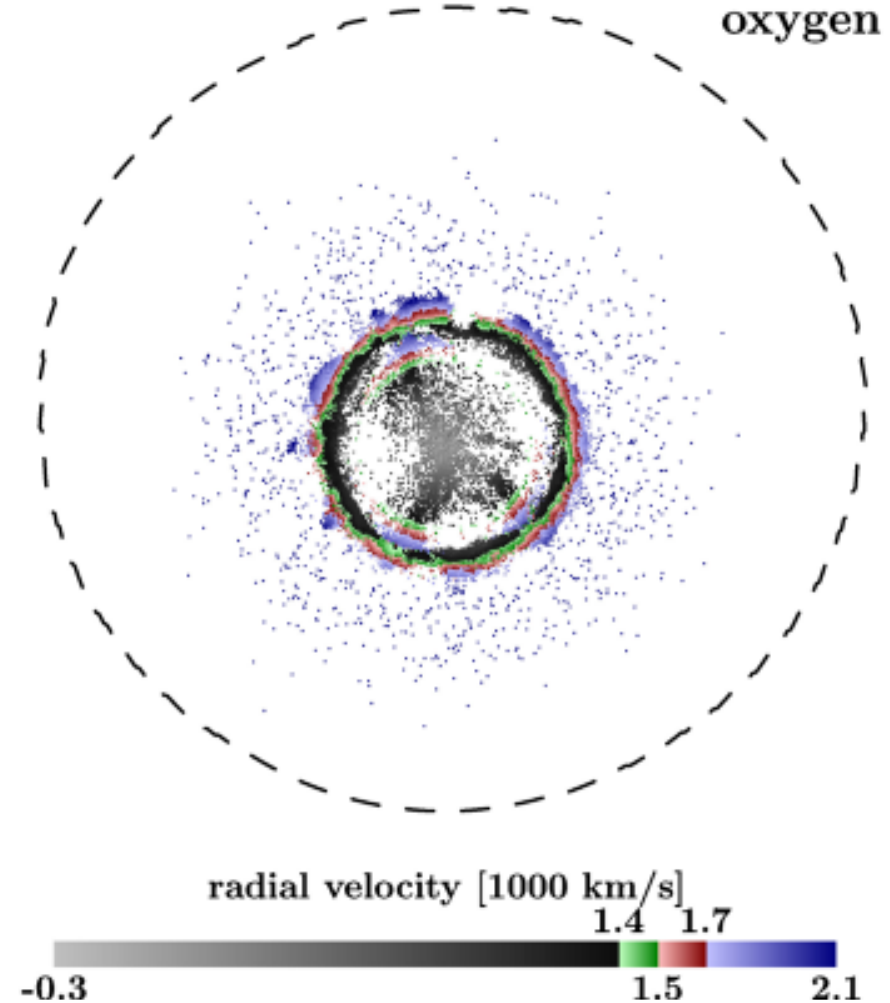}} 
\resizebox{0.245\hsize}{!}{\includegraphics*{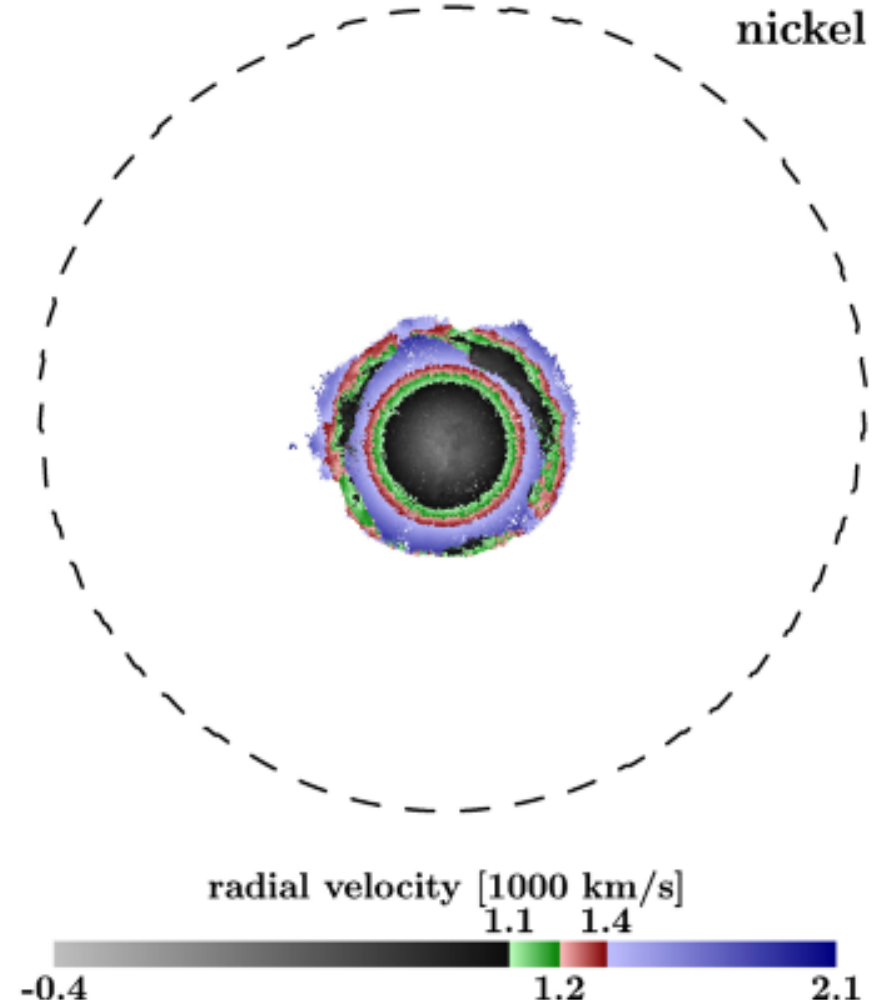}}\\
\bigskip\medskip
\resizebox{0.245\hsize}{!}{\includegraphics*{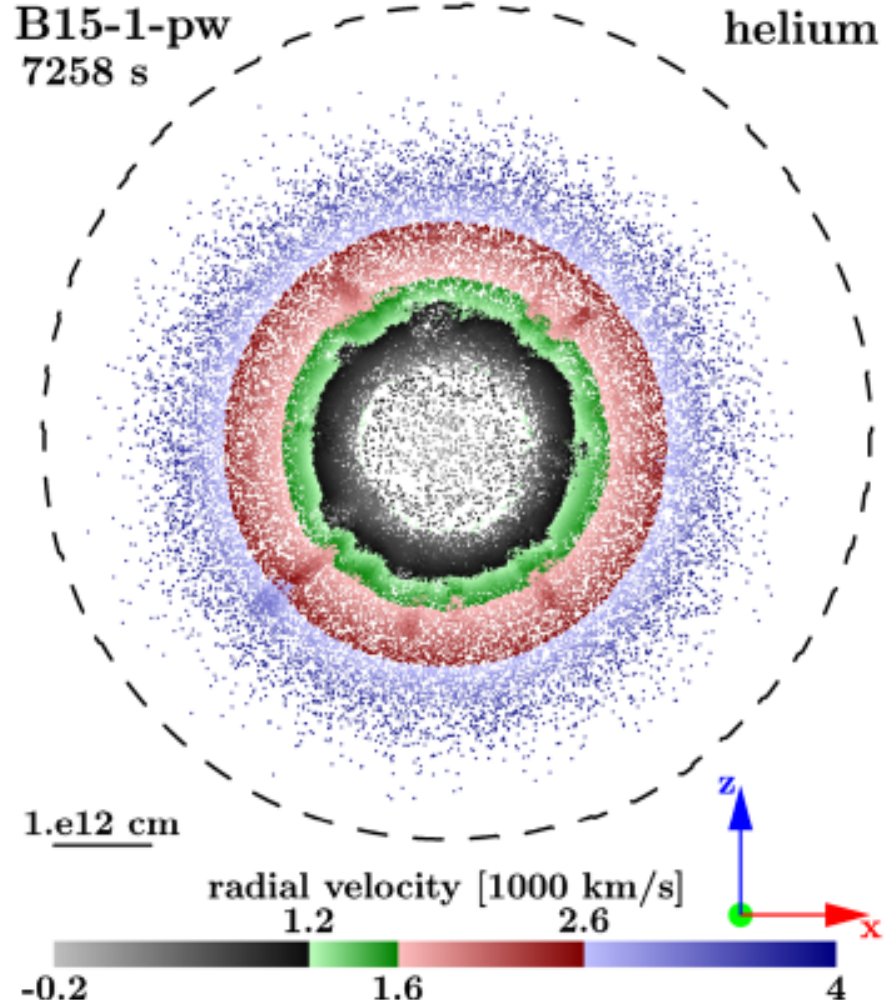}}
\resizebox{0.245\hsize}{!}{\includegraphics*{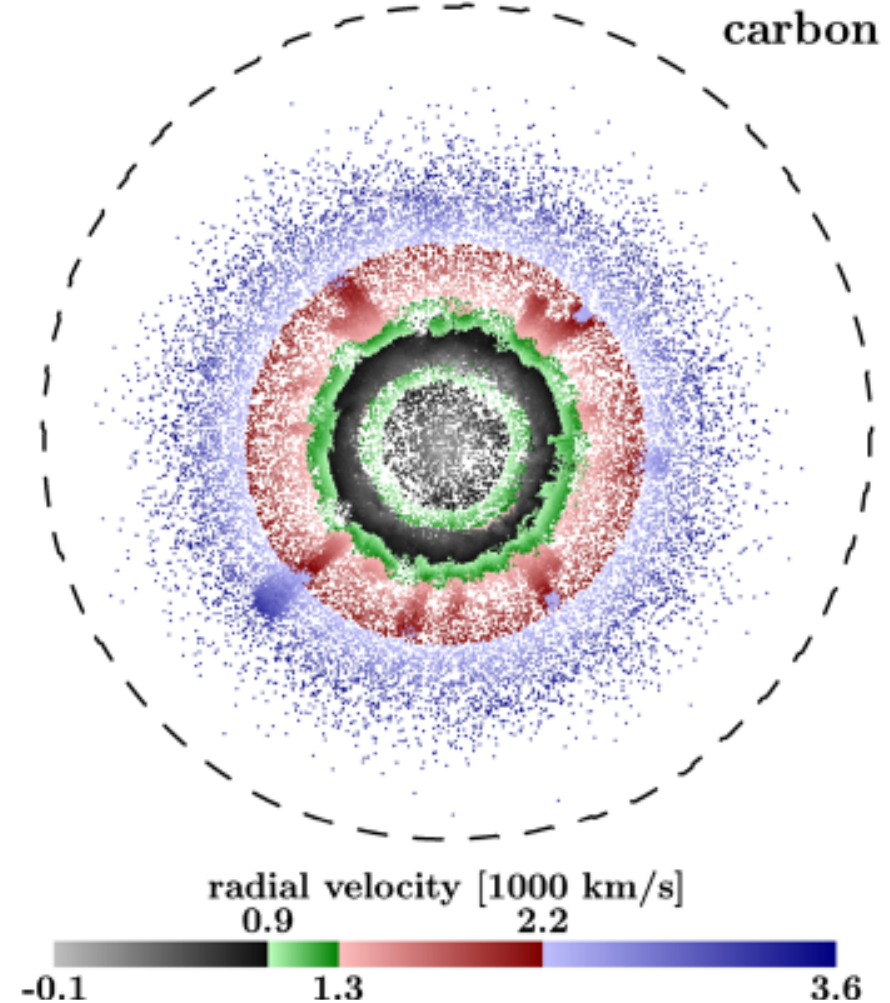}}
\resizebox{0.245\hsize}{!}{\includegraphics*{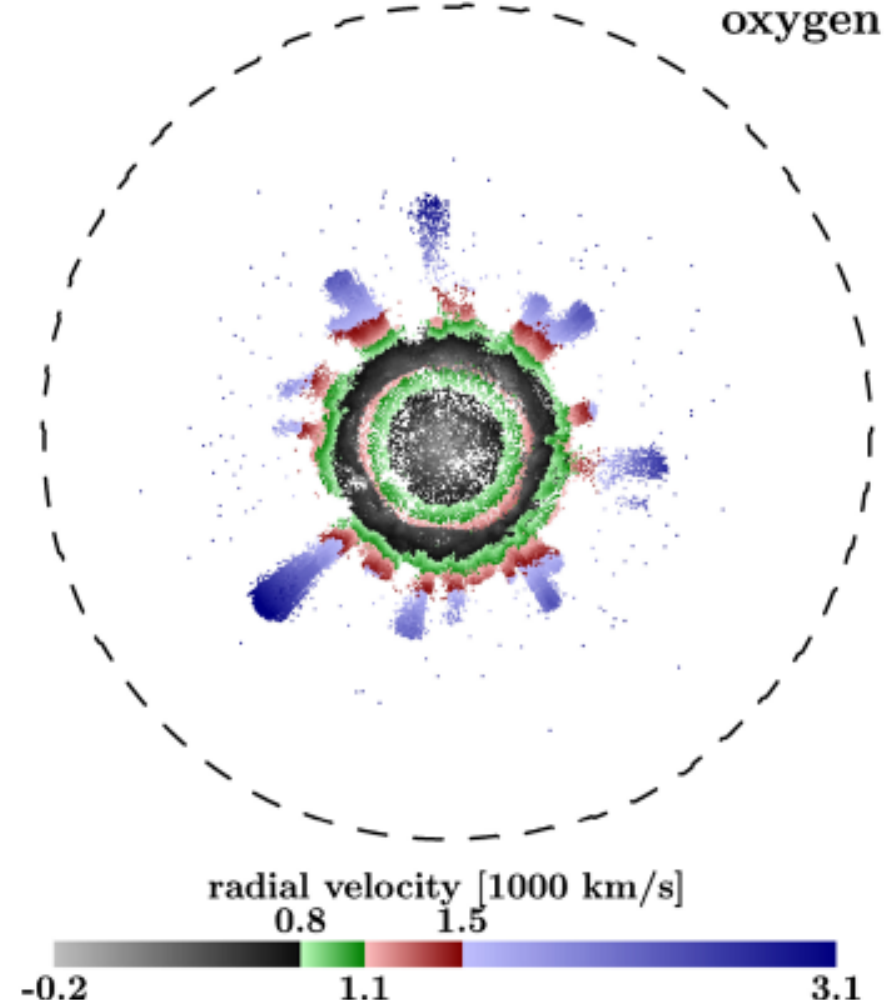}} 
\resizebox{0.245\hsize}{!}{\includegraphics*{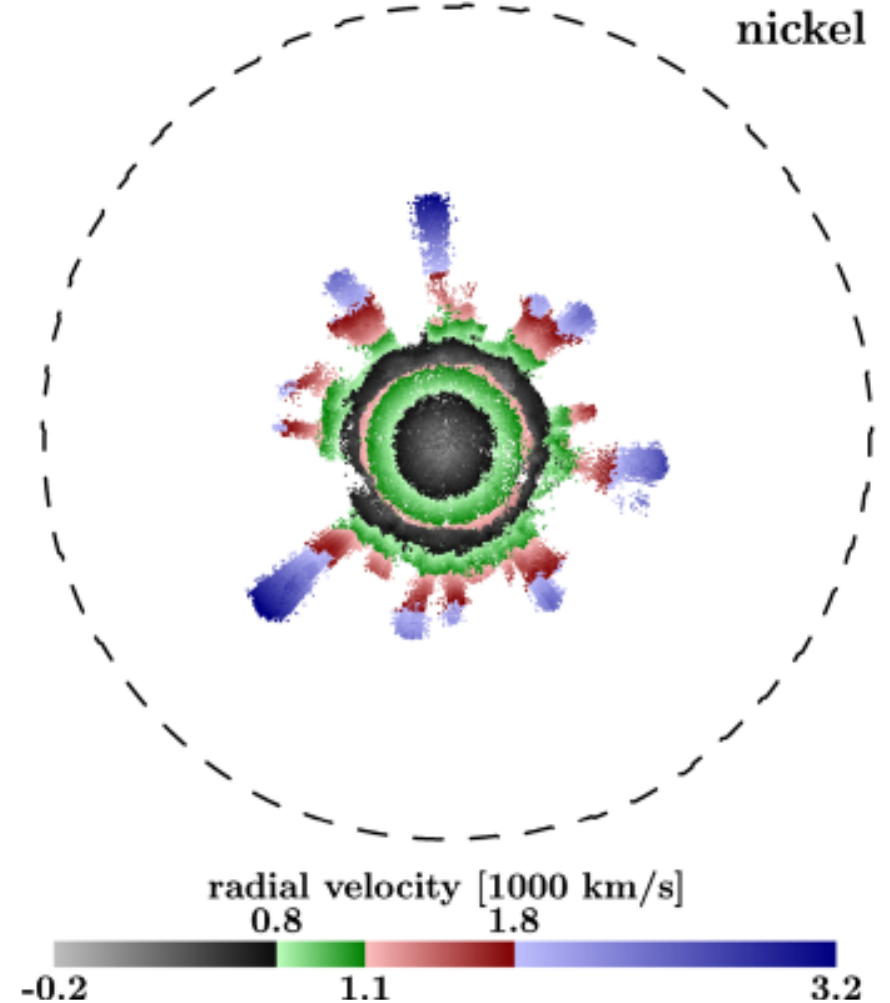}}
\caption{Particle representations of the spatial distributions of He,
  C, O, and Ni+X for models W15-2-cw, L15-1-cw, N20-4-cw, and B15-1-pw
  (from top) at the time of shock breakout. We show only particles
  within a slab given by $-0.05 R_\mathrm{s}^\mathrm{max} \leq y \leq
  0.05 R_\mathrm{s}^\mathrm{max}$. The dashed lines give the shock's
  location in the $y=0$ plane. The particles are colored according to
  their radial velocity, which is binned into four bins, each bin
  containing 25\% of the total mass of the element. Blackish particles
  represent the slowest 25\% by mass of an element, while bluish
  particles mark the fastest 25\%.  The black thin shell contains
  matter that was hit by the reverse shock, and thus moves slower than
  matter located inside this shell (reddish and greenish particles).
  The positive y-axis points away from the viewer.}
\label{fig:elements}
\end{figure*}
%

We first calculate the number of particles of species $i$ having a
radial velocity in the interval $[v_\mathrm{r}, v_\mathrm{r}+\Delta
  v_\mathrm{r})$ by
\begin{eqnarray}
 N_{v,i}(v_\mathrm{r}) = N_{\mathrm{tot},i}\,
                       \frac{\Delta M_i(v_\mathrm{r})}{M_i}
\end{eqnarray}
where $N_{\mathrm{tot},i}$ is the total number of particles of species
$i$, $\Delta M_i (v_\mathrm{r})$ is the amount of mass of species $i$
having a radial velocity in the range $[v_\mathrm{r},
  v_\mathrm{r}+\Delta v_\mathrm{r})$, and $M_i$ is the species' total
  mass. Then we determine in which grid cell a particle $n$
  resides. The probability of a particle having a radial velocity in
  the range $[v_\mathrm{r}, v_\mathrm{r}+\Delta v_\mathrm{r})$ and
    residing in a grid cell $G$ is
\begin{eqnarray}
 \mathcal P_{G,i} = \frac{ \Delta m_i(v_\mathrm{r}) }{
                           \Delta M_i(v_\mathrm{r}) }, 
\end{eqnarray}
where $\Delta m_i (v_\mathrm{r})$ is the mass of the species $i$ in
the grid cell $G$ having a radial velocity in the range
$[v_\mathrm{r}, v_\mathrm{r}+\Delta v_\mathrm{r})$. Using a set of
  uniformly distributed random numbers $\mathcal R_1$, we determine
  the grid cell $G$ the particle $n$ is to be assigned to. Knowing the
  grid cell $G$, where particle $n$ resides, we can calculate the
  coordinates $(r_n, \theta_n, \phi_n)$ of particle $n$ as
\begin{eqnarray}
  r_n     &=& r^L_G      + \mathcal R_2(n)\cdot\Delta r_G,\\
 \theta_n &=& \theta^L_G + \mathcal R_3(n)\cdot\Delta\theta_G,\\
 \phi_n   &=& \phi^L_G   + \mathcal R_4(n)\cdot\Delta\phi_G,
\end{eqnarray}
where $r^L_G$, $\theta^L_G$, and $\phi^L_G$ are the coordinates of the
left interface of the grid cell $G$, and $\Delta r_G$,
$\Delta\theta_G$, and $\Delta\phi_G$ are the sizes of the grid cell in
$(r, \theta, \phi)$ direction. Note that we used three additional sets
of uniformly distributed random numbers, $\mathcal R_2$, $\mathcal
R_3$, and $\mathcal R_4$, in calculating the coordinates of the
particle. The particle's radial velocity is given by the radial
velocity of grid cell $G$.

\begin{figure}
\centering
\resizebox{0.49\hsize}{!}{\includegraphics*{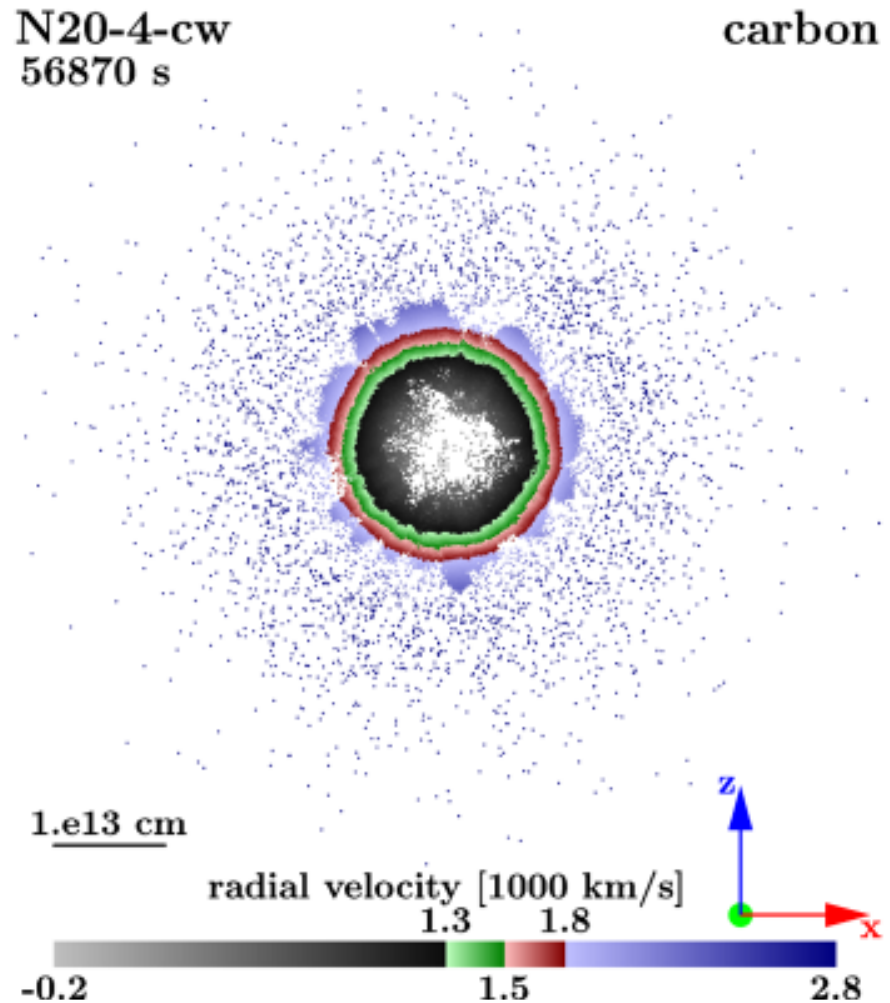}} 
\resizebox{0.49\hsize}{!}{\includegraphics*{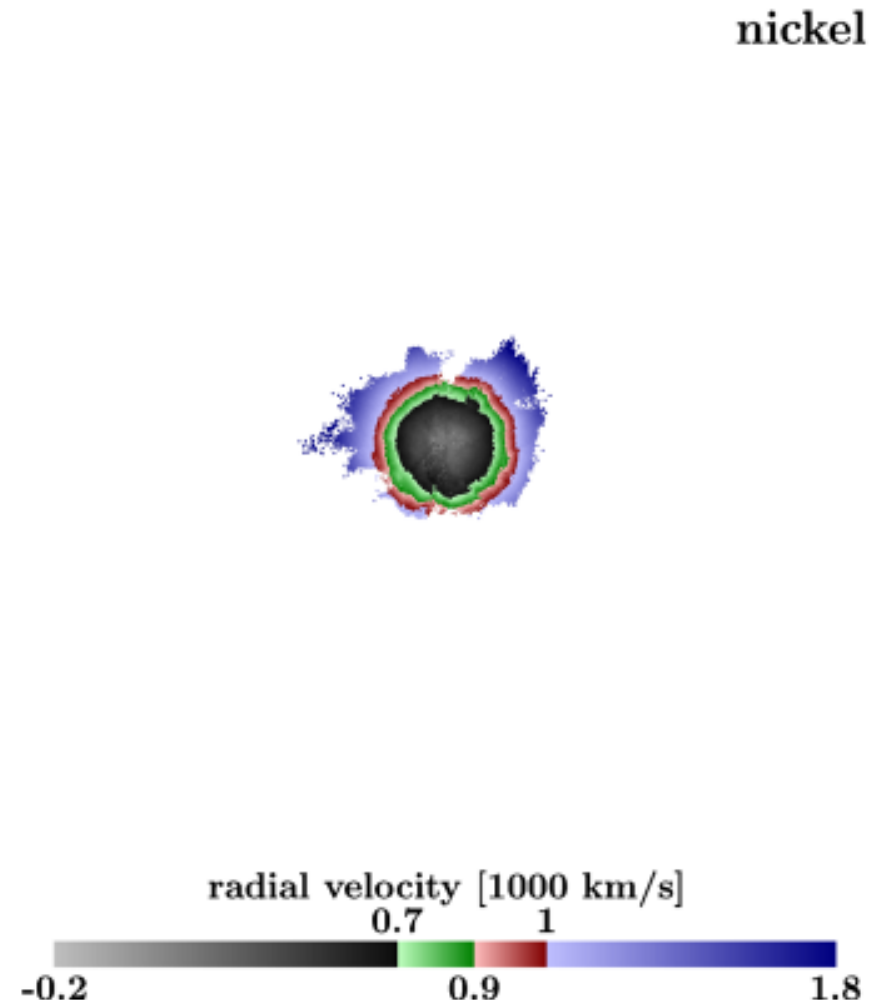}}\\
\resizebox{0.49\hsize}{!}{\includegraphics*{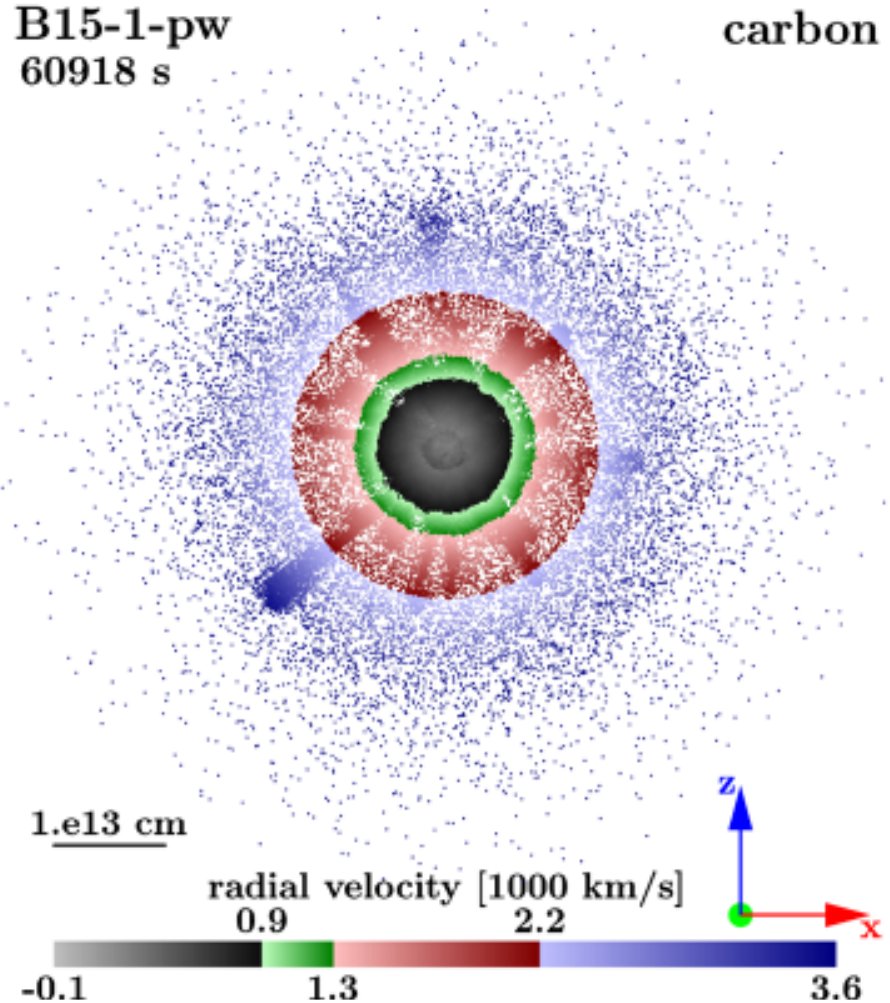}} 
\resizebox{0.49\hsize}{!}{\includegraphics*{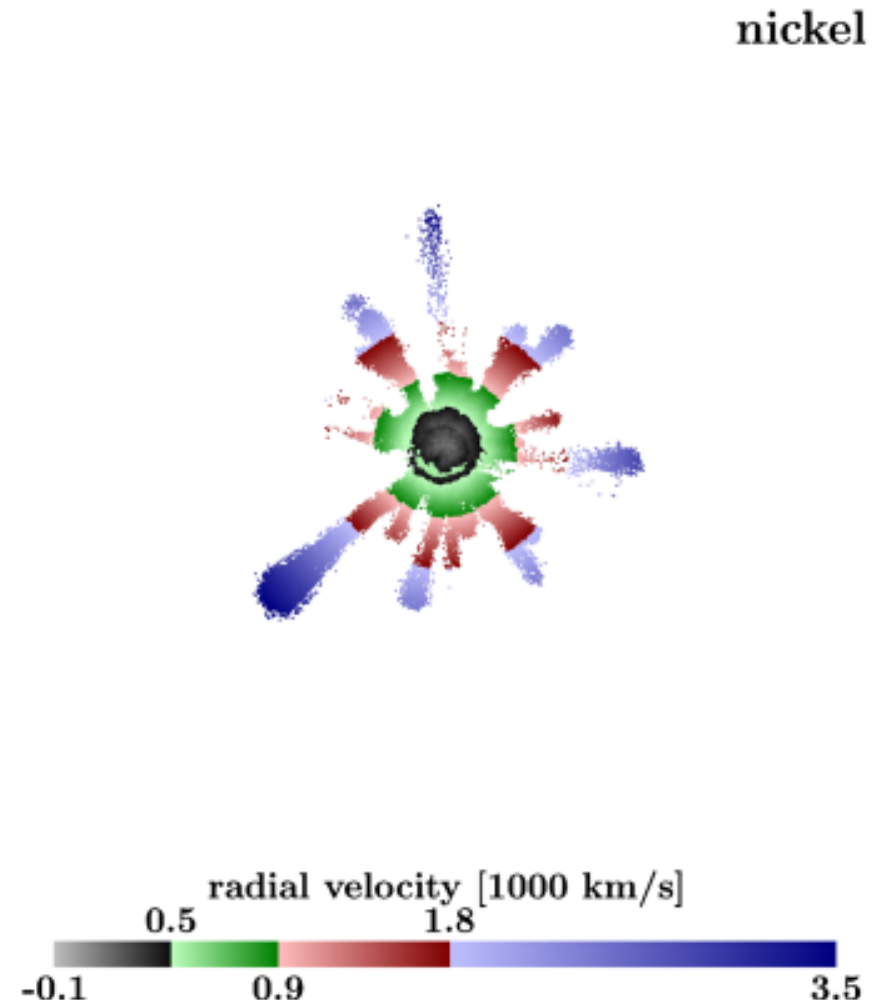}}
\caption{Same as Fig.\,\ref{fig:elements} but displaying only those
  particles that reside inside a sphere of radius $4\times
  10^{13}$\,cm and within a slab given by $-2\times
  10^{12}\,\mathrm{cm} \leq y \leq 2\times 10^{12}\,\mathrm{cm}$ for
  models N20-4-cw (top) and B15-1-pw (bottom) at 56870\,s and
  60918\,s, respectively.}
\label{fig:elements-late}
\end{figure}

We binned the radial velocities into four bins (not to be confused
with the velocity intervals discussed above!), each bin containing
25\% of the total mass of an element. To provide some qualitative
information of the mass distribution at a glance, we used different
color gamuts to show the variation of the velocity within each bin.
Because each color gamut contains 25\% of the total mass of a species,
blackish particles represent the slowest 25\% by mass of a species,
while blueish particles mark the fastest 25\% (see
Figs.\,\ref{fig:elements} and \ref{fig:elements-late}).

Fig.\,\ref{fig:elements} displays a slab of the particle distributions
of helium, carbon, oxygen, and nickel (including the tracer $X$) for
models W15-2-cw, L15-1-cw, N20-4-cw, and B15-1-pw. The slab is defined
by $-0.05R_\mathrm{s}^\mathrm{max} \le y \le
0.05R_\mathrm{s}^\mathrm{max}$, where $R_\mathrm{s}^\mathrm{max}$ is
the maximum radius of the SN shock.  In all models, the helium
distributions are almost spherically symmetric with only slight
imprints of asymmetries, while the distributions of the metals show
pronounced asymmetries.

A comparison of the distributions of carbon and oxygen between RSG and
BSG models is particularly interesting. In the RSG models the ejecta
that were compressed by the reverse shock (forming after the SN shock
had crossed the He/H interface) are strongly fragmented by RT
instabilities. As a result both the carbon and oxygen shell are
strongly distorted. This situation contrasts with that in the BSG
models, in particular model N20-4-cw, where carbon and oxygen retain a
more coherent spherical shell structure because of a less vigorous and
less long-lasting growth of RT instabilities at the C/O/He and He/H
interfaces in the BSG progenitors in comparison with the RSG
progenitors.  To confirm that no significant fragmentation of the
carbon shell occurs in the BSG models we show in
Fig.\,\ref{fig:elements-late} the carbon distribution of models
N20-4-cw and B15-1-pw at 56870\,s and 60918\,s, respectively, \ie at a
time when mixing seems to have ceased. We used here the same
visualization technique as in Fig.\,\ref{fig:elements}, but the slab
is defined by $-2\times 10^{12}\,\mathrm{cm} \le y \le 2\times
10^{12}\,\mathrm{cm}$, and only particles residing inside a radius of
$4\times 10^{13}$\,cm are shown.

In model B15-1-pw the spherical distributions of carbon and oxygen are
superimposed by asymmetries produced by elongated RT fingers of ejecta
which escaped the reverse shock below the He/H interface.
Figure\,\ref{fig:elements} implies that more than 50\% of the nickel
mass of this model is carried by the elongated structures (fingers)
with velocities larger than approximately $1100\,$km\,s$^{-1}$ (red
and blue particles in the bottom right panel).

The spatial locations of the He and metal ejecta (\ie He, C, O, and
Ni+X) also differ between RSG and BSG models
(Fig.\,\ref{fig:elements}). In the former ones they are mixed outward
close to the SN shock, while they stay far behind the shock in the BSG
models. This even holds for model B15-1-pw with its fast-moving RT
fingers that escaped the interaction with the reverse shock.

%
%


\section{Conclusions}
\label{sec:conclusions}

We have presented the results of three-dimensional hydrodynamic
simulations of core collapse supernovae comprising the evolution of
the shock wave and of the neutrino-heated metal-rich ejecta from shock
revival to shock breakout in four different progenitor stars: two
15\,$M_\odot$ RSG, a 20\,$M_\odot$ BSG, and a 15\,$M_\odot$ BSG.  The
simulations were initialized from a subset of the 3D explosion models
of \citet{Wongwathanaratetal10b, Wongwathanaratetal13}, which cover
the evolution from about 10\,ms up to 1.4\,s after core bounce.  The
explosions in these models were initiated by imposing suitable values
for neutrino luminosities and mean energies at the inner grid boundary
located at a finite, time-dependent radius. Neutrino transport and
neutrino-matter interactions were treated by the ray-by-ray gray
neutrino transport scheme of \citet{Schecketal06} including a slight
modification of the prescription for neutrino mean energies at the
inner grid boundary \citep{Uglianoetal12}.

At the end of the initial explosion simulations the SN shock had
reached a radius of $10^9$ to 2$\times1 0^9$\,cm, \ie it still resided
inside the C+O core of the progenitor star. Combining the (early-time)
3D explosion models of \citet{Wongwathanaratetal13} with the late-time
3D runs presented here allowed us to study the evolution of CCSN and
their ejecta for a set of progenitor models (and not just for one case
as in HJM10) in 3D for the first time from shortly ($\approx 10\,$ms)
after core bounce until many hours later.

To aid us with the analysis of the 3D simulations, we have also
calculated the linear Rayleigh-Taylor growth rates of 1D explosion
simulations performed with the angle-averaged (early-time) 3D
explosion models, which provide a qualitatively good criterion for the
expected growth in different layers of the progenitor star.  The
growth factors are large near the C+O/He and He/H interfaces, which is
expected from the density profiles of the progenitor stars that vary
strongly at these interfaces in all models.

We compared our simulations performed with the Yin-Yang grid to those
of \citet{Hammeretal10} and find good agreement concerning the amount
and extent of mixing and the ejecta morphologies, except that the peak
radial velocity of nickel is roughly 10\% slower and that the slowest
hydrogen moves slightly faster in our models. We attribute these
differences to the numerical resolution, which was slightly higher in
the \citet{Hammeretal10} models in polar regions. The results of two
specific 3D simulations, which we performed with 1$^\circ$ and
2$^\circ$ angular resolution, suggest that an angular resolution of
2$^\circ$ is sufficient for our goals, but lower angular resolution
tends to cause the mentioned velocity differences.  Our primary
intention was neither to determine details of the small-scale
structure nor the precise peak velocities of the ejecta, but to study
the dependence of the ejecta morphology on explosion energy and the
progenitor.

Our simulations show that the evolution of the SN shock and the
neutrino-heated ejecta within the stellar envelope is complex,
involving several types of hydrodynamic instabilities. After crossing
a composition interface, the deceleration of the SN shock leads to the
formation of RT unstable dense shells and the formation of a strong
reverse shock in the case of the He/H interface.  The neutrino-heated
ejecta propagating at some distance behind the SN shock may penetrate
these dense shells causing large-amplitude perturbations there, if
they are fast enough to catch up with the shell.  Whether or not an
interaction occurs between the reverse shock (propagating inward in
mass) and the fastest fraction of the iron-nickel-rich ejecta is
determined by the velocity of the latter relative to the SN shock.

We find that the relative velocity depends in turn on three important
features of the density profile of the progenitor star: the
compactness of the C+O core, the density structure of the helium
shell, and the density gradient at the He/H interface. We confirmed
this fact by performing three additional 3D simulations where we
replaced the outer C/O core, the helium and hydrogen shell of the two
RSG progenitors and of the 20\,$M_\odot$ BSG progenitor by that of the
15\,$M_\odot$ BSG progenitor.  These 3D simulations demonstrated: (i)
the iron/nickel-rich plumes do not have enough time to grow into
finger-like structures before they are slowed down by the reverse
shock in the more compact BSG stars because of the earlier onset of SN
shock deceleration and reverse shock formation, and (ii) both a
shallow density profile inside the helium shell and a small density
decrease at the He/H interface reduce the relative velocity between
the ejecta and the SN shock, and hence help the fastest
iron/nickel-rich ejecta to avoid the interaction with the reverse
shock.

Considering the results of our 3D simulations performed with four
different progenitor stars, we can categorize the morphology of the
iron/nickel-rich ejecta hours after the onset of the explosion into
three types.

For RSG progenitors (W15 and L15), the late time morphology of the
iron/nickel-rich ejecta is characterized by small RT fingers bunched
together into a few groups whose angular positions agree with those of
the few largest, fast-rising plumes of neutrino-heated ejecta, which
were generated by hydrodynamic instabilities during the shock revival
phase. This type of ejecta morphology is mainly due to the strong
acceleration of the SN shock at the He/H interface and its subsequent
even stronger deceleration inside the H envelope of the RSG
progenitors. The decelerating SN shock gives rise to a strongly RT
unstable stratification near the interface and the formation of a
strong reverse shock, while the acceleration of the SN shock increases
the spatial separation of SN shock and iron-nickel-rich ejecta. The
latter fact also implies that the distance between the reverse shock
and the ejecta is relatively large in the RSG models, \ie there is
more time for the fast plumes of iron/nickel-rich ejecta to grow into
extended fingers before they eventually are decelerated by the reverse
shock.

The second type of morphology is produced by models based on the N20
progenitor star, where the iron/nickel-rich ejecta can be described by
fragmented roundish structures, which are the result of two facts.
Firstly, the iron/nickel-rich ejecta propagate with a lower velocity
than the SN shock. Secondly, the only moderate drop of the density in
this model at the He/H interface causes the SN shock not to accelerate
much at this interface. The combination of both facts implies little
time for large non-spherical deformations of the iron/nickel-rich
ejecta to develop before they are compressed by the reverse shock,
which forms ahead of the ejecta near the He/H interface.

Models based on the B15 progenitor star exhibit a third type of
morphology. The iron/nickel-rich ejecta appear as a few, very distinct
elongated RT fingers penetrating quite large distances into the
fast-moving hydrogen envelope. These stretched RT fingers are those
parts of the iron/nickel-rich ejecta that propagated fast enough to
avoid a deceleration by the reverse shock. This was possible for this
progenitor, because the SN shock experienced a strong deceleration
inside its helium shell and only a small acceleration at its He/H
interface.

For all simulated models, except the one based on the N20 progenitor,
we find that there is a clear correlation between the asymmetries of
the iron/nickel-rich ejecta at late times and the early-time
asymmetries resulting from hydrodynamic instabilities generated during
the onset of the SN explosion. However, this is not too surprising,
because the growth of RT instabilities at composition interfaces in
our simulations is in fact seeded by large-scale, large-amplitude
perturbations caused by the blast asymmetries at early times rather
than by small-scale random perturbations. The latter perturbations
were used in simulations which cover the first second of the SN
explosion assuming spherical symmetry.  On the other hand, in the
simulated model based on the N20 progenitor, the early-time,
large-scale asymmetries of the iron/nickel-rich ejecta are greatly
diminished by the interaction between the ejecta and the reverse shock
occurring already in the helium shell of the star. Thus, the
iron/nickel-rich ejecta lose memory of the early-time asymmetries, \ie
we find no clear correlation of late-time and early-time asymmetries
of the iron/nickel-rich ejecta for this progenitor.

Concerning the maximum velocities of the iron/nickel-rich ejecta we
find that these vary between 3700 and 4400\,km\,s$^{-1}$ in our RSG
models, while they do not exceed 2200\,km\,s$^{-1}$ in our 20 solar
mass BSG model and 3400\,km\,s$^{-1}$ in our 15 solar mass BSG model.

Hydrogen is mixed less deeply into the metal core in the BSG models
than in the RSG ones so that there is no significant amount of
hydrogen at a velocity lower than 1000\,km\,s$^{-1}$ in the BSG
models, except for a low-velocity tail down to 500\,km\,s$^{-1}$ in
the model based on the B15 BSG progenitor. This difference between BSG
and RSG progenitors can be understood from the growth rates of the RT
instabilities, which are larger in RSG progenitors because these
progenitors possess a steeper density gradient at the H/He interface
than BSG progenitors, giving rise to a more strongly RT unstable layer
after the passage of the SN shock. Hence, there will be more and more
extended mixing of hydrogen into the inner parts of the ejecta, \ie to
lower velocities, in RSG progenitors.

In our 3D explosion models, we seeded the initial development of
nonradial instabilities in the SN core in the neutrino-heated region
by small (0.1\%) random (velocity) perturbations on the grid
scale. Hence, we ignored the possible existence of larger-scale and
larger-amplitude inhomogeneities in the progenitor core before the
onset of the collapse, which may arise due to vigorous convection
during pre-collapse oxygen and/or silicon burning
\citep{ArnettMeakin11}. Corresponding nonradial mass 
motions could affect explosion asymmetries, neutron-star kicks, and 
nickel mixing before core collapse as well as during the explosion.
In addition, \citet{CouchOtt13} showed that the revival of the stalled
supernova shock by neutrino heating could be triggered by nonradial
flows in the progenitor core, because perturbations in the Si/O
burning shells, when advected through the accretion shock, enhance
nonradial mass motions and turbulent pressure in the postshock 
region \citep{CouchOtt14,MuellerJanka14}. While such effects might be
of relevance in the context discussed in our paper, it is unclear
whether the relatively large initial perturbations that are needed to
make a sizable impact are realistic \citep{MuellerJanka14}. A
quantitatively meaningful assessment is currently not possible because
the perturbation pattern and amplitude in 3D stars prior to core
collapse must still be determined by multi-dimensional simulations of
the late stages of stellar evolution.

Concerning the final ejecta morphology of the supernova remnant, there
are two further potentially important effects that are not covere by
our present simulations.  Firstly, RT instabilities will occur at the
interface separating matter in the shocked stellar envelope from
shocked circumstellar matter after shock breakout
\citep{{Gawryszczaketal10}}.  This interface becomes RT unstable when
the supernova shock is decelerating in the circumstellar matter, \ie
when the less dense circumstellar matter is moving slower than the
shocked and denser matter of the stellar envelope ejected in the SN
\citep{Chevalieretal92}. Secondly, the energy release by the
radioactive decay of nickel into cobalt and iron can cause local
heating of the ejecta, which can modify the ejecta morphology. The
actual importance of both effects can only be clarified by extending
our simulations to much later times, which we plan to do next.


\begin{acknowledgements}
%
This work was supported by the Deutsche Forschungsgemeinschaft through
the Transregional Collaborative Research Center SFB/TR~7
``Gravitational Wave Astronomy'' (wwwsfb.tpi.uni-jena.de), the Cluster
of Excellence EXC~153 ``Origin and Structure of the Universe''
(www.universe-cluster.de), and by the EU through ERC-AdG
No.\ 341157-COCO2CASA. Computer time at the Rechenzentrum Garching
(RZG) is acknowledged.
\end{acknowledgements}

\bibliographystyle{aa}
\bibliography{longtime}

\end{document}